\documentclass[a4paper,11pt]{article}
\pdfoutput=1 

\usepackage{jcappub} 
\usepackage{aas_macros} 

\usepackage{fontawesome}
\usepackage[T1]{fontenc}
\usepackage{amsmath,bm}
\usepackage[nameinlink,capitalise]{cleveref}

\newcommand{\fNL}{f_{\rm NL}}

\newcommand{\aMpch}{h/\mathrm{Mpc}}
\newcommand{\dd}{\mathrm{d}}
\newcommand{\hi}{\textsc{Hi}}
\usepackage{hyperref}
\usepackage{url} 
\usepackage{graphicx}
\usepackage{subcaption}
\usepackage{booktabs}
\usepackage{float}
\DeclareMathOperator{\sinc}{sinc}

\title{Unbiased analysis of primordial non-Gaussianity: the multipoles of the full relativistic power spectrum}

\author[a,1]{Chris Addis,\note{Corresponding author.}}
\author[b]{S\^ecloka L. Guedezounme,}
\author[a]{Jessie Hammond,}
\author[a,b]{Chris Clarkson,}
\author[c,d]{Federico Montano,}
\author[c,d,e,b]{Stefano Camera,}
\author[f,b]{Sheean Jolicoeur,}
\author[b,h]{and Roy Maartens}

\affiliation[a]{Astronomy Unit, School of Physical \& Chemical Sciences, Queen Mary University of London, London E1 4NS, UK}
\affiliation[b]{Department of Physics \& Astronomy, University of Western Cape, Cape Town 7535, South Africa}
\affiliation[c]{Dipartimento di Fisica, Universit\`a degli Studi di Torino, Via P.\ Giuria 1, 10125 Torino, Italy}
\affiliation[d]{INFN -- Istituto Nazionale di Fisica Nucleare, Sezione di Torino, Via P.\ Giuria 1, 10125 Torino, Italy}
\affiliation[e]{INAF -- Istituto Nazionale di Astrofisica, Osservatorio Astrofisico di Torino, Strada Osservatorio 20, 10025 Pino Torinese, Italy}
\affiliation[f]{VIT Mauritius, Uniciti International Education Hub, 72448, Pierrefonds, Mauritius}
\affiliation[h]{National Institute for Theoretical \& Computational Sciences, Cape Town 7535, South
Africa}

\emailAdd{c.l.j.addis@qmul.ac.uk}
\emailAdd{seclokaguedezounme@gmail.com}
\emailAdd{j.r.hammond@qmul.ac.uk}
\emailAdd{chris.clarkson@qmul.ac.uk}
\emailAdd{federico.montano@unito.it}
\emailAdd{stefano.camera@unito.it}
\emailAdd{jolicoeursheean@gmail.com}
\emailAdd{rmaartens@uwc.ac.za}

\abstract{A major goal of ongoing and future cosmological surveys of the large-scale structure is to measure local type primordial non-Gaussianity in the galaxy power spectrum through the scale-dependent bias. General relativistic effects have been shown to be degenerate with this measurement and therefore one needs to consider a non-Newtonian approach. In this work, we develop a consistent framework to compute integrated effects, including lensing convergence, time delay, and integrated Sachs--Wolfe, along with the local relativistic projection and wide-separation corrections in the multipoles of the power spectrum.
We show that, for a \textit{Euclid}-like H$\alpha$-line galaxy survey and a MegaMapper-like Lyman-break galaxy survey, ignoring these effects leads to a bias on the best fit measurement of the amplitude of primordial non-Gaussianity, $\fNL$, of around $ 3\sigma$ and $ 20 \sigma$ respectively. When we include these corrections, the uncertainty in our knowledge of the luminosity function leads to further uncertainty in our measurement of $\fNL$. However, we show that this degeneracy can be partly mitigated by using a bright-faint multi-tracer analysis, where the observed galaxy sample is subdivided into two separate populations based on luminosity. This  provides a $15$--$20\%$ improvement on the forecasted constraints of local type $\fNL$. In addition, we present a novel calculation of the full multi-tracer covariance with the inclusion of wide-separation corrections. All of these results are implemented in the \textit{Python} code \textsc{CosmoWAP}.}

\begin{document}
\maketitle
\flushbottom

\section{Introduction}
\label{sec:intro}

Over the past few decades, there have been rapid experimental advances in our surveys of the large-scale structure of the universe. The next generation of spectroscopic galaxy surveys such as the ongoing Dark Energy Spectroscopic Instrument (DESI) \cite{desi} and \textit{Euclid} \cite{euclid_2} surveys as well as the future planned missions of SPHEREx \cite{Dore_SPHEREx}, Nancy Grace Roman Space Telescope (WFIRST) \cite{wfirst} and MegaMapper \cite{MegaMapper} will go a step further, providing an even higher number density of detected objects over larger volumes and redshift ranges \citep[see, for example, Figure 1 of][]{Schlegel_2022}. With this increased statistical power, it becomes critical to fully consider previously subdominant theoretical systematics.

A major goal for these surveys is to constrain Primordial non-Gaussianity (PNG), which can provide insight on the physics of the early universe \cite{inflation_2022}. Currently, the tightest constraints on local PNG, determined via the parameter $\fNL$, come from observations of the cosmic microwave background (CMB) by \textit{Planck} \cite{Akrami_2020} who found $\fNL=0.9 \pm 5.1$. Handily, local type PNG leaves an imprint in large-scale structure (LSS) not just in higher-order statistics, such as the bispectrum, but also in the power spectrum through the scale-dependent halo bias \cite{Dalal_2008,Slosar_2008,Matarrese_2008}. Next-generation galaxy surveys should provide competitive constraints with $\mathcal{O}(1)$ uncertainties using these observables \cite{Giannantonio_2012,FontRibera_2014,Amendola_2018,Mueller_2019}.

Observations of LSS on our past light-cone are subject to distortions along our line-of-sight (LOS); to account for this projection along our LOS, the standard Newtonian `Kaiser' \cite{Kaiser_1987,Hamilton_1998} treatment of redshift space distortions (RSD), which takes into account the main contribution from the peculiar velocity of the source, has usually been sufficient. However, with next-generation surveys probing larger scales and with increased precision, 
this simplified analysis could lead to biased conclusions. Linear relativistic corrections to galaxy number counts such as the relativistic Doppler and gravitational redshift corrections have been known for some time \cite{Macdonald_2009,Yoo_2010,Bonvin_2011,Challinor_2011}. As they are generally more prevalent on large scales, they have been shown to be degenerate with the local $\fNL$ signal both in the angular \cite{Bruni_2012,Camera_2015,Alonso_2015,Alonso_2015b,Viljoen_2021} and the 3D Fourier \cite{Jeong_2012,Wang_2020,Maartens_2021,Foglieni_2023,Jolicoeur_2024,Guedezounme:2024pbj} power spectrum and can lead to biases in our constraints on $f_{NL}$ for future surveys at $\mathcal{O}(1)$ if they are neglected. More recent analyses \cite{Wang_2020,Guedezounme:2024pbj} have also accounted for the sensitivity of these effects to the luminosity threshold for the survey through evolution, $b_e$, and magnification, $Q$, biases (\citealp[see][]{Maartens_2022} for an overview). The uncertainty in our knowledge of the luminosity function for a given survey therefore leads to greater uncertainty in the measurement of $\fNL$. 

Alternatively, detecting these relativistic projection effects themselves has attracted interest, as they are sensitive to the metric potentials, $\Phi$ and $\Psi$, in a different way to the Kaiser terms and, as such, can be used to construct tests of general relativity \cite{Lombriser_2013,Bonvin_2018,Bonvin_2020,Castello_2022,Blanco_2023,Castello_2025}. When detecting these effects, it is advantageous to use the multipoles of the 3D power spectrum (or $2$-point correlation function, $2$-PCF) rather than the angular equivalent, as not only does it allow us to access information from the full 3D spectroscopic map, but one can also isolate the leading order relativistic contribution in the odd-parity multipoles in a multi-tracer analysis \cite[e.g.][]{Macdonald_2009,Bonvin_2014_b}. Within the bright-faint split approach, as proposed by \cite{Bonvin_2014,Bonvin_2016,Gaztanaga_2017}, a single galaxy population is split according to luminosity into two independent populations, which can lead to improved constraining power on relativistic effects in the multi-tracer power spectrum \cite{Bonvin_2016,Gaztanaga_2017,Montano_2024a,Montano_2024b,Novara_2025,Montano_2026}. Since working with the multipoles of the Fourier power spectrum is also common for PNG analysis, developing a consistent framework to include these relativistic effects is essential.

Multi-tracer techniques have also been proposed to improve observational constraints on $\fNL$ by using highly biased samples \cite{Seljak_2009,Barreira_2023,Karagiannis_2024}, as for two tracers, $A$ and $B$, the constraining power is related to the difference in $b_1^X$, the linear bias, and $b^X_{\phi}$, the non-Gaussian bias, between both tracers. In the cosmic variance limit, this simplifies to $\propto |b_1^B b_{\phi}^A - b_1^A b_{\phi}^B|$ (or just $\propto |b_1^A - b_1^B|$ under the universality relation \cite{Slosar_2008, Barreira_2023}). Therefore, this motivates developing a fully relativistic multi-tracer framework whereby we can use highly biased tracers to not only maximise the $\fNL$ signal but also to better constrain our uncertainties in the luminosity function.

While integrated corrections (lensing convergence, time delay, and integrated Sachs-Wolfe, ISW) to the galaxy number counts have been commonly included in angular analyses, their inclusion in the Fourier power spectrum is less straightforward. Our goal in this paper is to consistently include these contributions in a $P(k)$ analysis of $\fNL$. One approach is to derive them starting from the $2$-PCF, \cite{Tansella_2018,Castorina_2022,Foglieni_2023} but this can be cumbersome and numerically expensive. The exact nature of these integrated effects in the 3D power spectrum, which, beyond the Limber approximation \cite{Limber_1953}, break local translation invariance\footnote{See \cref{sec:los_depend} for a full discussion and explanation of this.} in our local 2-point function is subtle, and has not yet been fully explored. Therefore, we adopt and expand upon the formalism presented in \cite{Noorikuhani:2022bwc}, where integrated effects can be directly computed in the Cartesian Fourier space, to allow for integrated effects to be correctly modelled at wide-separations. Further, we extend this framework to include integrated cross integrated, (I\texttimes I), contributions to the power spectrum multipoles.

The multipoles of the power spectrum can be measured from a given survey using the efficient and simple `Yamamoto' estimator \cite{Yamamoto_2005,Beutler_2014,Bianchi_2015,Scoccimarro_2015}. However, in order to compute this quantity theoretically it is convenient to define the `local' 2-point function/power spectrum, which is defined for a single LOS and therefore, one can define the local multipoles \cite{Zaroubi_1996,Hamilton_1998,Scoccimarro_2015,Castorina_2018, Addis_2025}\footnote{One can also consider this as: without translation invariance, the Fourier modes are no longer statistically independent. The position-dependent line-of-sight induces geometric mode-mode coupling. Consequently, the power spectrum is no longer the Fourier conjugate of the two-point correlation function, and the field's statistical properties are spread into off-diagonal terms of the covariance matrix rather than being contained solely in the diagonal.}. To include the full LOS information in this local region, where we are correlating a pair of points, one then typically relies on a perturbative expansion \cite{Yoo_2015,Reimberg_2016,Beutler_2020,Noorikuhani:2022bwc}. These corrections contain the part of the signal which breaks translation invariance in our local statistic and as such become relevant when considering large pair separations; therefore they are termed wide-separation corrections (WS). Here, the term wide-separation describes the corrections deriving from this symmetry breaking both angularly, due to RSD (termed wide-angle corrections), and radially, due to evolution (termed radial evolution corrections). As they arise on large scales, these corrections are degenerate with signals of local PNG and relativistic effects \cite{Noorikuhani:2022bwc,Paul:2022xfx,Beutler_2020,Guedezounme:2024pbj,Jolicoeur_2024} and therefore must be considered for a consistent analysis on large scales. Indeed, with Stage-IV surveys, all of these corrections become relevant not only due to the increased precision, but also as these surveys map increased volume, we gain access to ultra-large-scale modes.

At very large separations and low redshifts, this perturbative wide-separation expansion breaks down \cite[e.g.][]{Benabou_2024}, and alternative bases, such as the angular or the spherical Fourier-Bessel (sFB) power spectrum -- which retains the full radial information -- are more appropriate. Including relativistic terms in the sFB formalism is non-trivial \citep[see][for recent work]{Wen_2024,Semenzato_2025}. However, for the majority of ongoing and upcoming galaxy surveys, the power spectrum multipoles remain a practical choice of basis for analysis.

Thus this work expands on previous work, notably \cite{Noorikuhani:2022bwc,Addis_2025,Guedezounme:2024pbj}, by providing a consistent framework for computing these light-cone effects (including all relevant relativistic and wide-separation effects) in the multipoles of the Fourier power spectrum and being explicit in how this is connected to observable quantities (cf.\ \cref{sec:local_pk}). In particular, we focus on the integrated contributions and their contribution to the local power spectrum (\cref{sec:local_int}), and then construct forecasts (\cref{sec:forecast_setup}), on the impact of these effects on PNG for these effects for a few different survey setups (\cref{sec:survey_details}). We show that neglecting these effects can lead to a bias on the inference of local $\fNL$ and that when including these effects, the uncertainty of our luminosity function leads to a greater uncertainty in our determination of $\fNL$. We also show how constraints on $\fNL$ can then be improved with a multi-tracer approach by better constraining our evolution and magnification bias. We present a novel calculation of the analytic multi-tracer power spectrum multipole covariance on large-scales, with the inclusion of wide-separation corrections for the first time, \cref{sec:covariance}. All of these results, including forecast code, is included in the \textit{Python} code \textsc{CosmoWAP} \href{https://github.com/craddis1/CosmoWAP}{\faGithub} \cite{Addis_2025,Addis_inprep,cosmowap} with companion \textit{Mathematica} notebooks included in \textsc{MathWAP} \href{https://github.com/craddis1/MathWAP}{\faGithub} \cite{mathwap}. Therefore, readers focused on the underlying nature of integrated effects should begin with \cref{sec:local_pk}, whereas those primarily interested in the implications for PNG analysis may skip to \cref{sec:bias}.

\section{Local Power Spectrum}\label{sec:local_pk}

The following section first defines the power spectrum multipole estimator and its connection to the local power spectrum, then expresses the full galaxy number counts (focusing on integrated contributions) in  \cref{sec:number_counts} and how these integrals can be computed for $P(k)$ in \cref{sec:local_int}. \cref{sec:los_depend} discusses the implication of breaking translation invariance in the local power spectrum and then \cref{sec:local_png} outlines the local PNG contribution to the power spectrum. Finally, \cref{sec:survey_pk} explicitly connects the local power spectrum to the survey averaged power spectrum that we observe.

Throughout this work, we adopt the following Fourier notation
\begin{equation}
\int_{\bm{x}_{12}} = \int \dd^3 \bm{x}_{12} \text{ and } \int_{\bm{k}} = \int \frac{\dd^3 \bm{k}}{(2\pi)^3},
\end{equation}
such that we can define our Fourier and inverse Fourier transforms, respectively as
\begin{equation}
F(\bm{k})= \int_{\bm{x}}\, {\rm e}^{-i \bm{k}\cdot \bm{x}}\, f(\bm{x}) \text{ and } f(\bm{x})= \int_{\bm{k}}\, {\rm e}^{i \bm{k}\cdot \bm{x}}\, F(\bm{k}).
\end{equation}
We also refer to `local' (non-integrated) contributions, e.g. Kaiser and local relativistic terms, and `local' power spectrum - the region in which we do our wide-separation expansion.

Starting from the Yamamoto estimator \cite{Yamamoto_2005,Beutler_2014} for the multipoles of the power spectrum
\begin{equation}\label{eq:yamamoto}
\hat{P}^{ab}_{\ell}(k) =  \frac{2 \ell+1}{N_{k}}\int_{\mathcal{S}_1} \dd^3 \bm{q}_1 \int_{\mathcal{S}_2} \dd^3 \bm{q}_2  \,\delta_D(\bm{q}_{1}+\bm{q}_{2}) \int_{\bm{x}_1,\bm{x}_2} {\rm e}^{-i\,(\bm{q}_1\cdot\bm{x}_1+\bm{q}_2\cdot\bm{x}_{2})}\,\Delta^{a}_W(\bm{x}_1)\Delta^{b}_W(\bm{x}_2)\mathcal{L}_{\ell}(\hat{\bm{q}}_1 \cdot \hat{\bm{d}}),
\end{equation}
where our LOS, $\bm{d}$, is varying across the survey but is set to an `endpoint' LOS (either $\bm{x}_1$ or $\bm{x}_2$), for each local 2-point function region, such that the integrals in the estimator become separable and therefore computable with Fast Fourier Transforms (FFTs) \cite{Bianchi_2015,Scoccimarro_2015}. Here, $\Delta_W(\bm{x})=\Delta(\bm{x})\,W(\bm{x})$ represents the windowed field for a given survey window function, $W(\bm{x})$. The Fourier integrals represent discrete sums, $\int_{\mathcal{S}_i} \dd^3 \bm{q}_i \, F(\bm{q}_i) \equiv \sum_{\mathcal{S}_i} F(\bm{q}_i)$, over thin $k$-space shells, $\mathcal{S}_i$, of width $\Delta k$, centred at $k_i$, such that $\mathcal{S}_i \equiv \mathcal{S}(k_i|\Delta k)$ is the region of $k$-magnitudes contained by a given $k_i$-bin, $k_i - \Delta k/2 \leq k < k_i + \Delta  k/2$. Finally, $N_k$ represents the number of $k$-modes in each bin,
\begin{equation}
N_k = k_{\rm f}^{-3} \int_{\mathcal{S}_1} \dd^3 \bm{q}_1 \int_{\mathcal{S}_2} \dd^3 \bm{q}_2 \, \delta_D(\bm{q}_{1}+\bm{q}_{2}),
\end{equation}
where $k_{\rm f}$ is the fundamental frequency of the survey box and $N_k$ simplifies to $4\,\pi\,k^2\,\Delta k \,/k_{\rm f}^{3} $ in the continuous limit. For $\ell=0$, this then reduces to the FKP estimator \cite{Feldman_1994}.
If we consider the unwindowed `true' theory\footnote{This significantly simplifies our forecasts but we note that we could include the impact of the convolution with the survey window in the standard way -- by using Hankel transforms to include the survey window function multiplicatively in the 2-point function, which can then be related to the windowed multipoles of the power spectrum using another Hankel transform \cite{Wilson_2017}. }, then as shown in \cite{Scoccimarro_2015}, this estimator can be related to the definition of the `local' power spectrum.

Using the change of variables
\begin{equation}
\int \dd^3 \bm{x}_1 \int \dd^3 \bm{x}_2 \rightarrow \int \dd^3 \bm{d} \int \dd^3 \bm{x}_{12}
\end{equation}
where $\bm{x}_{12}=\bm{x}_1-\bm{x}_2$, and splitting the integration in each shell
\begin{equation}\label{eq:change_variables}
\int_{\mathcal{S}_i} \dd^3 \bm{q}_i = \int^{k_i + \Delta k/2}_{k_i - \Delta k/2} \, \dd q_i \,q_i^2 \int\dd\Omega_{q_i},
\end{equation}
where $\dd\Omega_q$ is the infinitesimal solid angle, the corresponding theoretical multipoles can be expressed in the familiar form:
\begin{equation}\label{eq:survey_pk}
\hat{P}^{ab}_{\ell}(k)=(2\,\ell +1)\int \frac{\dd \Omega_k}{4\,\pi}\int \dd^3 \bm{d} \, \left[ P^{ab}_{\rm loc}(k,\mu;d) \, \mathcal{L}(\mu)\right],
\end{equation}
where $\mu=\hat{\bm{k}}\cdot\hat{\bm{d}}$ specifies the LOS orientation of the $k$-vector, and we define the local power spectrum
\begin{equation}\label{eq:local_pk}
P^{ab}_{\text{loc}}(k,\mu ;d)=\int_{\bm{x}_{12}} {\rm e}^{-i\,\bm{k} \cdot \bm{x}_{12}} \,  \langle\Delta^a(\bm{x}_1)\,\Delta^b(\bm{x}_2)\rangle,
\end{equation}
which is defined at a single position, $\bm{d}$. This is then the Fourier transform of our `local' 2-point function $\langle \Delta(\bm{x}_1) \Delta(\bm{x}_2)\rangle \equiv \xi_{\rm loc}(\bm{x}_1,\bm{x}_2) \equiv \xi_{\rm loc}(\bm{x}_{12},\bm{d})$ which is simply ensemble average of our 2-point function over our local region; the common key assumption that makes the local power spectrum an easily computable quantity, is translation invariance, such that the statistics of the underlying dark matter density field at $\bm{x}_1$ and $\bm{x}_2$ are the same\footnote{This is equivalent to assuming the local plane-parallel, constant redshift limit.}. However, with large separations and integrated contributions computing the full local power spectrum that corresponds to the estimator is more subtle and we discuss this in the next section.

\subsection{Galaxy Number Counts and the Local Two-Point Function}\label{sec:number_counts}

\begin{figure}[ht]
\centering
\includegraphics[width=0.4\linewidth]{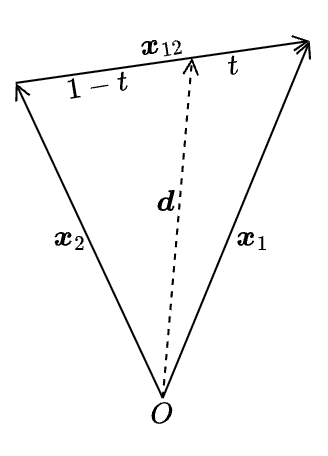}
\caption{Geometry of the local $2$-point function described by a single LOS, $\bm{d}$ where $t$ describes its choice along the vector $\bm{x}_{12}$. $t=0$ and $t=1$ are referred to as `endpoint' LOS while another common choice is the `midpoint' LOS, $t=1/2$.}
\label{vector_diagram}
\end{figure}

The number of galaxies in a given survey volume can be expressed in terms of standard terms `S' -- linear bias plus linear `Newtonian' RSD, which we describe as the Kaiser term, the non-integrated (local/source) relativistic projection `NI', and integrated `I' contributions\footnote{Note that we ignore the effect of observer terms in our analysis - e.g. \cite{BahrKalus_2021}.}
\begin{equation}\label{eq:NumCount}
\Delta(\bm{x}) = \Delta^{\rm S}(\bm{x}) + \Delta^{\rm NI}(\bm{x}) + \Delta^{\rm I}(\bm{x}).
\end{equation}

These position space quantities can then be defined from the simple Fourier space kernels,

\begin{equation} \label{IntNumCount}
\Delta(\bm{x}) = \int_{\bm{q}} \left[\mathcal{K}^{\rm S}(\bm{q},\bm{x}) + \mathcal{K}^{\rm NI}(\bm{q},\bm{x})\right]\,\delta_0(\bm{q}) + \int_0^{x} \dd r\int_{\bm{q}}  \, {\rm e}^{i\,\bm{q} \cdot \bm{r}} \,\mathcal{K}^{\rm I}(\bm{q},\bm{x}, \bm{r})\,\delta_0(\bm{q}),
\end{equation}
where $\delta_0$ is the linear matter overdensity, $\mathcal{K}^{\rm S}(\bm{q},\bm{x})$ is the standard RSD Kaiser kernel \cite{Kaiser_1987}, $\mathcal{K}^{\rm NI}(\bm{q},\bm{x})$, is the linear local relativistic projection kernel \cite{Yoo_2010,Bonvin_2011,Challinor_2011}; these kernels, along with the local PNG contribution, are included in \cref{ap:more_kernels} for completeness. The integrated contribution to the number counts contains the integrated Sachs-Wolfe effect (ISW), Shapiro time delay (TD), and lensing effects (L) instead are
\begin{subequations}\label{eq:int_kernel}
\begin{align}
\mathcal{K}^{\rm L}(\bm{q},\bm{x}, r)& = 3\,D(r)\,(\mathcal{Q} - 1)\,\Omega_m(r)\,\mathcal{H}^2(r)\,\frac{(x - r)\,r}{x} \,\left[ 1 - (\hat{\bm{q}} \cdot \hat{\bm{x}})^2 + 2\,i\,\frac{(\hat{\bm{q}} \cdot \hat{\bm{x}})}{r\,q} \right],
\\
\mathcal{K}^{\rm TD}(q,x, r) & = 6\,D(r)\,(\mathcal{Q} - 1)\,\frac{\Omega_m(r)\,\mathcal{H}^2(r)}{q^2\,x},\\
\mathcal{K}^{\rm ISW}(q,x, r) & = 3\,D(r)\,\left[ b_e - 2\,\mathcal{Q} + \frac{2\,(\mathcal{Q} - 1)}{x\, \mathcal{H}} - \frac{\mathcal{H}'}{\mathcal{H}^2} \right]\, \frac{\Omega_m(r)\,\mathcal{H}^3(r)\,(f(r) - 1)}{q^2},
\end{align}
where $\mathcal{H}$ the conformal Hubble parameter, $D$ is the linear growth function, $f$ is the linear growth rate and $'$ refers to a derivative with respect to conformal time ($\mathcal{Q}$ and $b_e$ are defined below). Functions of $r$ here are therefore evaluated along the integrand from observer to source while the other functions are evaluated at the redshift of the source, $x$, e.g. $\mathcal{Q}\equiv\mathcal{Q}(x)$; note that the integrand itself is pure cosmology and independent of tracer but has bias dependence only in the prefactors. The total integrated contribution to the galaxy number counts can then be written as
\begin{equation}
\mathcal{K}^{\rm I}(\bm{q},\bm{x}, \bm{r}) = \mathcal{K}^{\rm L}(\bm{q},\bm{x}, r) + \mathcal{K}^{\rm TD}(q,x, r) + \mathcal{K}^{\rm ISW}(q,x, r).
\end{equation}
\end{subequations}

We could also re-express these kernels in terms of scalar quantities if $\bm{x}\equiv\bm{d}$, such that these kernels can be expressed in terms of $q, \, \mu, \, d, \, r$.
These kernels are dependent on the luminosity function through $\mathcal{Q}$ and $b_e$ which are defined with respect to the luminosity cut for a flux limited survey,

\begin{equation}
b_e = -\frac{\partial \, {\rm ln} \,n_g}{\partial \, {\rm ln}(1+z)}\bigg|_c, \, \, \mathcal{Q}= - \frac{\partial \, {\rm ln} \, n_g}{\partial \, {\rm ln} \, L}\bigg|_c,
\end{equation}
where $L$ is the luminosity, $n_g$ is the co-moving number density and $|_c$ refers to an evaluation at the flux cut; see \cite{Maartens_2022} for a more detailed overview.

Therefore, the local $2$-point function, can be expressed an expansion over all contributions such that,

\begin{equation}\label{eq:local_2point}
\xi_{\rm loc}(\bm{x}_1,\bm{x}_2) \equiv \langle \Delta(\bm{x}_1)\,\Delta(\bm{x}_2) \rangle = \sum_{\alpha, \beta \in \{\rm S, NI, I\}} \langle \Delta^\alpha_g(\bm{x}_1)\,\Delta^\beta_g(\bm{x}_2) \rangle.
\end{equation}

In the next section we derive the I\texttimes I contribution, which has been previously been neglected, to the local power spectrum. A brief summary of the I\texttimes S contribution in our framework is also included in \cref{ap:local_IS}. The local relativistic, Kaiser and local PNG terms are also computed \citep[e.g.][]{Beutler_2020,Noorikuhani:2022bwc} with the inclusion of wide-separation (wide-angle and radial evolution) corrections.

\subsection{Integrated Contributions to the Local Power Spectrum}\label{sec:local_int}

The I\texttimes I contribution to the local $2$-point correlation can be expressed as

\begin{equation}
\begin{split}
\langle \Delta^{\rm I}(\bm{x}_1)\,\Delta^{\rm I}(\bm{x}_2) \rangle &= \int_{\bm{q}} \int_0^{x_1} \dd r_1 \int_0^{x_2} \dd r_2 \, {\rm e}^{ i\,\bm{q} \cdot (\bm{r}_1-\bm{r}_2)}\,\mathcal{K}^{\rm I}(\bm{q},\bm{x}_1, r_1)\,\mathcal{K}^{\rm I}(-\bm{q},\bm{x}_2, r_2)\,P (q)\;,
\end{split}
\end{equation}
where $P(k)$ is our linear matter power spectrum and we have a double integral from the observer to source for $r_1$ and $r_2$. Note the negative $\bm{q}$ in the $\bm{x}_2$ kernel, arising as $\bm{q}_2 \rightarrow - \bm{q}$, which impacts the odd $\mu$ terms.

Similar to as one would in the non-integrated case, we can then switch our coordinates using $\bm{r}_1=\frac{r_1}{x_1}\,\bm{x}_1, \, \bm{r}_2=\frac{r_2}{x_2}\,\bm{x}_2$ and $\bm{x}_1 = \bm{d} + t\,\bm{x}_{12}$, $\bm{x}_2 = \bm{d} - (1-t)\,\bm{x}_{12}$ to re-express the exponential in terms of $r_1,\, r_2,\,\bm{d}$ and $\bm{x}_{12}$ as
\begin{equation}
\begin{aligned}
{\rm e}^{i\,\bm{q} \cdot (\bm{r}_{1}-\bm{r}_{2})} = &\, {\rm e}^{i\,\bm{q} \cdot (\frac{r_1}{x_1}\,\bm{x}_{1}-\frac{r_2}{x_2}\,\bm{x}_{2})}\\
=&\,{\rm e}^{i\,(\frac{r_1}{x_1}-\frac{r_2}{x_2})\,(\bm{q} \cdot \bm{d})}{\rm e}^{i\,(\frac{r_1}{x_1}\,t+(1-t)\,\frac{r_2}{x_2})\,(\bm{q} \cdot \bm{x}_{12})}.
\end{aligned}
\end{equation}

It is then convenient to introduce the new integration variables, $y_1 = r_1/x_1$ and $y_2=r_2/x_2$, such that the local 2-point function can be written as
\begin{equation}
\begin{split}\label{eq:full_2point}
\langle \Delta^{\rm I}(\bm{x}_1)\,\Delta^{\rm I}(\bm{x}_2) \rangle & = \int_{\bm{q}} \int_0^{1} \dd y_1 \int_0^{1} \dd y_2 \, (x_1)(x_2) \, {\rm e}^{ i\,\left(y_1-y_2 \right)\,(\bm{q} \cdot \bm{d})} \, {\rm e}^{ i\,\left(y_2 + t(y_1-y_2)\right) (\bm{q} \cdot \bm{x}_{12})} \\
& \hspace{10em} \times \mathcal{K}^{\rm I}(\bm{q}, \bm{x}_1, x_1 \, y_1)\,\mathcal{K}^{\rm I}(-\bm{q}, \bm{x}_2, x_2 \,y_2)\,P (q).
\end{split}
\end{equation}
To form the local power spectrum with a single Fourier transform over $\bm{x}_{12}$, we then perform our wide-separation series expansion in $\epsilon = x_{12}/d$ about the point $\bm{d}$, such that we can remove the $\bm{x}_{12}$ dependence from the integral. Here, we just assume the zeroth order term, $\epsilon \rightarrow 0$,\footnote{This is equivalent to enforcing translation invariance/statistical homogeneity in the statistics of the local $2$-point function or assuming the plane-parallel, constant redshift limit in the case of the local contributions but not so when we have integrated contributions due to them breaking translation invariance outside the Limber approximation.} but we could Taylor expand up to any order and compute wide-separation corrections, using the standard framework \cite{Noorikuhani:2022bwc,Addis_2025}. Using the definition of the local power spectrum, \cref{eq:local_pk}, the I\texttimes I contribution to the local power spectrum can be written as a Fourier transform of our local $2$-point function
\begin{equation}
\begin{split}
P^{\rm I \times I}_{\text{loc}}(\bm{k};\bm{d}) & = \int_0^{1} \dd y_1 \int_0^{1} \dd y_2 \int_{\bm{q}} {\rm e}^{i\,\left(y_1-y_2\right)\,(\bm{q} \cdot \bm{d})} \int_{\bm{x}_{12}} {\rm e}^{ -i\,\left\{\bm{k} - \left[
y_2+t(y_1-y_2)\right]\,\bm{q} \right\} \cdot \bm{x}_{12}} \\
& \hspace{10em} \times d^2\,\mathcal{K}^{\rm I}(\bm{q}, \bm{d}, d \, y_1)\,\mathcal{K}^{\rm I}(-\bm{q}, \bm{d}, d \, y_2)\,P (q).
\end{split}
\end{equation}

As there is now no $\bm{x}_{12}$ dependence left in the kernels, the $\bm{x}_{12}$ integral therefore becomes a Dirac-delta. Then, using
\begin{equation}
\delta_D^3 \left(\bm{k} + A \, \bm{q} \right) = A^{-3}\,\delta_D \left( A^{-1}\,\bm{k} + \bm{q}\right)
\end{equation}
and defining the function $G(y_1, y_2) = [y_2 + t\,(y_1-y_2)]$, we can then simplify such that
\begin{equation}
\begin{split}
P^{\rm I \times I}_{\text{loc}}(\bm{k}; \bm{d}) & = \int_0^{1} \dd y_1 \int_0^{1} \dd y_2 \int_{\bm{q}} {\rm e}^{i\,\left(y_1-y_2\right)\,(\bm{q} \cdot \bm{d})}\, d^2 \,G(y_1, y_2)^{-3}\,\delta_D \left(\bm{q} - G(y_1, y_2)^{-1}\,\bm{k} \right) \\
& \hspace{16.4em} \times \mathcal{K}^{\rm I}(\bm{q}, \bm{d}, d\,y_1)\,\mathcal{K}^{\rm I}(-\bm{q}, \bm{d}, d\,y_2)\,P (q).
\end{split}
\end{equation}
Closing the Dirac-deltas, the I\texttimes I contribution to the local power spectrum can then be expressed as a double integral,
\begin{equation}
\begin{split}
P^{\rm I \times I}_{\text{loc}}(k,\mu; d) & = \int_0^{1} \dd\,y_1 \int_0^{1} \dd\,y_2 \, {\rm e}^{ i\,\left(y_1 - y_2 \right)\,k\,\mu\, d/G(y_1, y_2) }\, d^2 \, G(y_1, y_2)^{-3} \, \mathcal{K}^{\rm I}(G(y_1, y_2)^{-1}\, k,\mu,d,y_1) \\ & \hspace{11.7em} \times \mathcal{K}^{\rm I}(G(y_1, y_2)^{-1}\, k,-\mu,d,y_2) \, P (G(y_1, y_2)^{-1}\, k),
\end{split}
\label{eq:II_final}
\end{equation}
where we now express the kernels in terms of scalar quantities.

The angular $\mu$ integral can then be computed analytically to get the integrated contribution to the local multipoles but the $y_1, \, y_2$ integral needs to be computed numerically, the details of which are discussed in \cref{ap:numerics}. Note that as $G(y_1, y_2)$ has a range of $(0,1)$ over the $2$D integrand, the integral includes all $k$-scales above the measured $k$-mode. We consider the impact of non-linear modelling and $k_{\rm max}$ in \cref{sec:nonlinear_effects}.

\subsubsection{LOS Dependence of the Local Power Spectrum}\label{sec:los_depend}

The theoretical integrated contribution to the local power spectrum (\cref{eq:II_final}) is therefore dependent on the LOS choice in the local 2-point function, due to the $t$-dependence of the $G(y_1,y_2)$ factor. This is only removed if we consider a Limber approximation\footnote{After the $\mu$ integral the $(y_1-y_2)$ exponential can be written as a spherical Bessel function - we could then assume the Limber approximation such that $$j_0\left( k\,d (y_1-y_2)/G(y_1,y_2)\right)\rightarrow \delta_D( k\,d (y_1-y_2)/G(y_1,y_2)),$$ which we expect to be valid at small scales and large redshifts. In this flat-sky limit, we have translation invariant integrated contributions and thus no $t$-dependence.} and indeed one can see only the diagonal part, where $y_1=y_2$, of the I\texttimes I integrand (see \cref{fig:II_integrand_ream}) is translation invariant. For the off-diagonal parts, the integrand correlates different redshifts which breaks the $y_1 \leftrightarrow y_2$ symmetry and as such this contribution is not translation invariant and one can therefore consider this to be a wide-separation type effect (i.e. it derives from breaking local translation invariance at large separations), even though this arises even in the zeroth order term in the $x_{12}/d$ expansion. The size of this $t$-dependence in the local power spectrum is dependent on the separation and scales that are being correlated and is present in all wide-separation type terms.

The Yamamoto estimator, \cref{eq:yamamoto}, measures the power spectrum averaged over the survey volume. For the monopole, $\ell=0$, this does not assume a particular LOS, it simply averages over LOS orientation as we average over the survey volume. After integrating our theoretical local power spectrum monopole over all LOS in our survey, $\bm{d}$, this survey averaged quantity will be independent of the LOS choice we assumed in our local 2-point function, as long as $x_{12}/d$ expansion is valid.

Beyond the monopole, the LOS choice also dictates the vector about which we decompose our multipoles - see discussion of LOS choice in \cref{sec:forecast_setup} - which induces additional LOS dependence in the observed power spectrum multipoles and as such these survey averaged multipoles are indeed dependent on the LOS choice, unlike in the monopole. Therefore, for $\ell>0$ the LOS choice is important and indeed, as the natural choice of LOS for the Yamamoto estimator is an endpoint LOS (see \cref{vector_diagram}), it is important one calculates the corresponding endpoint LOS theory.

\begin{figure}
\centering
\includegraphics[width=1\textwidth]{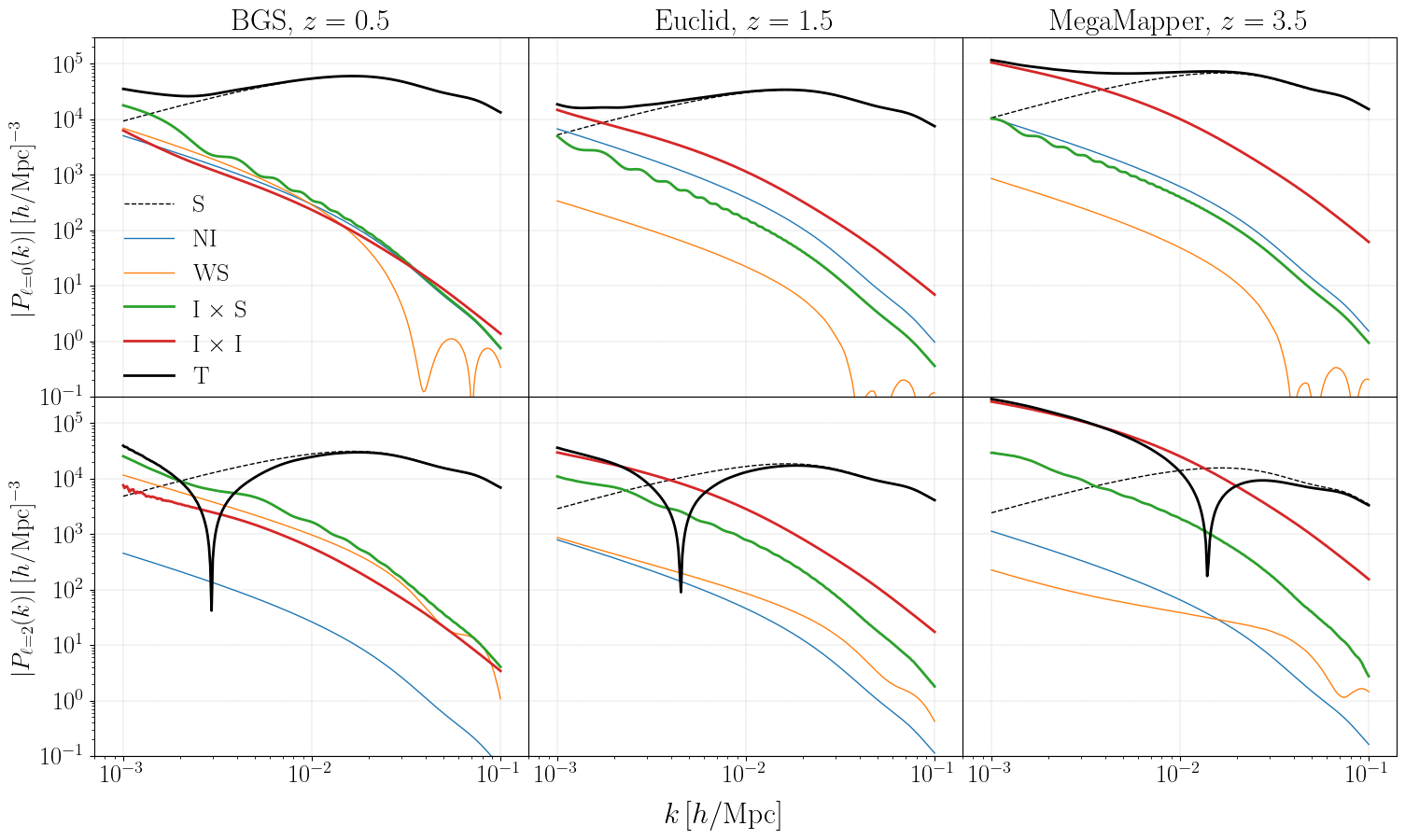} 
\caption{Monopole, $t=1/2$, (top row) and the quadrupole, $t=0$, (bottom row) of the power spectrum for each term over three different survey setups at set redshifts, with the standard Kaiser power spectrum shown in black dashed.}
\label{fig:mono_quad}
\end{figure}

A major consequence of this is that the leading order integrated terms generate non-zero, imaginary, odd-parity multipoles to the auto-power spectrum for $t\neq1/2$ \footnote{$t=1/2$ preserves the $y_1 \leftrightarrow y_2$ symmetry and as such the sum of off-diagonal odd multipole signals cancels.}. While this is a novel result, we reiterate that this signal should be considered a wide-separation type contribution - akin to the more well studied dipole from wide-separation corrections to the Kaiser contribution. The integrated contributions for the even and odd multipoles are plotted in \cref{fig:mono_quad} and \cref{fig:dipole_mt} (see \cref{sec:survey_details} for the survey specifications used to produce these results) and we further discuss the nature of the LOS dependence on large scales in \cref{ap:los_depend}.

\subsection{Local PNG Contribution}\label{sec:local_png}

The local-type PNG contribution to the power spectrum is a non-integrated term driven by the scale-dependent halo bias. Dropping the explicit comoving distance (redshift) dependence in the kernels, the local-type PNG kernel can be expressed as
\begin{equation}
    \mathcal{K}_{\rm NG}(k, x) = D \,  b_{\phi}/\mathcal{M}(k).
\end{equation}
Here $\mathcal{M}$ is the scaling factor between the primordial scalar power spectrum and the late-time matter power spectrum, the non-Gaussian bias $b_{\phi}$ is given, assuming the universality relation, by
\begin{equation}
    b_{\phi} =2 \, \delta_c \, \fNL \, (b_{1} - 1),
\end{equation}
and we assume $\delta_c=1.686$, the critical density for spherical collapse.

\subsection{Survey Power Spectrum Multipoles}\label{sec:survey_pk}

To get the theoretical expectation of the power spectrum multipoles we measure with the estimator, we must average our local power spectrum over our survey volume. In the standard literature, we can extract the local multipoles by averaging over $\mu$ the local power spectra with the appropriate Legendre polynomial
\begin{equation}\label{eq:mu_integral}
P^{ab}_{\ell, \text{loc}}(k;d) = \frac{(2\,\ell+1)}{2}\int^{1}_{-1} d \mu \, P^{ab}_{\rm loc}(k,\mu,d)\,\mathcal{L}_{\ell}(\mu),
\end{equation}
and finally we can obtain the theoretical power spectrum multipole of the survey by averaging the local power spectrum multipoles over the survey volume
\begin{equation}\label{eq:d_integral}
\hat{P}^{ab}_{\ell}(k) = \int_{V} \frac{d^3 \bm{d}}{ V} P^{ab}_{\ell, \text{loc}}(k;d),
\end{equation}
as in \cref{eq:survey_pk}. However, to correctly do this averaging over $\bm{d}$ for wide-separations we need to properly account for the LOS dependence in our 2-point function in our averaging - by pair counting over all the points we are correlating; for different LOS choices we sample over slightly different redshifts. Further consideration of this averaging is left to \cref{ap:los_depend}.

\begin{figure}
\centering
\includegraphics[width=0.8\textwidth]{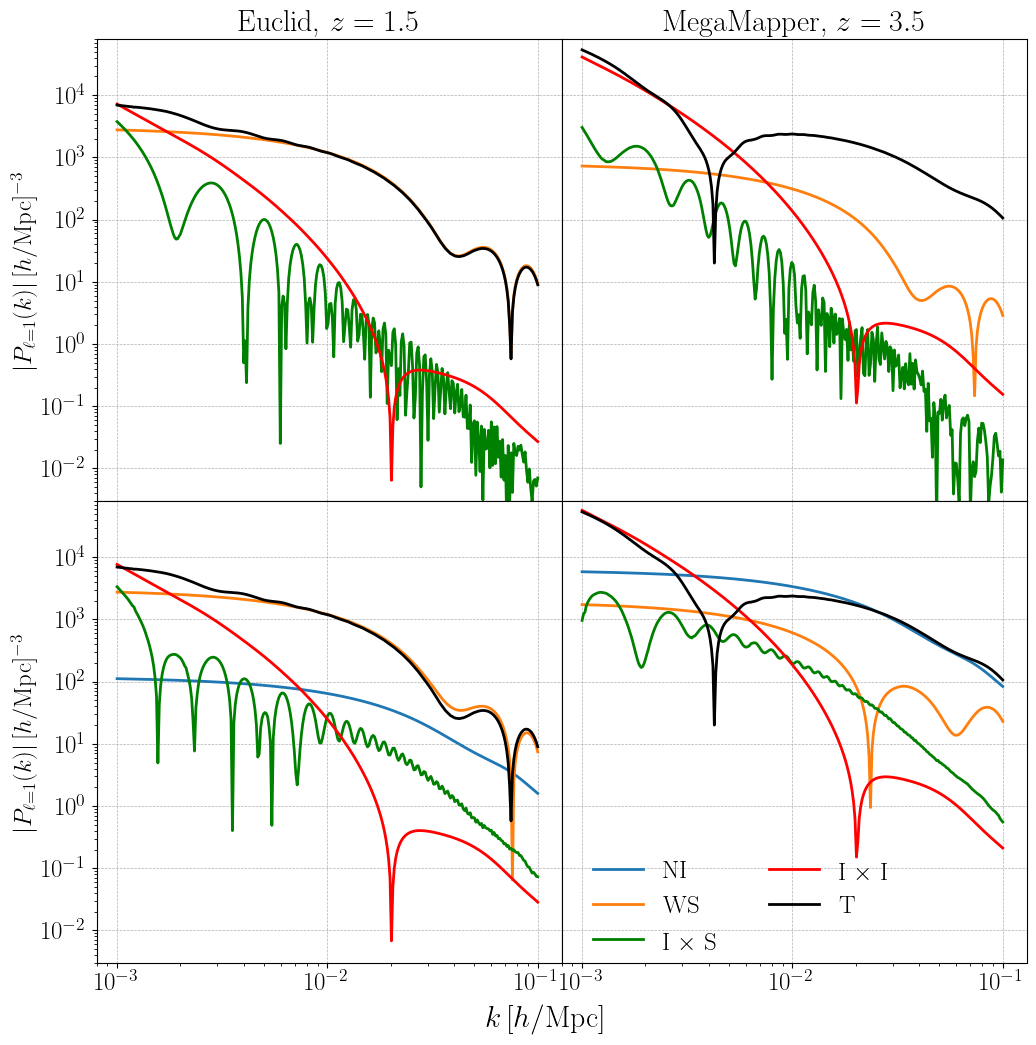} 
\caption{Top panel: Imaginary single-tracer dipole contribution, $t=0$, for \textit{Euclid} (left) and MegaMapper-like (right) surveys. Bottom panel: Bright-faint split cross-power spectrum dipole with a splitting flux of $F_s = 3\times10^{-16} \, [{\rm erg \, \,cm}^{-2} s^{-1}]$ for the Euclid-like H$\alpha$ survey and a splitting magnitude of $m_s=24$ for the MegaMapper-like LBG survey. Full details of the survey specifications and the bright-faint split are covered in \cref{sec:survey_details}.}
\label{fig:dipole_mt}
\end{figure}

\section{Forecast Setup}\label{sec:forecast_setup}

This section provides details of the setup we use for forecasts -- forecast results are presented in \cref{sec:results}. In particular, it overviews the choices of cosmology, LOS, free/nuisance parameters and bins, as well as how we forecast the constraints and biases with Fisher matrices and MCMC. Full details on the survey specifications are provided in \cref{sec:survey_details} and full details of the covariance calculation are included in \cref{sec:covariance}.

Throughout this study, we assume a fiducial cosmology from the Planck results \cite{Planck_2018} and we also adopt Planck constraints as priors on our cosmological parameters. We also assume a fiducial value of $\fNL=0$. Bias modelling for each survey is outlined in \cref{sec:survey_details}. To include the uncertainty on our bias parameters ($b_1$, $\mathcal{Q}$, $b_e$) we introduce and marginalize over bias amplitude parameters, $\alpha_{b_x}$, such that $b_1 \to \alpha_{b_1} \, b_1$, $\mathcal{Q} \to \alpha_{\mathcal{Q}}\,\mathcal{Q}$, $b_e \to \alpha_{b_e}\,b_e$. Therefore, we assume we know the redshift evolution of each bias parameter but not its overall amplitude.

In the multi-tracer case (see \cref{sec:multi_tracer} for details) we now marginalise over double the number of nuisance parameters -- as we model both the bright and faint samples, e.g., $b^B_1 \to \alpha^B_{b_1} \, b^B_1, \,b^F_1 \to \alpha^F_{b_1} \, b^F_1$. While the bright and faint biases are correlated to the total sample, we could also include the additional information from the total sample in our analysis \cite[e.g.][]{Blanco_2024} (such that one would have three auto-correlations and three cross-correlations for each multipole) without adding any extra free parameters, as the biases of the total sample are fully constrained from the bright and faint values -- see \cref{eq:mt_bias}, but we leave this extension to future work. 

To model the scale-dependent bias induced by local type PNG (see \cref{ap:more_kernels}), we make the standard assumption of universality of the mass function \cite{Slosar_2008}, such that $b_{\phi}$ can be linearly related to the linear bias $b_1$\footnote{Note that this is a simplistic assumption and is not expected to hold for real-life galaxies in most scenarios \cite{Barreira_2020,Barreira_2022b,PNGUNIT_2024,Fondi_2024}. While one could account for this additional uncertainty with the introduction of new nuisance parameters, we expect this to mainly affect the uncertainty on the determination of $\fNL$, but to have less  impact on core focus of this work -- the degeneracy of  $\fNL$ with light-cone effects (since it does not change the $k$-dependence).}. We do not account for any uncertainty or try to constrain $b_{\phi}$; therefore, we are actually constraining $b_{\phi}\, \fNL$ not $\fNL$. This is a non-trivial degeneracy to break without strict theoretical priors on $b_{\phi}$ \cite{Barreira_2022} which motivates using different approaches, such as higher-order statistics like the bispectrum \cite[e.g.][]{Tellarini_2016} or multi-tracer techniques \cite[e.g.][]{Seljak_2009,Barreira_2023} to help break this degeneracy.

The LOS choice in our theory is informed by two things \cite[for a more detailed overview see][]{Addis_2025}. First, it defines the point about which we perform our wide-separation expansion (this is relevant for the integrated contributions - see \cref{sec:los_depend}) to approximate the full local 2-point function, and second, it sets the vector about which we decompose our multipoles. When we measure our power spectrum over one survey, we naturally include the full LOS information of each point (we only do a perturbative expansion in our theory to approximate this) and therefore the choice of LOS in our estimator only affects our multipole decomposition. As the Yamamoto estimator is computable with separable FFTs only in the case of an endpoint LOS ($t=0 \text{ or } t=1$), we use $t=0$ as our LOS for the $\ell>0$ multipoles. However, the monopole is not decomposed with respect to a LOS (there is no LOS dependence in the monopole estimator); it is simply the average over all LOS orientation angles, $\mu$. Therefore, we choose the midpoint LOS $t=1/2$ for the monopole in our analysis as it is a natural choice of LOS about which to Taylor expand our theory.

Our full data vector, $D$, is a sum over redshift bins (which we assume are independent -- one would expect this to be a reasonable assumption due to the thick redshift bins considered in our analysis), where for each redshift bin we have a data vector, $D_z$, containing the power spectrum multipoles $\{\ell_i,...,\ell_n\}$:
\begin{equation}
D_z = \{P_{\ell_i}, ... , P_{\ell_n}\}.
\end{equation}
For each multipole, in $k$-space, we impose bin widths of $\Delta k = k_{\rm f}$ with a maximum scale of $k_{\rm f}$, where $k_{\rm f}$ is the fundamental frequency of the survey ($k_{\rm f} = (2 \pi)/V^{1/3}$) and, for simplicity, a cut-off scale at $k_{\rm max} =0.15\: [h/ {\rm Mpc}]$, unless stated otherwise. We also impose an additional effective $k_{\rm min}$ by excluding Fourier modes where $k < 2 \pi/x(z_{\rm min})$ for the minimum redshift included in the bin, as above these scales we expect to the perturbative wide-separation expansion to have broken down. 

In our single tracer analysis our usual data vector for each bin is $\ell=0,2,4$ and in the full multi-tracer analysis, the data vector for each multipole is larger, $P_{\ell_i=\text{even}} = \{P_{\ell_i}^{BB}, P_{\ell_i}^{BF}, P_{\ell_i}^{FF}\}$ as we have both our bright, $B$ and faint, $F$, samples. In addition, in the multi-tracer case one can now also consider the odd multipoles $(\ell=1,3)$ as well, however the auto-correlations are still zero and therefore we only include the cross-correlations for the odd multipoles, $P_{\ell_i= \text{odd}} = P_{\ell_i}^{BF}$. 

To quantify both the constraints on a particular parameter, in this case $\fNL$, and the bias on the maximum likelihood value if we ignore particular effects, we can either adopt a Fisher matrix formalism, \citep[e.g. see][]{Kitching_2009}, or we can sample the full likelihood function directly with a full Markov chain Monte Carlo (MCMC) pipeline. Fisher matrices are fast and easily reproducible but they only compute the likelihood at the fiducial point in parameter space and they assume Gaussianity in the parameter covariance. Therefore, for large parameter biases and non-Gaussian, degenerate posteriors we would expect Fisher matrices to not fully describe the true nature of the parameter space. Therefore, we initially adopt an MCMC formalism but when we include integrated effects in our modelling then this becomes significantly more computationally expensive and so we revert to Fisher forecasts, though we note that the MCMC posteriors appear largely Gaussian. A brief summary of both methods is provided below.

\paragraph{MCMC}

Our likelihood function is given by
\begin{equation}
log(\mathcal{L}) \propto (D^{\rm True}(\theta_0)-D(\theta))^{\dagger} \textsf{C}^{-1} (D^{\rm True}(\theta_0)-D(\theta)),
\end{equation}
where $D^{\rm True}(\theta_0)$ is the full `True' theory evaluated at the fiducial parameter, $\theta_0$ and $D(\theta)$ is our incomplete theory where we have neglected some contributions. Therefore, we evaluate $D(\theta_i)$ for each sample in parameter space, $\theta_i$.

We use the MCMC sampler implemented in the public code \textsc{Cobaya} \cite{Lewis_2002,Lewis_2013,Torrado_2021}. To efficiently sample over the cosmological parameter space, we use the power spectrum emulators from \textsc{CosmoPower} \cite{Mancini_2022}, which in the case of the non-linear power spectrum is emulated from results of \textsc{HMCode} \cite{Mead_2021}. For all chains the tolerance for convergence is set by the Gelman-Rubin parameter $R-1 = 0.001$.

\paragraph{Fisher Matrices}

The Fisher information matrix is defined as 
\begin{equation}
F_{ij} = \left(\frac{\partial D}{\partial \theta_i}\right)^{\dagger}(\textsf{C}^{-1})_{ij}\, \frac{\partial D}{\partial \theta_j},
\end{equation}
where we have partial derivatives of the data vector with respect to each parameter. 

If we use an incomplete model, $D(\theta)$, in our analysis where we have neglected some correction to our observable then our parameter inference will be biased from `True' value, or in this case fiducial value, $\theta_0$. Here our data vector, $D$, is evaluated solely at the fiducial values in parameter space, $\theta_0$. With the assumption of a Gaussian likelihood, the predicted bias in the maximum likelihood value, with respect to the true value of a given parameter $\theta_i$, is given by \cite{Kim_2004,Taylor_2007,Bernal_2020}
\begin{equation}
\Delta\theta_i = (F_{ij})^{-1} B_j
\end{equation}
where 
\begin{equation}
B_j = \left(\frac{\partial D}{\partial \theta_j}\right)^{\dagger} (C^{-1})(D^{\rm True}(\theta_0) - D),
\end{equation}
and $D^{\rm True}(\theta_0)$ is our complete theory model. Note that we use the conditional errors, not marginalised errors, to compute the bias.

\section{Results and Discussion}\label{sec:results}

This section is structured as follows: first we discuss the nature of the integrated contribution to the power spectrum \cref{sec:integrated_results}, secondly we explore bias in the determination of $\fNL$ if one neglects different corrections in their analysis, \cref{sec:bias} and lastly we discuss the forecasted constraints on $\fNL$ when we need to account for uncertainty in our luminosity function, \cref{sec:fnl_constraints}, and how this situation improves if we consider a multi-tracer approach, \cref{sec:multi_tracer_result}. The forecasting setup and methods are described in \cref{sec:forecast_setup}, while the specifications used for each survey are described in \cref{sec:survey_details}. Additional results for each integrated contribution are contained in \cref{ap:components} and additional graphs are included in \cref{ap:more_plots}.

\subsection{Integrated Contributions to the Power Spectrum}\label{sec:integrated_results}

Here we briefly discuss the nature of the different light-cone corrections to the power spectrum multipoles; though we stress these results are not only scale-, tracer- and redshift-dependent but also on the specific bias modelling adopted.

\paragraph{Even Multipoles} The absolute values of the integrated contributions to the even multipoles are plotted alongside the other terms in \cref{fig:mono_quad}. For the lower redshift DESI-like BGS sample one can observe that the integrated light-cone contributions (I\texttimes I and I\texttimes S) are comparable to the local, local relativistic (NI) and wide-separation (WS) corrections (note this is dominated by the wide-angle part for low redshifts), with the same $\approx 1/k^2$ dependence. For the higher redshift \textit{Euclid}-like H$\alpha$ and MegaMapper-like LBG samples, the I\texttimes I is the leading light-cone correction and indeed, it is even larger, particularly in the MegaMapper case, than the Kaiser contribution for ultra-large, post-equality, scales. The I\texttimes S contribution is a similar order of magnitude to the local relativistic contribution in the monopole but is roughly an order of magnitude larger in the quadrupole. Most of the contributions are positive, except for the local relativistic term and the wide-angle corrections to the Kaiser terms.

Wide-separation corrections (WS) here are generally subdominant to the other light-cone corrections beyond $z=1$; the wide-angle correction to the Kaiser terms have a $1/(kd)^2$ dependence so at higher redshifts the wide-separation contribution is predominantly caused by radial evolution corrections, however this remains subdominant to the other light-cone effects. Generally, the most important wide-separation contribution here is the first order correction to the $(\mathcal{H}/k)$ odd parity local relativistic term.

\paragraph{Odd Multipoles} In the single tracer case, the odd multipoles are purely generated by the contributions to the local power spectrum which break translation invariance, see top panel of \cref{fig:dipole_mt} -- this includes the off-diagonal parts of the integrated contributions. These contributions, as expected peak on large scales, and while the wide-separations corrections to Kaiser terms are more dominant for \textit{Euclid}, the integrated contributions become more relevant at higher redshifts.

However, when we cross-correlate two samples with different biases, this break in symmetry generates a non-zero imaginary contribution to the odd multipoles even in both the plane-parallel constant redshift and Limber approximations. \cref{fig:dipole_mt} shows the dipole $(\ell=1)$ for bright-faint splits for \textit{Euclid} and MegaMapper-like surveys. For the local terms, the wide-separation contributions, are dominated by the first order $(x_{12}/d)$ corrections to the Kaiser term. These are of the same order of magnitude as the leading order $(\mathcal{H}/k)$ local relativistic effects, wide-separation tends to be larger for \textit{Euclid} but smaller for MegaMapper, and they also have the same general $1/k$ dependence. The integrated contributions have a steeper $k$ dependence and so are particularly relevant on ultra-large scales as well as at higher redshifts. The I\texttimes I term in particular, is suppressed and has a negligible impact to the odd multipoles above $k\approx0.01 \, \aMpch$ but theoretically would be leading contribution to the dipole at $k=0.001 \,\aMpch$.

\subsubsection{Non-linear Contributions to the Integrated Effects}\label{sec:nonlinear_effects}

\begin{figure}
\centering
\includegraphics[width=0.7\linewidth]{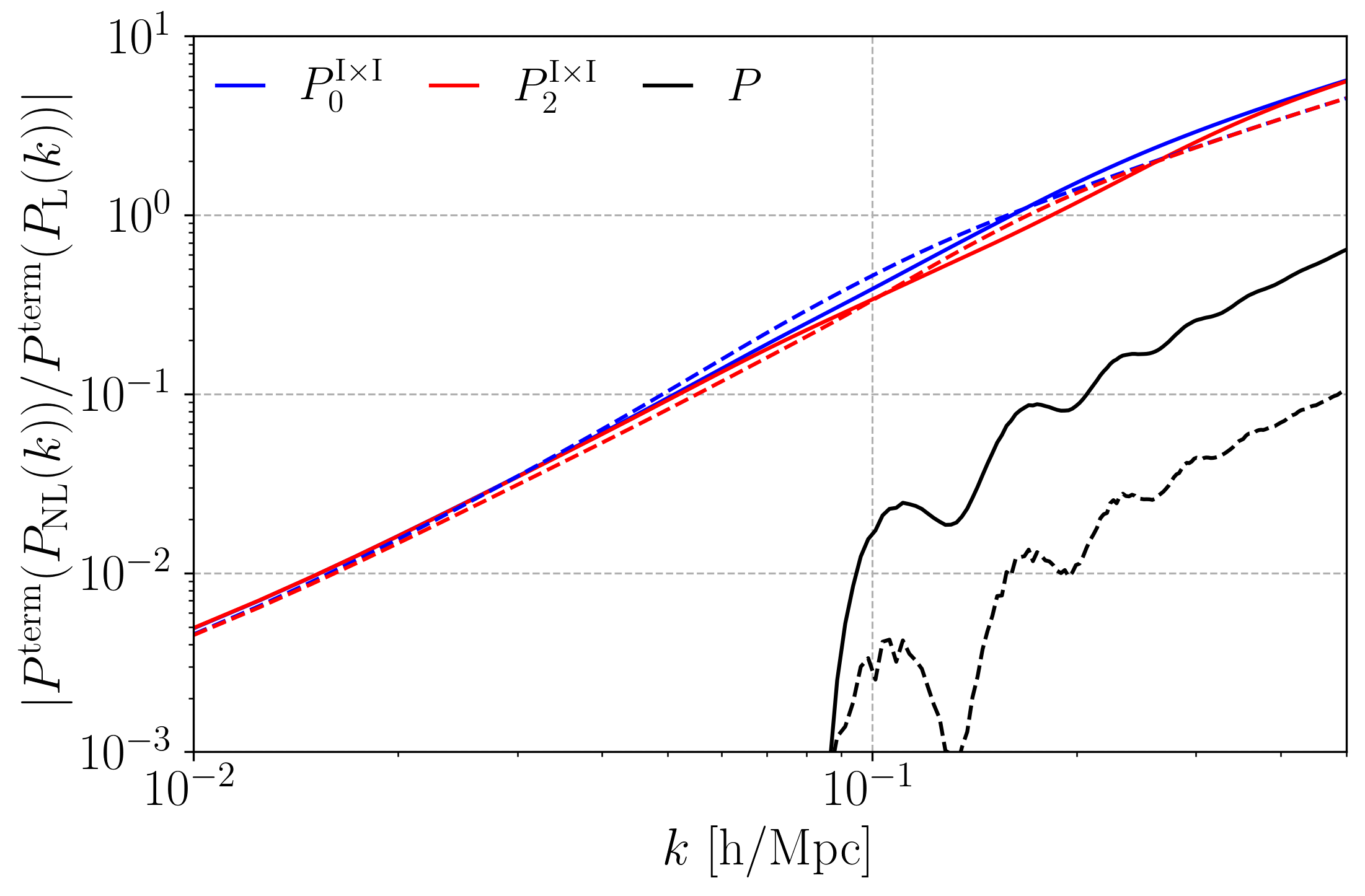} 
\caption{Ratio of power spectrum contributions when using a non-linear $P(k)$ (HMcode) modelling compared to using linear $P(k)$. The solid lines represent a $z=1$ \textit{Euclid}-like survey while the dashed lines represent a MegaMapper-like LBG survey at $z=3$. This ratio is shown for the I\texttimes I term in blue for the monopole and red for the quadrupole, while the ratio for all the other terms (all terms except I\texttimes I) is depicted in black. Outside of the I\texttimes I contributions, the ratio is simply proportional to the ratio of the non-linear to the linear power spectra sampled at the source redshift and measured $k$-scale.}
\centering
\label{fig:nonlin_pk}
\end{figure}

As the integrand in \cref{eq:II_final}, is a function of $k/G(y_1,y_2)$ -- rather than just $k$ -- where $G(r_1,r_2,d)$ has a range of $(0,1)$, the integrand is dependent on all $k$-modes where $k' < k$. The I\texttimes I contribution is therefore sensitive to non-linear effects even when considering large linear scales. To explore the nature of this non-linear dependence, we consider a simple model of including non-linear effects where we replace the linear matter power spectrum with the non-linear one, $D(x_1)D(x_2)P(k)\rightarrow P_{\rm NL}(k,x_1,x_2)$\footnote{The non-linear power spectrum we use here is obtained from HMCode \cite{Mead_2021}}, where $x_1,\, x_2$ could be $d$, or $r_1, \, r_2$ if we are considering an integrated term (to quantify the non-linear power spectrum from two unequal redshifts of the density field, as in the integrated case, we use the mean redshift). Although this is a simplified treatment of including non-linear effects, it illustrates their impact on the integrated contributions at even very mildly non-linear scales. The ratio of each contribution using a non-linear $P_{\rm NL}(k)$ compared to a linear $P(k)$ is simply the ratio of the two power spectra at the source redshift (\cref{fig:nonlin_pk}). This holds for nearly all terms, including the I\texttimes S contribution, since this is dominated by the part of the integrand at the source -- see the discussion in \cref{ap:numerics}. However, for the I\texttimes I contribution, there is a significant increase in the signal even at low $k$:  we find  a $\sim\!10\%$ boost in the signal from the non-linear power spectrum at $k=0.05 \, \aMpch$, for both MegaMapper and \textit{Euclid}. This non-linear boost is of the order of $100\%$ at $k=0.15$ and also appears somewhat independent of redshift (as \textit{Euclid} and MegaMapper receive similar boosts) compared to power spectrum itself which becomes more linear at higher redshifts; this behaviour for the I\texttimes I can be understood as the integrand is sensitive to smaller scales as it samples the lower redshift power spectrum closer to observer. We consider the potential bias from the non-linear boost to the I\texttimes I term in \cref{ap:nonlinear_effects}.

\subsection{Bias on Inference of PNG}\label{sec:bias}

We now quantify the bias (shift) in the inferred value of the local type $f_{\rm NL}$ that arises when specific light-cone systematics are neglected in the analysis. The forecast setup and the details of the computation of the bias on the best fit value is described in \cref{sec:forecast_setup}. When forecasting the bias in this section we do so to mimic a naive analysis that completely neglects these corrections and so therefore, we also do not include the effect of these correction on the covariance; the covariance is simply the Kaiser sample variance plus shot-noise. Therefore the error on $\fNL$ only changes when we investigate the bias from neglecting different effects as we are sampling different areas of parameter space.

\begin{figure}[H]
\centering
\begin{subfigure}{\linewidth}
    \centering \includegraphics[width=0.75\linewidth]{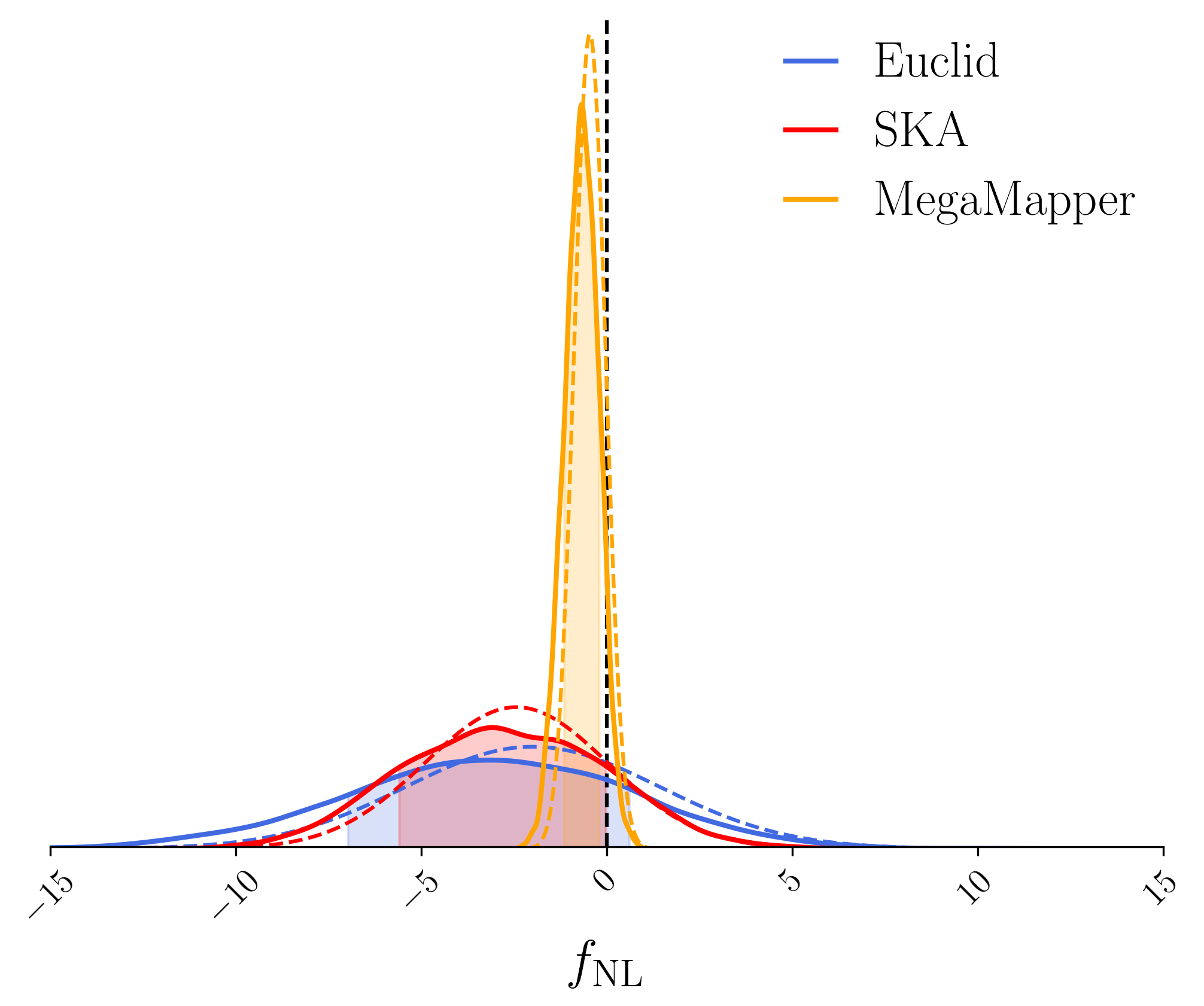}
    \label{sfig:fnl_bias}
\end{subfigure}
\vspace{1cm}
\begin{subfigure}{\linewidth}
    \centering
    \includegraphics[width=0.75\linewidth]{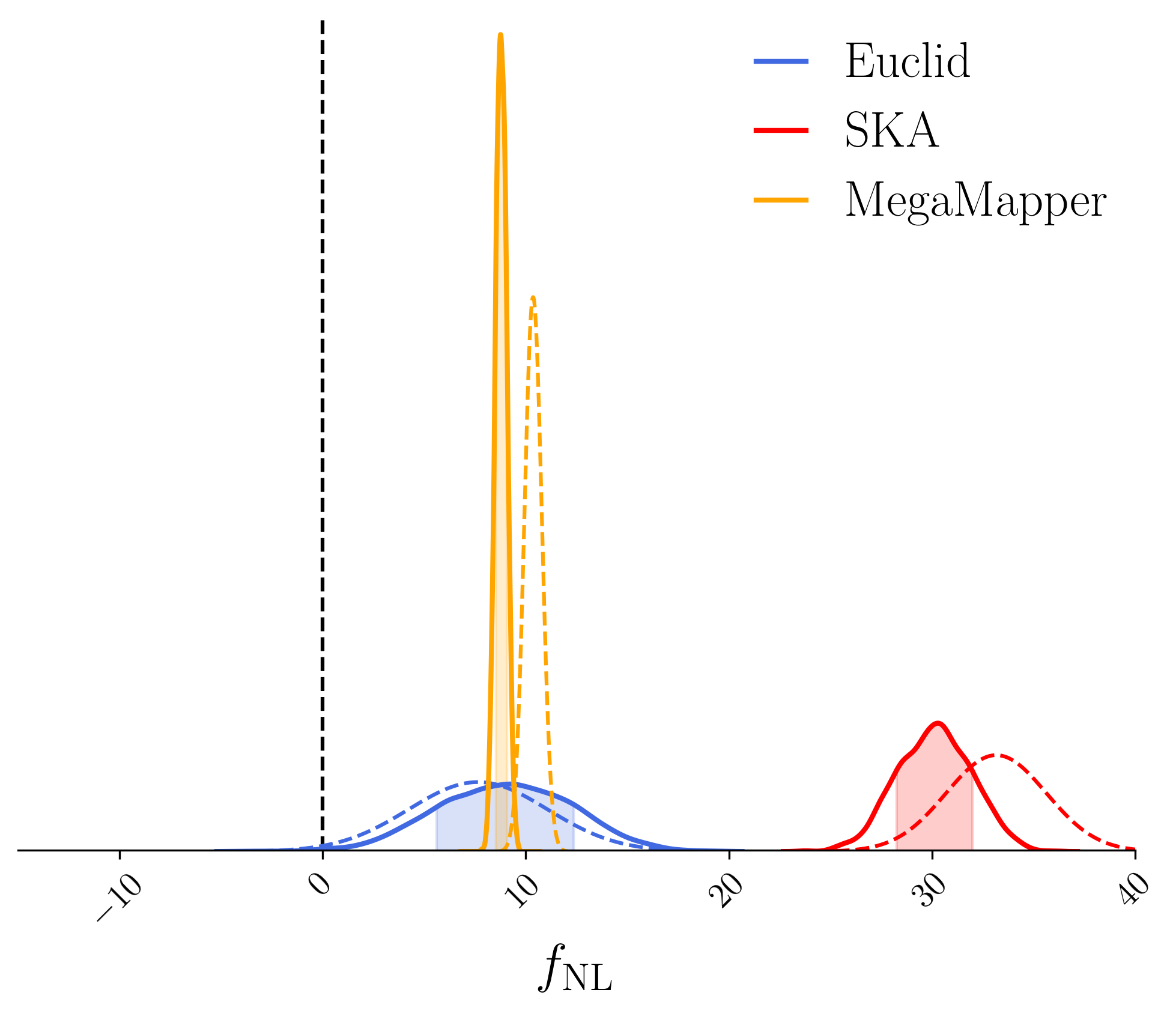}
    \label{sfig:second}
\end{subfigure}
\caption{PDF of the forecasted constraints on local $\fNL$ for a fiducial value of $\fNL=0$, after marginalising over $\alpha_{b_1}$, using a MCMC analysis (solid lines) and a Fisher analysis (dashed lines). Top panel considers the case where local relativistic effects and wide-separation corrections are ignored while the bottom panel considers the case where local relativistic, wide-separation and the integrated contributions are all neglected. }
\label{fig:fnl_bias_1d}
\end{figure}

\cref{fig:fnl_bias_1d} shows the bias induced by local relativistic and wide-separation terms is of $\Delta\fNL \sim \mathcal{O}(1) \, \sigma$ for \textit{Euclid}, SKA and MegaMapper-like surveys after marginalising over $\alpha_{b_1}$ (values are printed in \cref{tab:bias_res}). For the BGS and Roman-like surveys, this bias is within the survey error bars; constraints on $\fNL$ are quite closely linked to the observed volume, as larger volumes not only allows for a greater number of $k$-modes (bin width is set by $k_{\rm f}$ and therefore the volume) but also because the $\fNL$ signal is greater on large scales. The constraints for BGS and Roman are therefore weakened by their small redshift range and observed sky fraction, respectively. While this bias is dominated by the pure local relativistic contribution, the most important wide-separation term is the 1st order wide-separation corrections to the $\mathcal{H}/k$ odd local relativistic contribution which then enters the even multipoles. Wide-angle effects are more relevant at low redshift with a bias of $\Delta\fNL \approx\mathcal{O}(1)$ for the DESI-like BGS survey from neglecting wide-separation effects despite different contributions of the wide-separation term shifting $\fNL$ in opposing directions.

The constraints on $\fNL$ are tighter for SKAO2 and MegaMapper than they are for \textit{Euclid}, and this can largely be attributed to the higher redshift ranges and sky fractions observed by these surveys such that they contain a higher number of modes. We can also see that for these relatively small deviations from the fiducial values the Fisher matrix approach provides a fair description of the true posterior as it generally agrees with the MCMC samples.

However, when we also consider the impact of neglecting integrated contributions, I\texttimes I and I\texttimes S, the bias goes from small negative values of $\fNL$ to large positive values. We can understand this behaviour if we consider each contribution to the even multipoles of the power spectrum (\cref{fig:mono_quad}), the local relativistic (and wide-separation) contributions are generally negative, suppressing power at large scales while the integrated contribution is positive, boosting power on large scales. The bias is then predominantly caused by the large I\texttimes I contribution, and indeed if one considers the bias in each individual redshift bin, one can see a general increase with redshift as the integrated contributions are comparatively larger. The particularly large bias in the SKAO2 case can be attributed to the ratio of $\mathcal{Q}$ and $b_1$. The $\fNL$ signal is proportional to $b_1$ but as the lensing term is proportional to $\mathcal{Q}$, we can see the bias is related to the ratio $\mathcal{Q}/b_1$, which is larger for SKAO2 than the equivalent redshift \textit{Euclid}-like H$\alpha$ survey, as the SKAO2 \(\hi\) galaxy catalogue is a less biased sample of the underlying dark matter distribution (it has lower values of $b_1$).

The shift on the bias is summative for each contribution in the case of the Fisher analysis, but this is no longer true when using an MCMC analysis. Also, as the posterior is several $\sigma$ away from the fiducial values, we can see the linear Fisher approximation breaks down and does not properly represent the true posterior. 

\begin{table}[tbp]
\centering
\caption{Constraints on local type $\fNL$, for a fiducial $\fNL=0$ if we neglect selected theoretical systematics in our MCMC forecast, after marginalising over the linear bias amplitude parameter $\alpha_{b_1}$.}
\renewcommand{\arraystretch}{1.6}
\begin{tabular}{|l|c|c|c|c|c|}
\hline
Light-cone terms &DESI BGS & Roman & \textit{Euclid} H$\alpha$ & MegaMapper & SKAO2\\
\hline
NI & $-3.5^{+20.8}_{-21.7}$ & $-2.2^{+8.5}_{-8.4}$& $-3.2^{+4.0}_{-4.0}$ & $-0.64^{+0.49}_{-0.49}$ & $-3.0^{+2.8}_{-2.9}$ \\
NI$+$WS & $-1.8^{+20.6}_{-20.5}$& $-1.6^{+8.5}_{-9.0}$&$-3.2^{+3.8}_{-3.8}$ & $-0.68^{+0.47}_{-0.48}$ &  $-2.8^{+2.8}_{-2.8}$ \\
NI$+$WS$+$I & $4.9^{+20.3}_{-21.4}$ & $3.3^{+8.4}_{-8.9}$& $9.0^{+3.3}_{-3.4}$ & $8.8^{+0.28}_{-0.27}$ & $30.2^{+1.8}_{-1.9}$ \\
\hline
\end{tabular}
\label{tab:bias_res}
\end{table}

\begin{figure}[ht]
\centering
\includegraphics[width=1\textwidth]{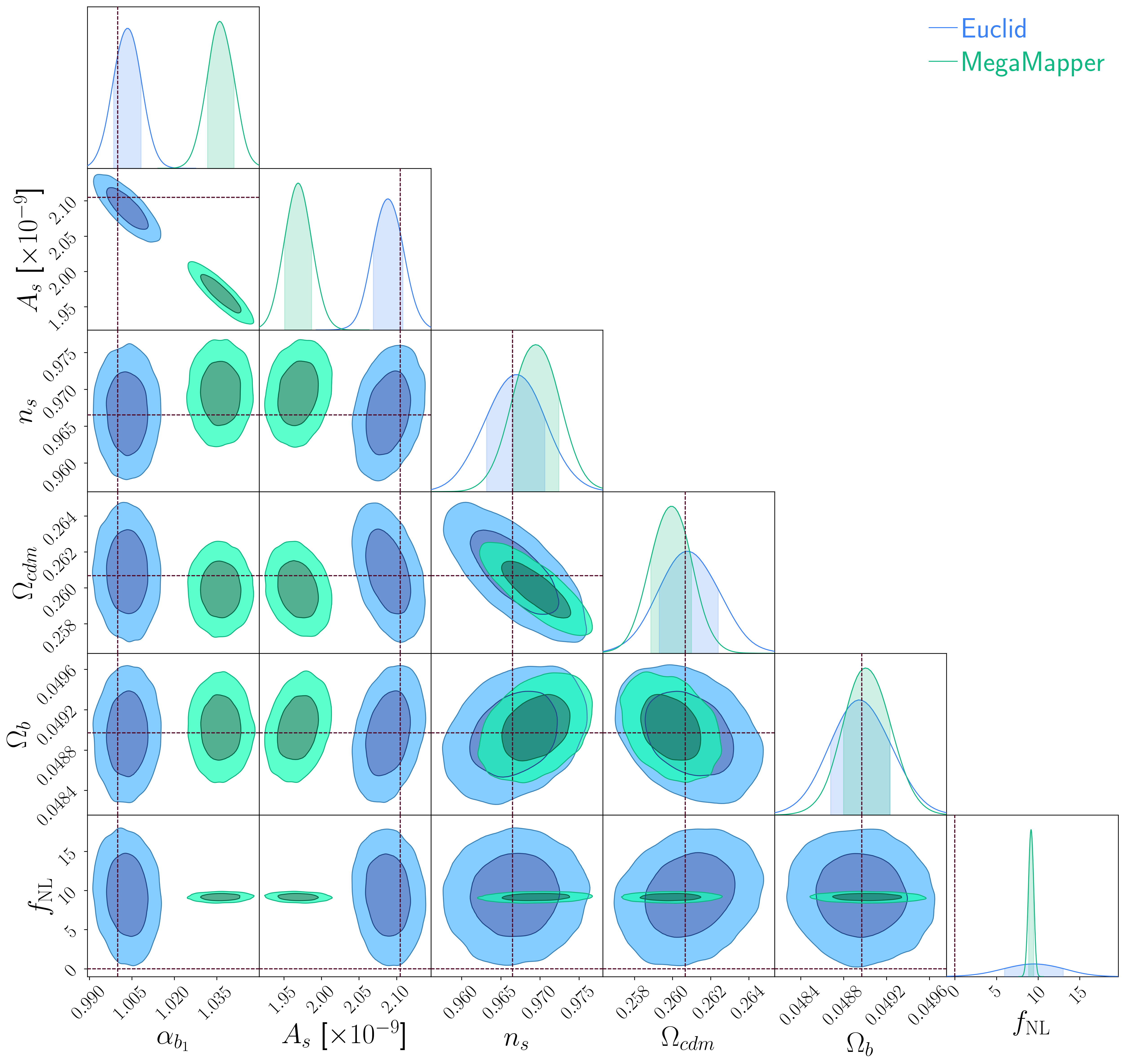} 
\caption{Forecasted marginalised and joint parameter constraints for a \textit{Euclid}-like H$\alpha$ survey (blue) and a MegaMapper-like LBG survey (green), if we neglect integrated, local relativistic and wide-separation effects in our analysis. We assume Planck priors on cosmological parameters and solid colour representing the $1\sigma$ constraints and the lighter region denoting the $2\sigma$ constraints.}
\label{fig:full_MCMC}
\end{figure}

We can also consider the impact of neglecting light-cone effects when we sample over cosmological parameter space. \cref{fig:full_MCMC} shows the full MCMC chains where we do not assume a fixed cosmology, for \textit{Euclid} (blue) and MegaMapper-like (green) surveys. The cosmological parameters ($\Omega_{\rm b}$, $\Omega_{\rm cdm}$, $A_s$, $n_s$) have priors drawn from the \textit{Planck} 2018 constraints, \cite{Planck_2018} and therefore both \textit{Euclid} and MegaMapper have smaller difference in cosmological parameter constraints, as these are bounded by tight \textit{Planck} priors. However, $A_s$, even with Planck priors, is biased from the fiducial value in the MegaMapper case. The $\fNL$ constraints, $\fNL = 9.39^{+3.37}_{-3.49}$ for the \textit{Euclid}-like sample and $\fNL = 9.15^{+0.29}_{-0.29}$ for the MegaMapper-like sample, are largely consistent to the case where we assume a fixed cosmology. If we forgo priors on cosmology (\cref{fig:full_MCMC_noprior}), then our constraining power on $\fNL$ weakens slightly for the \textit{Euclid}-like (MegaMapper-like) sample, $\fNL=10.01^{+3.53}_{-3.76}$ ($\fNL=9.42^{+0.31}_{-0.33}$), due to the broader constraints on cosmological parameters. In addition, for the MegaMapper-like sample the cosmological parameter is also biased for both $\Delta(A_s) = 9.3 \, \sigma$ and $\Delta(n_s) = 1.4 \, \sigma$. Therefore, the biases and errors on our determination of $\fNL$ appear robust to our choice of free parameters.

\subsubsection{Comparison with Guedezounme et al.}\label{sec:comparison}

Our findings regarding the bias on local-type $\fNL$ when neglecting relativistic effects contrast with those of Guedezounme et al.\ \cite{Guedezounme:2024pbj}. However, this discrepancy is almost entirely driven by the I\texttimes I contribution. Guedezounme et al.\ report a shift in $\fNL$ that falls within the forecasted survey errors for MegaMapper and SKAO2-like galaxy surveys (see Table 4 in \cite{Guedezounme:2024pbj}). If we ignore the I\texttimes I contribution, our results are highly consistent with theirs (accounting for a sign change due to differing definitions of the `shift')\footnote{The forecast setups also differ slightly in  the bias modelling, choice of $k_{\rm max}$, $k$-bins, and the inclusion of radial evolution effects, among other minor differences.}. Indeed, when only the local relativistic and I\texttimes S terms are included, the bias remains small, partially due to cancellations between these two contributions. Ultimately, it is the I\texttimes I term that drives the large positive biases observed in our analysis.

\subsection{Constraints on Local PNG}\label{sec:fnl_constraints}

The constraints on $\fNL$ in the previous section are comparable with previous analyses \cite[e.g.][]{Giannantonio_2012,FontRibera_2014} which have largely ignored the impact of light-cone systematics. However, as we have shown that neglecting these effects leads to a sizeable bias in our inference of $\fNL$ and other cosmological parameters, we need to include these systematics in our analysis. This impacts our constraints on $\fNL$ as the relativistic terms are dependent on the survey luminosity function through the evolution and magnification bias; therefore uncertainty in our luminosity function leads to uncertainty in our determination of $\fNL$. We account for this uncertainty by marginalising over the amplitude of the bias parameters, $\alpha_{\mathcal{Q}},\alpha_{b_e}$. We do not include any prior constraints on $\alpha_{\mathcal{Q}}$ and $\alpha_{b_e}$.

In addition, these light-cone effects also contribute to the sample variance (see \cref{sec:covariance}) and here the integrated contributions are relevant on larger scales and higher redshifts. This extra positive contribution therefore leads to a reduction in SNR for local type $\fNL$. In these forecasts, we include the same terms in the covariance as we do in the signal. 

\begin{figure}
\includegraphics[width=\linewidth]{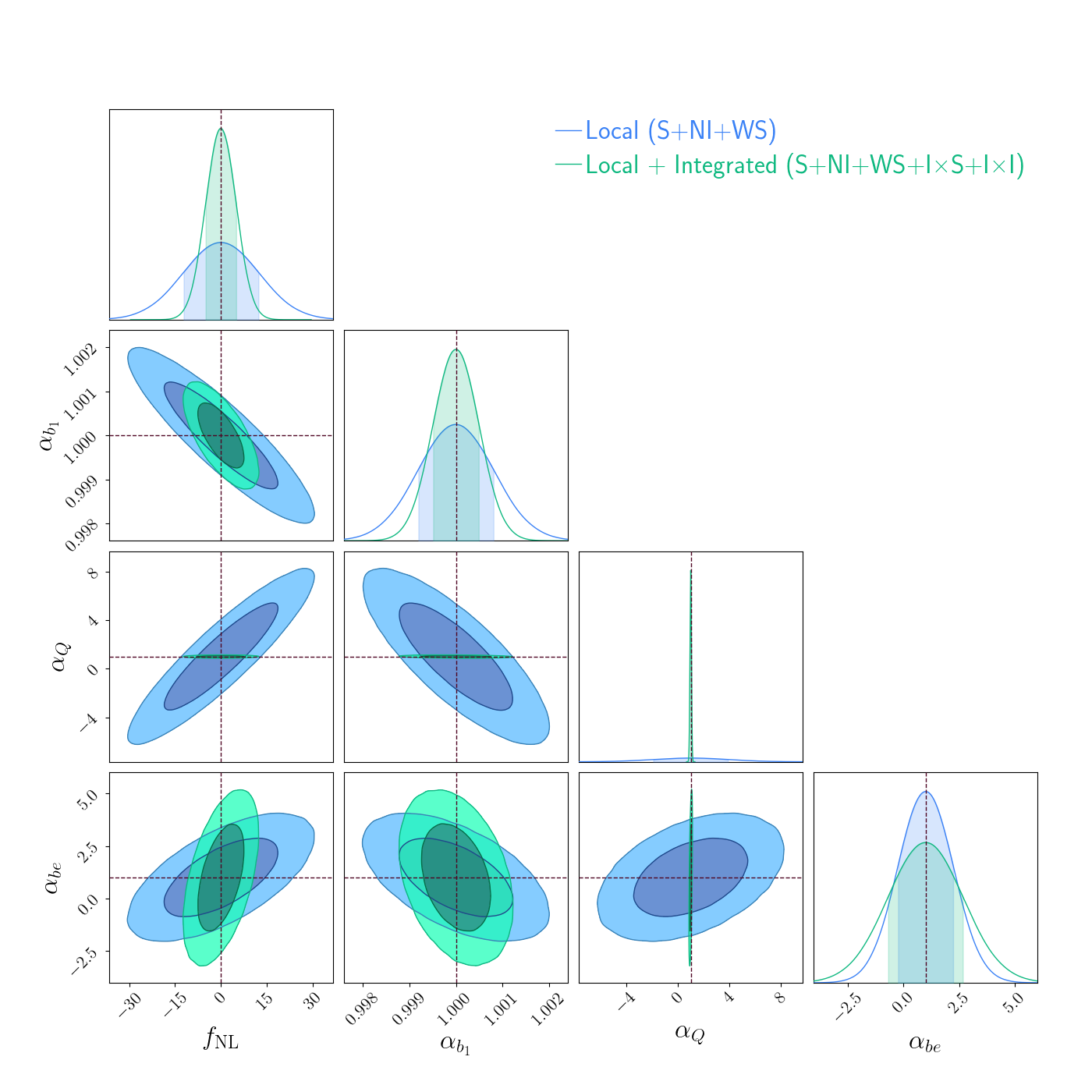} 
\caption{Fisher forecasted marginalised and joint constraints on local type $\fNL$ and bias amplitude parameters for a \textit{Euclid}-like H$\alpha$ survey. Ellipses are green where we only include local corrections (S+NI+WS) and blue where we also include integrated contributions (S+NI+WS+I\texttimes S+I\texttimes I). The constraints on $\alpha_{Q}$ are difficult to resolve visually on this scale, as the inclusion of integrated contributions tightens them by three orders of magnitude.}
\label{fig:localint_fnl}
\centering
\end{figure}

Previous studies, such as \cite{Wang_2020,Guedezounme:2024pbj}, have considered the impact of the uncertainty in the luminosity function on the constraints of $\fNL$, however these have not included the impact of integrated corrections. In that scenario, both $\mathcal{Q}$ and $b_e$ are poorly constrained by the power spectrum and therefore there is a large increase in the uncertainty on $\fNL$. However, with the additional inclusion of integrated effects, the large lensing contribution yields far tighter constraints on $\alpha_{\mathcal{Q}}$ -- which helps to break the degeneracy between it and $\fNL$, arising from the $(\mathcal{H}/k)^2$ local projection terms. Thus the inclusion of the integrated terms yields tighter constraints on $\fNL$ than in the local effects only case. This is demonstrated for a \textit{Euclid}-like survey in \cref{fig:localint_fnl}, with only local corrections, $\sigma(\fNL)= 12.0$, in blue and when including integrated effects, $\sigma(\fNL)= 4.5$, in green.

In order, to understand the impact of additional contributions to the covariance and the added nuisance parameters $\alpha_{\mathcal{Q}}, \alpha_{b_e}$, we can forecast the uncertainty on $\fNL$ with and without the additional corrections to the covariance. However, we find that the additional corrections to the covariance has a minimal impact on the constraints on $\fNL$ despite it leading to noticeably worse constraints on other bias parameters, notably $\alpha_{\mathcal{Q}}$, compared to the Newtonian covariance case. Marginalised constraints on $\fNL$ for a full relativistic analysis are shown in \cref{tab:constaint_res}.

\begin{table}[tbp]
\centering
\caption{Marginalised constraints on $\fNL$ for a full relativistic analysis - including the uncertainty in our luminosity function.}
\renewcommand{\arraystretch}{1.6}
\begin{tabular}{|l|c|c|c|c|c|}
\hline
&DESI BGS & Roman & \textit{Euclid} H$\alpha$ & MegaMapper & SKAO2\\
\hline
$\sigma (\fNL)$ &  $24.3$ & $10.1$ & $4.6$  &  $0.49$  &   $2.9$  \\
\hline
\end{tabular}
\label{tab:constaint_res}
\end{table}

\subsubsection{Multi-Tracer Analysis}\label{sec:multi_tracer_result}

\begin{figure}[H]
\centering
\begin{subfigure}{\linewidth}
    \centering \includegraphics[width=0.77\linewidth]{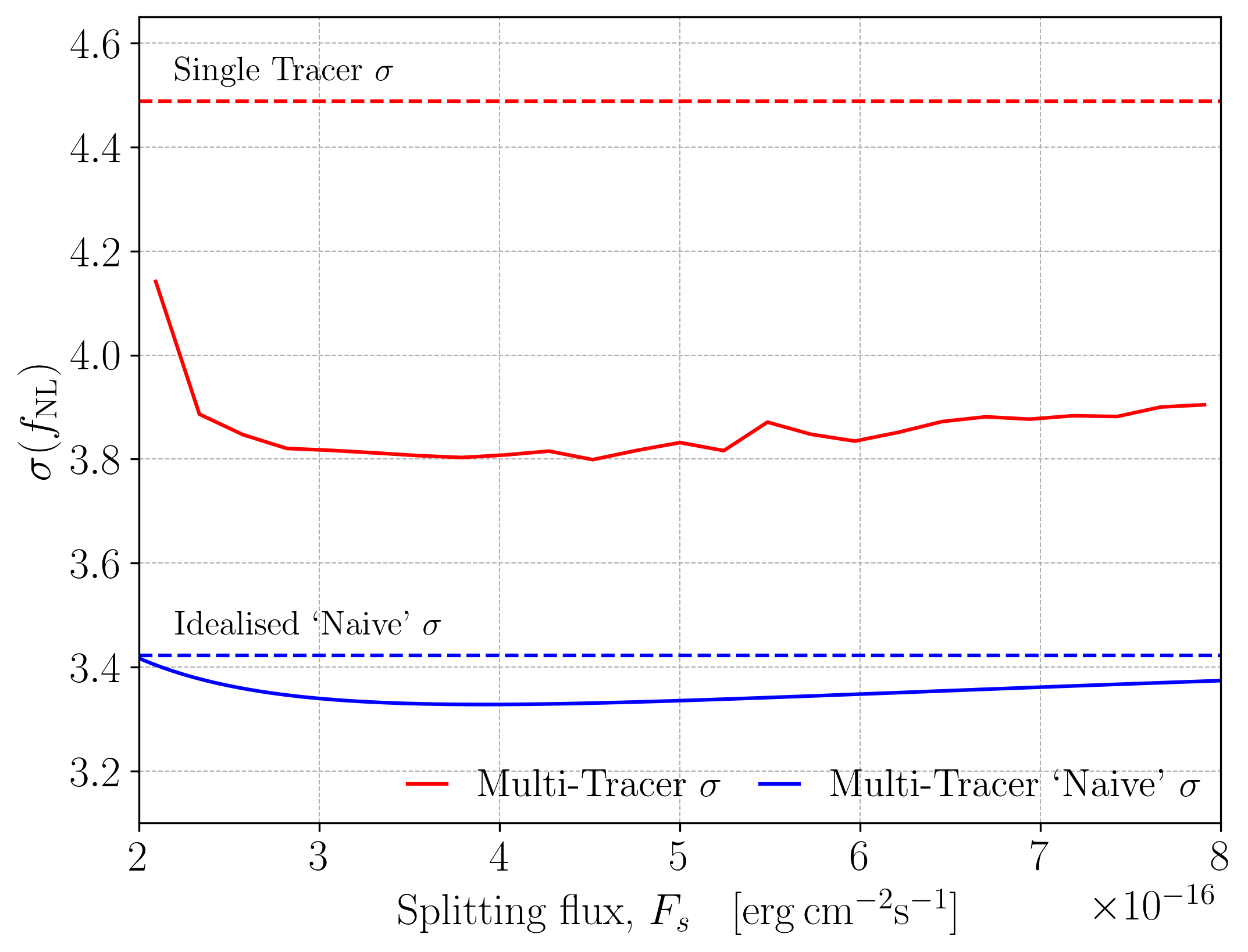}
\end{subfigure}
\vspace{1cm}
\begin{subfigure}{\linewidth}
    \centering
    \includegraphics[width=0.8\linewidth]{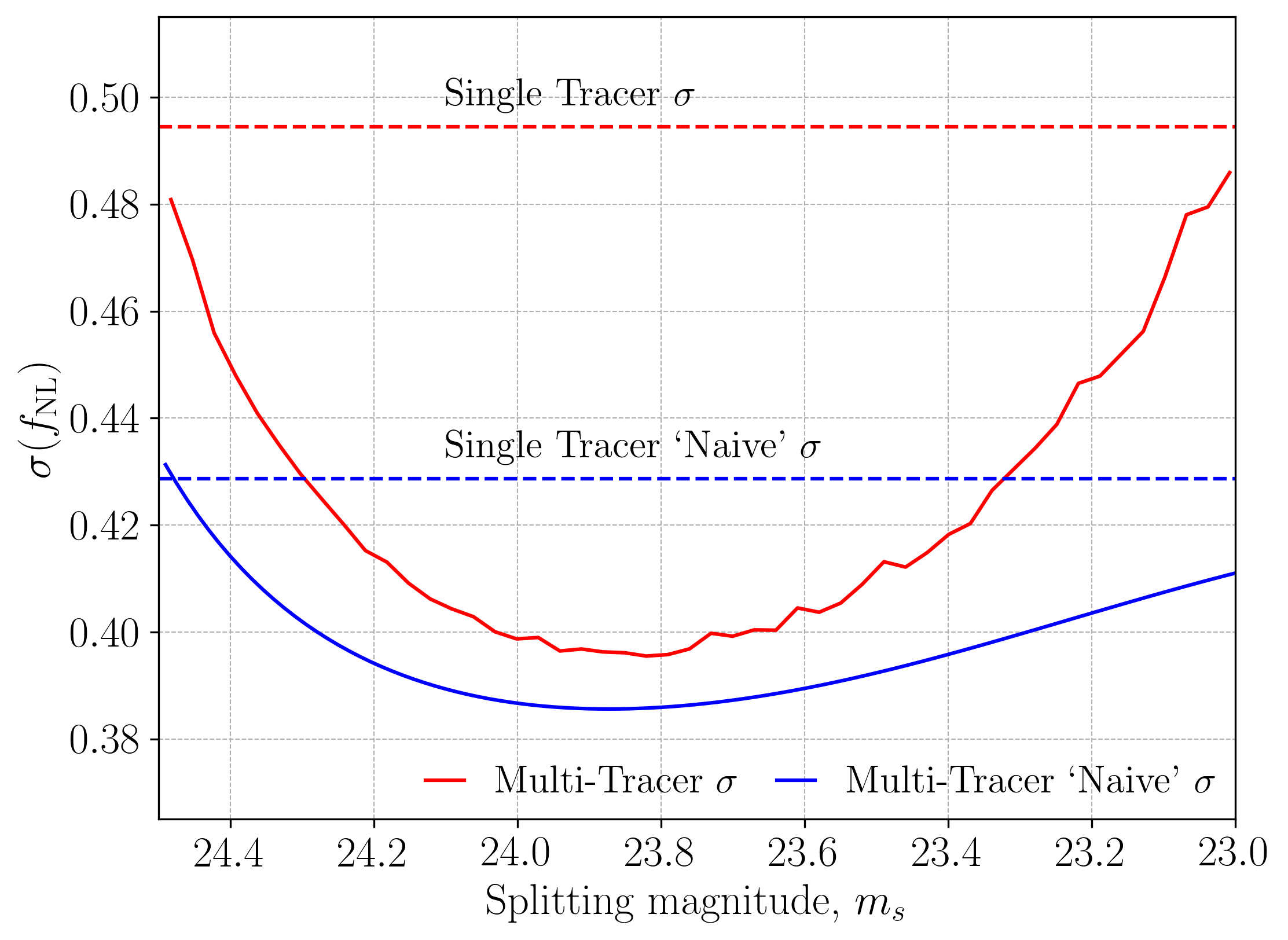}
\end{subfigure}
\caption{Fisher forecasted errors on local $\fNL$ for multi-tracer (solid lines) and single tracer (dashed) analyses. The `naive' analysis where one ignores light-cone corrections, including in the covariance, is shown in blue and the case where we include these corrections and the additional nuisance parameters is shown in red. Top panel is for a bright-faint split of a \textit{Euclid}-like H$\alpha$ survey plotted as a functions of the flux split, $F_s$ and the bottom panel is for a MegaMapper-like LBG survey plotted as a function of apparent magnitude split, $m_s$.}
\label{fig:fnl_mt}
\end{figure}

As our constraints on $\fNL$ are now dependent on how well we know both $\mathcal{Q}$ and $b_e$, constraints on $\fNL$ will improve if we have more information about the survey luminosity function. This motivates the consideration of a multi-tracer analysis, for which there are non-cancelling odd multipoles, which are sourced, in part, by the leading order $\mathcal{H}/k$ local relativistic corrections and further the I\texttimes S contribution is also non-negligible here - for example see \cref{fig:dipole_mt}. Therefore, the signal in the odd multipoles while not containing a local $\fNL$ signal, can help break the degeneracy between $\mathcal{Q}$ and $b_e$, and $\fNL$; improved constraints on $b_e$ is the main gain here as $\mathcal{Q}$ is already well constrained by the large I\texttimes I contribution, for example see \cref{fig:localint_fnl}. In addition, the use of highly biased tracers can enhance the $\fNL$ though the $b_{\phi}$ dependence \cite{Seljak_2009,Barreira_2023,Karagiannis_2024}. We do however have more free parameters as we have to model biases for both the bright and faint populations. Details of the multi-tracer analysis are outlined in \cref{sec:forecast_setup} and \cref{sec:multi_tracer}.

\cref{fig:fnl_mt} shows a comparison of the forecasted constraints of $\fNL$ for a few different scenarios: first, for the idealised `naive' case (blue) where $\mathcal{Q}$ and $b_e$ are fixed and there are no relativistic or wide-separations corrections to the covariance; therefore this is what one would forecast if they ignored light-cone effects. Secondly, for the single tracer case where we account for uncertainty in $\mathcal{Q}$ and $b_e$ and include all the effects discussed in this work both in the signal and covariance, for $\ell=0,2,4$ and lastly for the multi-tracer case where we now include not only odd multipoles $(\ell=1,3)$ but also bright-bright, bright-faint and faint-faint auto- and cross-correlations, plotted over splitting flux/magnitude.

We can see that, adopting a bright-faint split approach, when we are modelling for light-cone effects, tightens constraints on local type $\fNL$ by approximately $15-20 \%$ for both \textit{Euclid} and MegaMapper-like surveys, for a sensible flux/magnitude cut, despite the fact we are marginalising over double the number of nuisance parameters. Therefore, this approach mitigates some of the uncertainty induced from parametrising the luminosity function to model relativistic effects, and indeed, for the MegaMapper-like survey the constraints are tighter than in the naive case, where we completely neglect relativistic corrections, for a sensible choice of the magnitude split. In part, the better constraints using the bright-faint split approach with MegaMapper can be explained by the fact that at high redshifts we are using highly biased tracers, which can amplify the $\fNL$ signal.

\section{Summary and Conclusions}

The multipoles of the 3D power spectrum remains a popular basis for LSS analysis for several reasons; they are theoretically straightforward, preserve the full 3D information of the field and have an efficient estimator. However, its modelling on large scales does present a few challenges. A couple of which have been the focus on in this work; firstly, until recently integrated effects were challenging to include directly in the 3D power spectrum framework, secondly, the inclusion of wide-separation corrections which arise as we correlate objects on our past light-cone. 

Accounting for these corrections is essential for two primary goals of large-scale structure cosmology: performing a robust and unbiased analysis of PNG, and testing general relativity on the largest scales through the detection of relativistic effects.

We therefore present a framework in which to include integrated effects---lensing, ISW and time delay--- and in particular the I\texttimes I term, in the multipoles of the 3D Fourier power spectrum. This contribution requires the computation of numerical integrals from observer to source and is dominated by the lensing contribution for all but the smallest redshifts, and indeed at large redshifts and ultra-large scales (so for largest $k$-bins for MegaMapper) it can be the dominant contribution to the even multipoles of the power spectrum. We forecast that the lensing contribution should be detectable for \textit{Euclid}, MegaMapper and SKAO2-like surveys, however the ISW and time delays are unlikely to be detectable in the multipoles of the Fourier power spectrum. In contrast to previous studies, primarily due to the I\texttimes I contribution (e.g. see \cref{sec:comparison}), we have also shown that, for upcoming spectroscopic galaxy surveys, such as \textit{Euclid} and MegaMapper, that by neglecting these terms our best fit determination of local type $\fNL$ will be biased, by roughly $ 3 \, \sigma$ and $ 20 \, \sigma$, respectively, from the true values and therefore these effects have to be included in an analysis. This result, while novel in the 3D power spectrum, also may be considered to be expected considering how lensing contributes to the angular power spectrum.

Therefore one has to include integrated relativistic effects, along with the local effects, in their modelling in order to perform an unbiased analysis of $\fNL$. However, to do this one needs to account for the uncertainty in the survey luminosity function through the evolution and magnification biases. To illustrate this, we again focus on a local $\fNL$ analysis, in which the uncertainty in our bias functions leads to greater uncertainty ($\approx 30 \%$ for a \textit{Euclid}-like H$\alpha$ survey and $\approx 20 \%$ for a MegaMapper-like LBG survey) in the determination of $\fNL$. Multi-tracer approaches, like a bright-faint split where we split our observed galaxy sample based on observed luminosity, can help improve constraints here. First, by increasing the relative local PNG signal by using highly biased tracers, and secondly, by giving us better constraints on the uncertainty in our luminosity function - particularly from the odd multipoles.

To enable these forecasts, we present a novel calculation of the full multipole multi-tracer covariance, including relativistic corrections, \cref{sec:covariance}. We have for the first time, computed wide-separation corrections to the covariance which, while generally subdominant for the even multipole covariance, can be relevant on large scales for the odd-parity multipoles. Note that this is the wide-separation correction to the covariance not just the inclusion of the corrections from the power spectrum into the sample variance. This therefore involves computing wide-separation corrections to a 4-point function. We also have included integrated effects in the covariance and due to them breaking translation invariance we need to define within with respect to the LOS of our full 4-point function (\cref{ap:integrated_covariance}). Here, like in the signal, the integrated contribution becomes important on large scales and at higher redshifts. In the cross-power spectrum, odd multipoles, however, are sourced by odd imaginary contributions to the power spectrum, as well as shot noise.

In this work, we have omitted the effect of the convolution with the survey window function, which will have a significant impact the large scales considered here. The window convolution will both dampen the power on large scales as well as mix up the multipole decomposition, such that the even contribution will enter the odd multipoles, including in the covariance. The implementation and study of this needs to be considered in future work. We also have only considered linear perturbation theory and, while most of our forecasts are dominated by large scale effects, the inclusion of non-linear corrections, particularly for the I\texttimes I term which integrates over all $k$-modes above the measured $k$ (see \cref{sec:nonlinear_effects}), will be relevant in a realistic analysis. Additionally, all of our results are heavily dependent on the modelling assumptions we use for each survey, and in particular the choice of luminosity function.

Our results are implemented in the publicly available \textit{Python} package \textsc{CosmoWAP} \href{https://github.com/craddis1/CosmoWAP}{\faGithub}, including all of forecasts. A short tutorial notebook reproducing selected results is also provided. We plan to extend the formalism laid out here to include the shear-shear power spectrum and to higher-order statistics, such as the bispectrum.

\acknowledgments
CA and JH are supported by studentships from the UK Science and Technology Facilities Council (STFC). 
FM and SC acknowledge support from the Italian Ministry of University and Research (\textsc{mur}), PRIN 2022 `EXSKALIBUR – Euclid-Cross-SKA: Likelihood Inference Building for Universe's Research', Grant No.\ 20222BBYB9, CUP D53D2300252 0006, from the Italian Ministry of Foreign Affairs and International
Cooperation (\textsc{maeci}), Grant No.\ ZA23GR03, and from the European Union -- Next Generation EU. CC is supported by STFC grant ST/X000931/1. SLG and RM are supported by the South African Radio Astronomy Observatory and the National Research Foundation (grant no. 75415). SJ is supported by VIT Mauritius.

\section*{Code Availability Statement}
\textsc{CosmoWAP} is publicly available at \url{https://github.com/craddis1/CosmoWAP} and the companion repository \textsc{MathWAP} is available at \url{https://github.com/craddis1/MathWAP}. We also release a small tutorial notebook that recreates some of the results shown in this work.
\textsc{CosmoWAP} is built on top of the public code \href{https://github.com/lesgourg/class_public}{\textsc{class}} \cite{Lesgourgues2011,blas2011}, and the following \texttt{python} packages and libraries: \href{https://numpy.org/}{\textsc{numpy}} \cite{harris2020}, \href{https://scipy.org/}{\textsc{scipy}} \cite{2020SciPy}, \href{https://matplotlib.org/}{\textsc{matplotlib}} \cite{hunter2007}, \href{https://samreay.github.io/ChainConsumer/ChainConsumer}{\textsc{ChainConsumer}} \cite{Hinton_2016}, \href{https://cobaya.readthedocs.io/en/latest/}{\textsc{cobaya}} \cite{Torrado_2021} and \href{https://alessiospuriomancini.github.io/cosmopower/}{\textsc{CosmoPower}} \cite{Mancini_2022}.

\appendix

\section{Survey Details}\label{sec:survey_details}

We consider several different ongoing and future galaxy surveys, covering different tracers and redshift ranges, using redshift bins of constant width: a three-bin, 15,000 $\rm{deg}^2$ DESI-like BGS $(0.05<z<0.6)$, a five-bin, 15,000 $\rm{deg}^2$ H$\alpha$ \textit{Euclid}-like survey $(0.9<z<1.8)$, a five-bin, 2000 $\rm{deg}^2$ H$\alpha$ Roman-like survey $(0.5<z<2)$, a ten-bin, Stage V 20,000 $\rm{deg}^2$ MegaMapper-like Lyman-Break galaxy (LBG) survey over $(2<z<5)$ as well as a ten-bin, futuristic Phase-2 SKA (SKAO2) 30,000 $\rm{deg}^2$ \(\hi\) galaxy survey over $(0.1<z<2)$. For each tracer, we adopt an existing luminosity function from the literature, which we can use to calculate the evolution and magnification biases for a given flux or magnitude cut. The details of the luminosity function and the linear bias modelling for each survey is described below.

\paragraph{Euclid}

We adopt the `Model $3$' H$\alpha$ luminosity function \cite{Pozzetti_2016,Euclid_2020} and use a flux cut of $F_c = 2\times10^{-16} \, [{\rm erg \, \,cm}^{-2} s^{-1}]$. In addition, we also adopt a model of linear bias, $b_1(x,z)$, as a function of redshift, $z$ and luminosity $x$ of \cite{Pan_2020} (Table 2). These fits derive from the semi-analytic galaxy formation GALFORM \cite{Cole_2000,Gonzalez_2017}. The cumulative linear galaxy bias above a certain luminosity flux $x_c$ is then given by
\begin{equation}\label{eq:biasluminosity}
b_i(\ge x)= \frac{\int_{x_c}^{+\infty} \dd x \, b_i^h(x,z) \phi(x,z)}{\int \dd x \, \phi(x,z)}
\end{equation}
where $\phi(x,z)$ is the H$\alpha$ luminosity function.

\paragraph{Roman}

For the Roman-like H$\alpha$ survey our modelling is similar to the \textit{Euclid}-like H$\alpha$ survey but instead we adopt the `Model' 1 luminosity function from \cite{Pozzetti_2016} with a flux cut of $F_s = 1\times10^{-16} \, [{\rm erg \, \,cm}^{-2} s^{-1}]$. Our linear bias modelling is consistent with the H$\alpha$ model we use with the \textit{Euclid}-like survey but with a different luminosity function.

\paragraph{MegaMapper}

We consider some idealised MegaMapper-like Lyman Break Galaxy (LBG) survey following \cite{Wilson_2019}, where we assume a UV Schechter luminosity function \cite{Schechter_1976,Ono_2018} with an apparent magnitude cut, $m_c =24.5$. For linear clustering bias, we adopt a redshift and apparent magnitude, $m$, dependent function also from \cite{Wilson_2019},
\begin{equation}\label{eq:megamapperbias}
b(z,m) = A(m)(1+z) + B(m)(1+z)^2
\end{equation}
where $A(m) = -0.98(m-25) + 0.11$ and $B(m) = 0.12(m-25) + 0.17$. The bias model for the survey for a given apparent magnitude cut is then derived using \cref{eq:biasluminosity}.

\paragraph{BGS}
For the DESI-like BGS specifications, we assume a linear bias \cite{desi} 
\begin{equation}
b_1(z) = 1.34/D(z),
\end{equation}
and we adopt the BGS luminosity functions detailed in \cite{Maartens_2022} with an apparent magnitude cut, $m_c =20$. This is a Schechter type luminosity function, using values from \cite{Loveday_2012} with a K-correction following \cite{Jelic-Cizmek_2021}.

\paragraph{SKAO2}

For the SKAO2 \(\hi\) galaxy survey, we use a linear bias model from \cite{Yahya_2015,Bull_2016}
\begin{equation}
b_1(z) = 0.554 \, \rm{e}^{0.783 \, z}.
\end{equation}
Expressions for evolution and magnification bias as well as the number density are interpolated from Table 2 in \cite{Maartens_2022}.

\subsection{Bright-Faint Split}\label{sec:multi_tracer}

We consider a bright-faint split, as proposed by \cite{Bonvin_2014,Bonvin_2016,Gaztanaga_2017}, whereby the single galaxy population is split according to luminosity into two independent samples. Defining a flux split, $F_s$, the faint sample then corresponds to the population of galaxies with an observed flux above the detector threshold, $F_c$, but below $F_s$ while the bright catalogue just constitutes galaxies with an observed flux above $F_s$. Therefore, while treatment of the bright catalogue is relatively simple, as it can be considered as a regular survey with a flux cut at $F_s$, the faint catalogue biases depend on both the total and bright quantities.

For the faint populations, following \cite{Ferramacho_2014}, the linear bias, $b_1^F$, is given by
\begin{subequations}\label{eq:mt_bias}
\begin{align}
\label{eq:mt_b1}
b^F_1 & = \frac{n^T b^T_1 - n^B b^B_1}{n^F}
\end{align}
while following, \cite{Bonvin_2023,Blanco_2024}, the evolution, $b_e^F$, and magnification, $\mathcal{Q}^F$, biases can be expressed as 
\begin{align}
\label{eq:mt_Q}
\mathcal{Q}^F & = \frac{n^T}{n^T-n^B}\mathcal{Q}^B- \frac{n^B}{n^T-n^B}\mathcal{Q}^T
\\
\label{eq:mt_be}
b_e^F & = \frac{\partial \ln (n^T-n^B)}{\partial \ln (1+z)}.
\end{align}
\end{subequations}

We consider bright-faint splits primarily for the \textit{Euclid}-like H$\alpha$ and the MegaMapper-like LBG surveys, for which luminosity-dependent models of $b_1$ are available. Unless otherwise specified, we use splits of $F_s = 3\times10^{-16} \, [{\rm erg \, \,cm}^{-2} s^{-1}]$ and $m_s =24$, respectively.

\section{Cross-Multipole Multi-Tracer Covariance}\label{sec:covariance}

While it is common---and often advantageous---to compute the covariance from an ensemble of numerical simulations \cite[e.g.][]{Hamilton_2006}, there are several motivations for analytic covariances in some cases. Indeed, on ultra-large scales one not only needs larger and more computationally expensive simulations, but one also needs to consider the impact of large-scale systematics, such as relativistic and wide-separation corrections. Here, therefore we calculate the analytic covariances to isolate and study each individual contribution to covariance. Including relativistic effects from a suite of simulations/mocks will require them to include these effects, either through ray-tracing \cite[e.g.][]{Breton_2019,Adamek_2016} or some other method, which is a non-trivial addition to the covariance pipeline. 

The covariance for the even parity part of the power spectrum has been well studied \cite{Scoccimarro_1999,Chan_2016,Wadekar_2020}; however, while a few studies have considered the odd parity (multi-tracer) covariance \cite{Beutler_2020,Macdonald_2009}, these covariances are, in general, less well understood, and so here we provide a derivation for a full multipole analysis.

\begin{figure}
\centering
\includegraphics[width=0.49\linewidth]{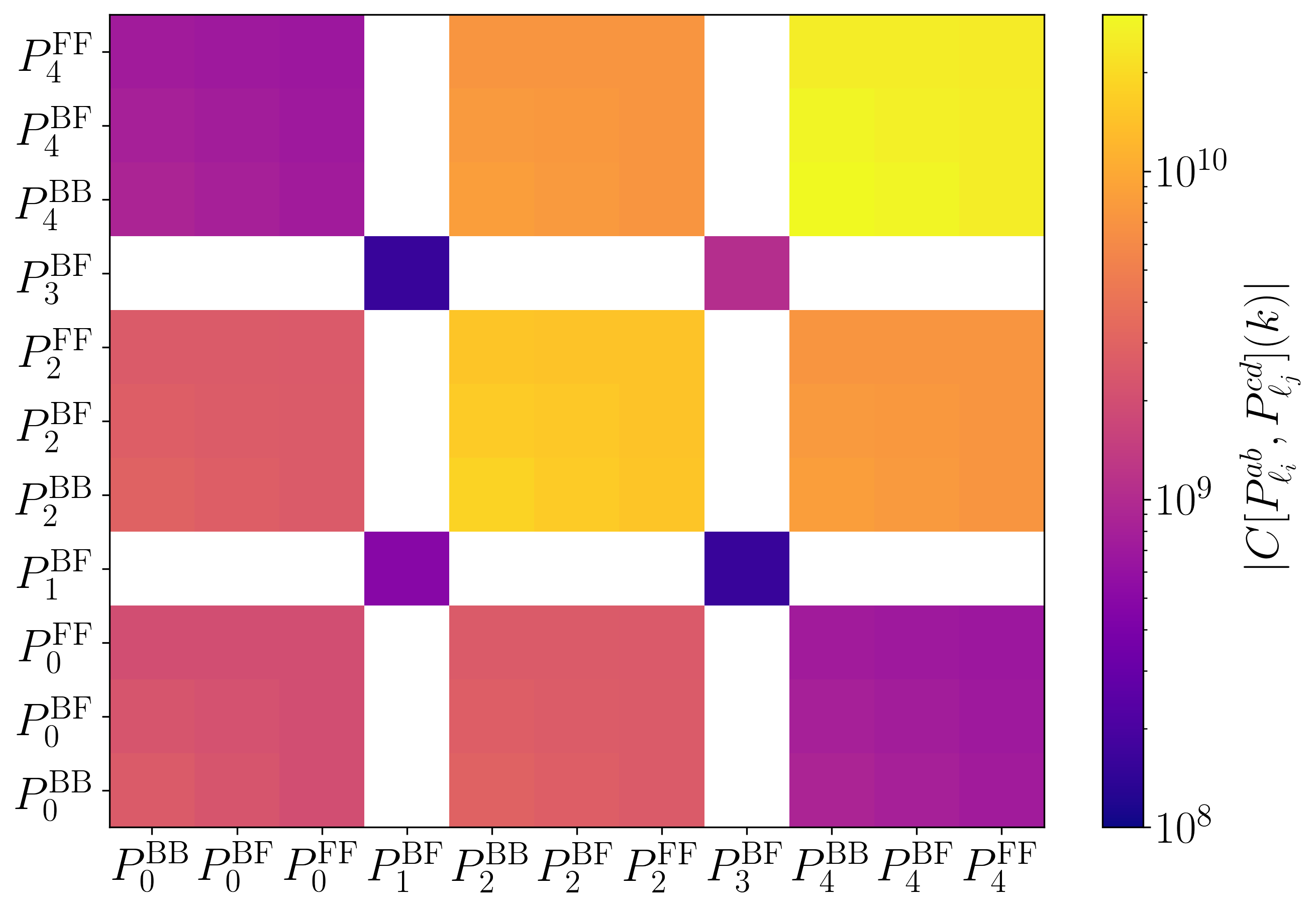}  
\includegraphics[width=0.49\linewidth]{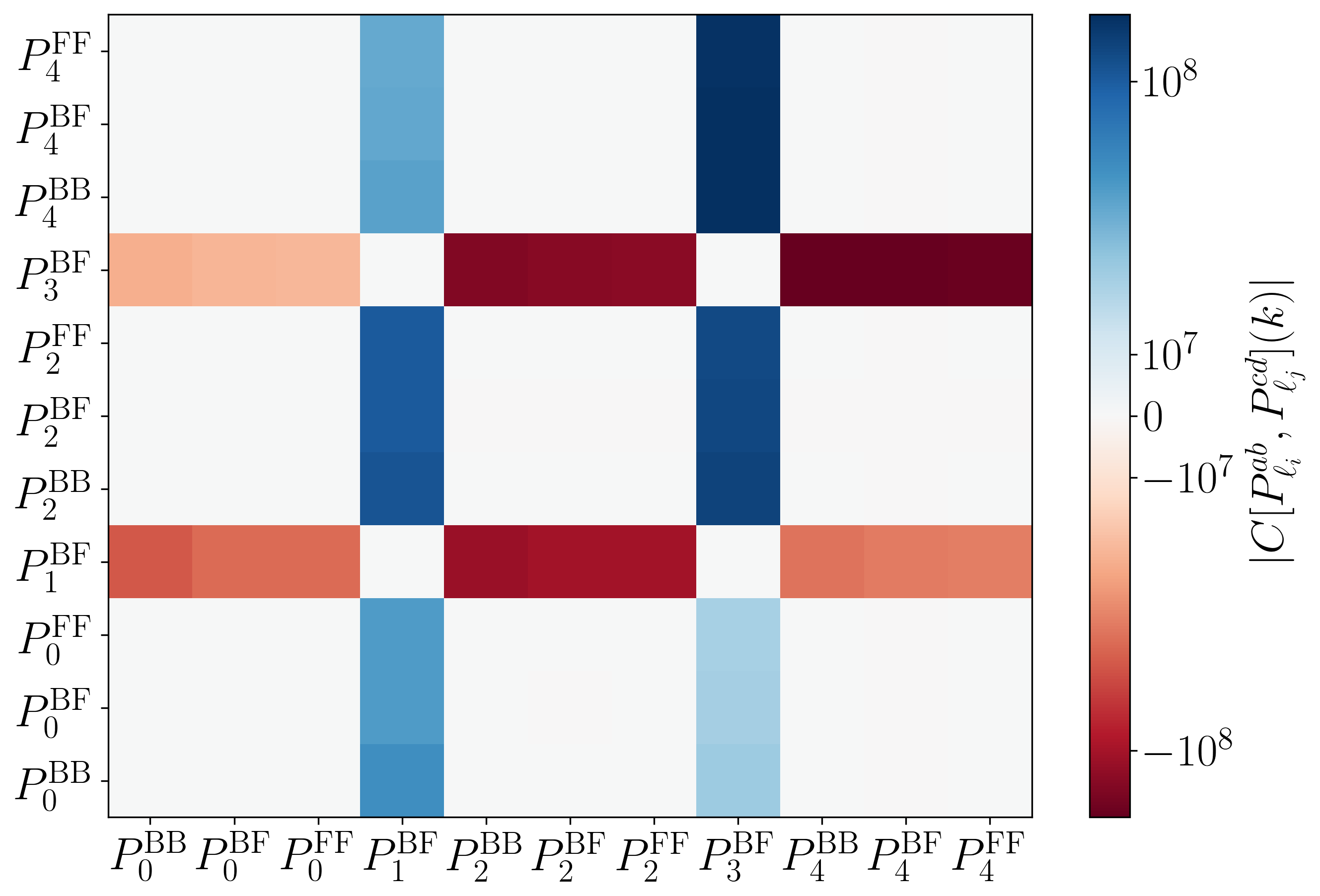} 
\caption{Real (left) and imaginary (right) parts of the full multipole covariance matrix for a multi-tracer analysis. We assume a bright-split of a \textit{Euclid}-like survey with $F_s=3\times10^{-16} \, [{\rm erg \, \,cm}^{-2} s^{-1}]$, for a redshift bin, $(1<z<1.2)$ and a $k$-bin centred at $k=0.018 \, [\aMpch]$, including all effects except wide-separation, to the covariance.} \label{fig:cov}
\end{figure}

So for a given data vector in a specific redshift bin, we can express the full covariance matrix as a block matrix for each multipole pair
\begin{equation}
\textsf{C} =
\begin{pmatrix}
{\textsf{C}}[P_{\ell_i}, P_{\ell_i}] & ... & {\textsf{C}}[P_{\ell_i}, P_{\ell_n}] \\
\vdots & \ddots & \vdots \\
{\textsf{C}}[P_{\ell_n}, P_{\ell_i}] & ... & {\textsf{C}}[P_{\ell_n}, P_{\ell_n}]
\end{pmatrix},
\end{equation}
where each cross/auto-multipole covariance block can then be written as a sub-matrix. For even-even multipoles, this is a $3 \times 3$ sub-matrix,
\begin{equation}
{\textsf{C}}(P_i, P_j) = 
\begin{pmatrix}
\textsf{C}[P^{XX}_{\ell_i},P^{XX}_{\ell_j}] & \textsf{C}[P^{XY}_{\ell_i},P^{XX}_{\ell_j}] & \textsf{C}[P^{YY}_{\ell_i},P^{XX}_{\ell_j}]\\
\textsf{C}[P^{XX}_{\ell_i},P^{XY}_{\ell_j}] & \textsf{C}[P^{XY}_{\ell_i},P^{XY}_{\ell_j}] &
\textsf{C}[P^{YY}_{\ell_i},P^{XY}_{\ell_j}]\\
\textsf{C}[P^{XX}_{\ell_i},P^{YY}_{\ell_j}] & \textsf{C}[P^{XY}_{\ell_i},P^{YY}_{\ell_j}] &
\textsf{C}[P^{YY}_{\ell_i},P^{YY}_{\ell_j}]
\end{pmatrix}
\end{equation}
but in the odd-odd case, since we just consider the cross-power spectrum, we have
\begin{equation}
{\textsf{C}}(P_{i=\rm{odd}}, P_{j=\rm{odd}}) = \textsf{C}[P^{XY}_{\ell_i},P^{XY}_{\ell_j}].
\end{equation}
Odd-even and even-odd covariance matrices are therefore of shape $1 \times 3$ and $3 \times 1$ respectively, e.g.
\begin{equation}
{\textsf{C}}(P_i, P_{j=\rm{odd}}) = 
\begin{pmatrix}
\textsf{C}[P^{XX}_{\ell_i},P^{XY}_{\ell_j}] & \textsf{C}[P^{XY}_{\ell_i},P^{XY}_{\ell_j}] & \textsf{C}[P^{YY}_{\ell_i},P^{XY}_{\ell_j}]\\
\end{pmatrix}.
\end{equation}
See \cref{fig:cov} for an overview of the full covariance for a given $k$-bin.

\subsection{Covariance Matrix Element}

Here, we compute the covariance of the of the multi-tracer power spectrum multipoles Yamamoto estimator, \cref{eq:yamamoto}. As we are considering the covariance of complex variables, the covariance matrix element is defined with a conjugation
\begin{equation}
\textsf{C}[\hat{P}^{ab}_{\ell_i},\hat{P}^{cd}_{\ell_j}](\bm{k}, \bm{q}) = \langle \hat{P}^{ab}_{\ell_i}(\bm{k}) \, \hat{P}^{cd}_{\ell_j}(\bm{q})^* \rangle.
\end{equation}
It is then useful to define the functions
\begin{equation}\label{eq:F_est}
F_{\ell_i}(\bm{k}) = \int_{\bm{x}} \, {\rm e}^{-i\,\bm{k}\cdot\bm{x}}\mathcal{L}_{\ell_i}(\hat{\bm{k}} \cdot \hat{\bm{x}}) \Delta(\bm{x}),
\end{equation}
such that, ignoring weights and windows, we can write the covariance element in terms of a Fourier space $4$-point correlation function
\begin{equation}
\begin{aligned}
\textsf{C}[P^{ab}_{\ell_i},P^{cd}_{\ell_j}](\bm{k},\bm{q}) = (2 \ell_i +1)(2\,\ell_j+1)\int \frac{d\Omega_{k}}{4\,\pi} \int \frac{d\Omega_{q}}{4\,\pi} \Big[ & \langle F^a_{\ell_i}(\bm{k}) F^b_0(-\bm{k}) F^c_{\ell_j}(\bm{q})^* F^d_0(-\bm{q})^* \rangle \\
& - \langle F^a_{\ell_i}(\bm{k}) F^b_0(-\bm{k}) \rangle \langle F^c_{\ell_j}(\bm{q})^* F^d_0(-\bm{q})^* \rangle\Big].
\end{aligned}
\end{equation}
Note that $F_{\ell_i}(-\bm{k}) =(-1)^{\ell_i}F_{\ell_i}(\bm{k})^*$.

Breaking the four-point function and assuming Gaussianity (ignoring the connected trispectrum contribution), as well as assuming that the tracers are independent (e.g. see \cite{Smith_2009}), the covariance element simplifies to
\begin{equation}\label{eq:full_cov}
\begin{aligned}
\textsf{C}[P^{ab}_{\ell_i},P^{cd}_{\ell_j}](\bm{k},\bm{q}) = (2 \ell_i +1)(2\,\ell_j+1)\int \frac{d\Omega_{k}}{4\,\pi} \int \frac{d\Omega_{q}}{4\,\pi} &\Big[\langle F^a_{\ell_i}(\bm{k}) F^c_{\ell_j}(\bm{q})^* \rangle \langle F^b_0(-\bm{k}) F^d_0(\bm{q})\rangle \\
& \quad + \langle F^a_{\ell_i}(\bm{k}) F^d_0(\bm{q}) \rangle \langle F^b_0(-\bm{k}) F^c_{\ell_j}(\bm{q})^* \rangle \Big].
\end{aligned}
\end{equation}

This is the full covariance element for our estimator, the full calculation including wide-separation corrections to this quantity is outlined in \cref{sec:WS_cov}, but, for simplicity and understanding, the plane-parallel constant redshift case is detailed below.

\subsubsection{Plane-parallel Constant Redshift Limit}
Assuming the plane parallel limit $\hat{\bm{d}}=d_z$, enables one to take the legendre outside the integral,
\begin{equation}
F^a_{\ell_i}(\bm{k}) = \mathcal{L}_i (\hat{\bm{k}} \cdot \hat{\bm{x}}) F^a_0 (\bm{k})
\end{equation}
and assuming the full plane-parallel constant redshift limit, such that the four points we are correlating have the same underlying statistics,
\begin{equation}
\begin{aligned}
\textsf{C}[P^{ab}_{\ell_i},P^{cd}_{\ell_j}]&(k,q) = (2 \ell_i +1)(2\,\ell_j+1) \int \frac{\dd \Omega_{k}}{4\,\pi} \int \frac{\dd \Omega_{q}}{4\,\pi} \, \mathcal{L}_i (\hat{\bm{k}} \cdot \hat{\bm{d}})\,\mathcal{L}_j (\hat{\bm{q}} \cdot \hat{\bm{d}}) \\ & \times \big[ \hat{P}^{ac}(\bm{k},\bm{d})\langle F^b_0(-\bm{k}) F^d_0 (\bm{q}) \rangle \delta_D (\bm{k}-\bm{q})+\hat{P}^{ad}(\bm{k},\bm{d}) \langle F^b_0(-\bm{k})F^c_0(-\bm{q}) \rangle \delta_D (\bm{k}+\bm{q})\big]
\end{aligned}
\end{equation}
where the power spectrum includes shot noise contribution in the autocorrelation case assuming we have non-overlapping samples
\begin{equation}
\hat{P}^{ab}(\bm{k},\bm{d}) = P^{ab}_{\rm loc}(\bm{k},\bm{d}) + \frac{\delta^K_{a,b}}{n_a}.
\end{equation}
where $n_a$ is the number density of our given sample.

Simplifying this equation by switching an angular integral, as in \cref{eq:change_variables}, to an integral over $\bm{q}$ and closing the Dirac deltas, leads to a general expression for each covariance element
\begin{equation}\label{eq:cov_component}
\boxed{
\begin{aligned}
\textsf{C}[P^{ab}_{\ell_i},P^{cd}_{\ell_j}](k) = \frac{(2 \ell_i +1)(2\,\ell_j+1)}{N_k} \int \frac{\dd \Omega_{k}}{4\,\pi} \, \mathcal{L}_i (\mu) \big[ & \mathcal{L}_j (\mu) \hat{P}^{ac}(k,\mu)\hat{P}^{bd}(k,\mu)^* \\ &+ \mathcal{L}_j(-\mu) \hat{P}^{ad}(k,\mu)\hat{P}^{bc}(k,\mu)^* \big].
\end{aligned}}
\end{equation}

\paragraph{Odd Multipole Covariance}\label{ap:odd_multipole_cov}

To examine the odd multipole covariance, we can split our density field into real (even), $R$, and imaginary (odd), $I$, parts, such that the local power spectrum can be expressed as (ignoring shot-noise)

\begin{equation}
P^{ab}_{\rm loc}(k,\mu) = (R^a R^b + I^a I^b + i(I^a R^b - I^b R^a)) P(k),
\end{equation}
and therefore, using \cref{eq:cov_component}, one can see that the odd multipole diagonal covariance is proportional to 
\begin{equation}
\textsf{C}[P^{XY}_{\ell_i=\text{odd}},P^{XY}_{\ell_j=\text{odd}}](k) \propto \left(P^{XX}P^{YY}-P^{XY}P^{XY}\right),
\end{equation}
which can then be expressed in terms of the real and imaginary parts such that

\begin{equation}
\textsf{C}[P^{XY}_{\ell_i=\text{odd}},P^{XY}_{\ell_j=\text{odd}}](k) \propto 2(R^X I^Y)^2 + 2(I^X R^Y)^2 -  4\, R^X R^Y I^X I^Y
\end{equation}
This is simply the imaginary part of the power spectrum squared (as in \cite{Macdonald_2009}) and is zero if we neglect relativistic corrections. It is also even, real and always positive which is necessary for the covariance matrix to be positive semi-definite. The shot noise contribution to the covariance is also non-zero as it does not cancel because there is no shot-noise in the $P^{XY}$ case. For odd-even components, the (off-diagonal) covariance is both odd and imaginary and $\textsf{C}[P^{XY}_{\ell_i},P^{XY}_{\ell_j}](k)=\textsf{C}[P^{XY}_{\ell_j},P^{XY}_{\ell_i}](k)^*$ which is necessary for the covariance matrix to be hermitian.

\paragraph{Relativistic Corrections to the Sample Variance}

\begin{figure}[ht]
\centering
\includegraphics[width=0.49\linewidth]{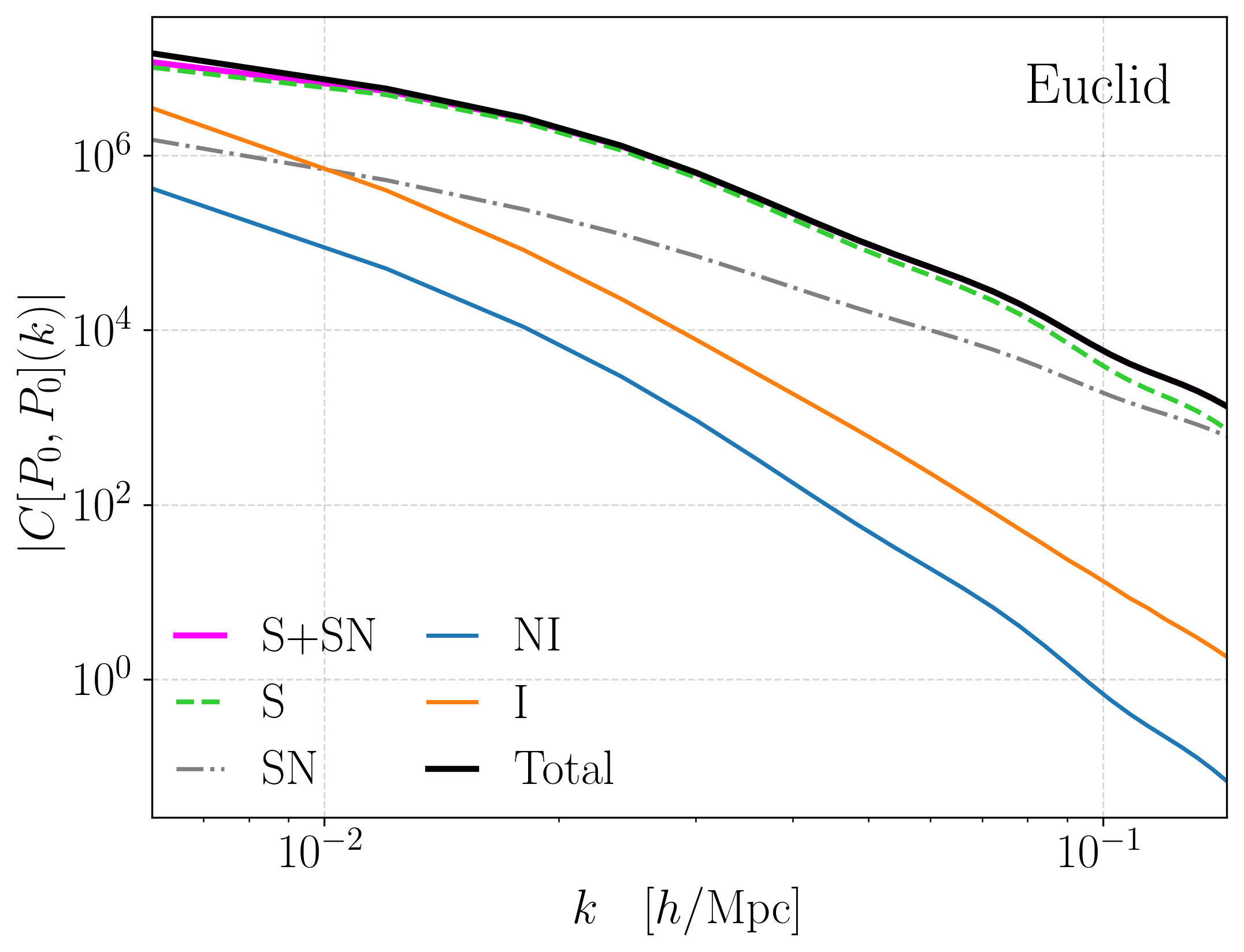}
\includegraphics[width=0.49\linewidth]{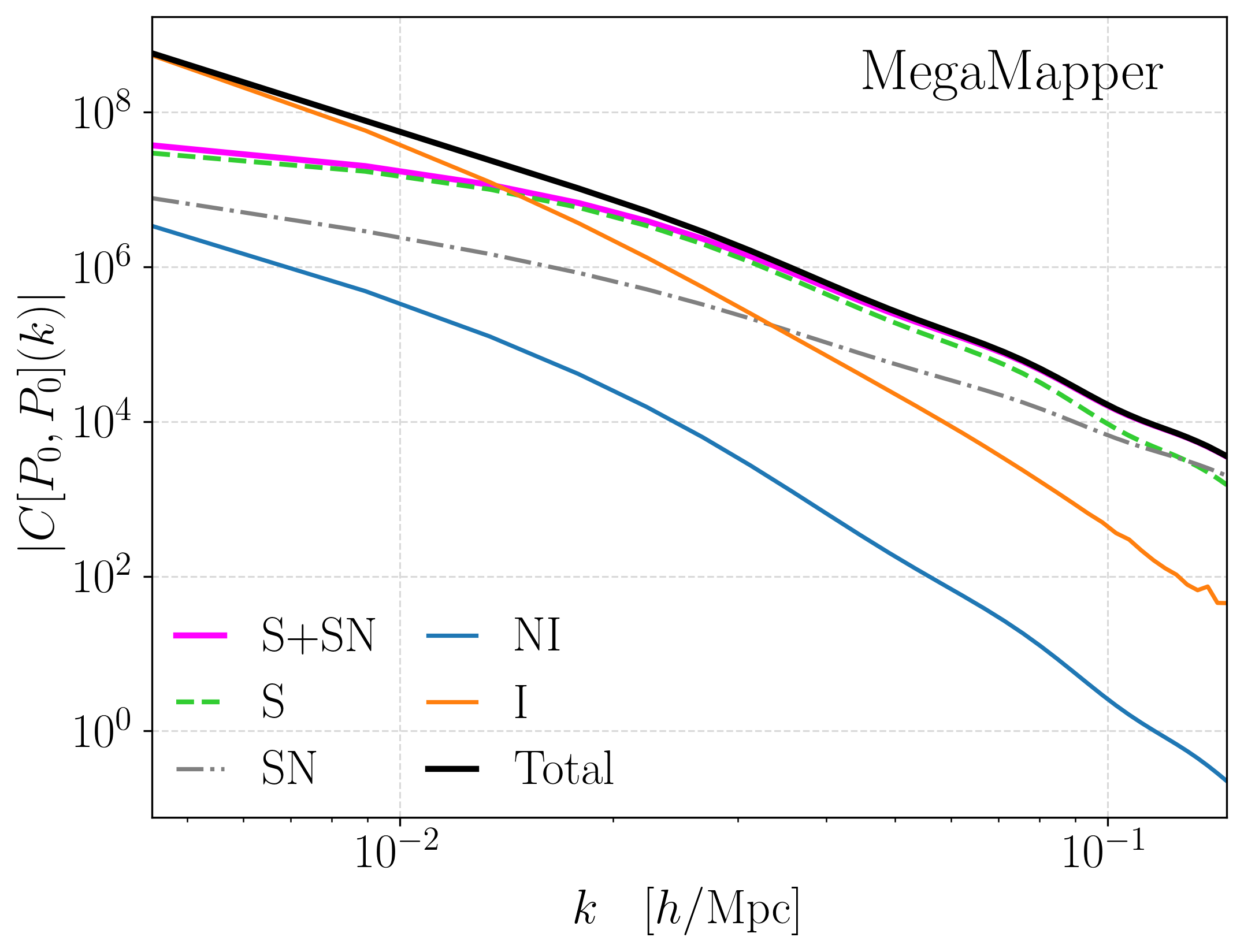}
\caption{Contributions to the monopole covariance for a \textit{Euclid}-like redshift bin $(1.26<z<1.44)$ left and a MegaMapper-like redshift bin $(3.5<z<3.8)$, where I represents all integrated contributions and SN is the shot-noise contribution. As we have cross-contributions in the covariance each additional corrections contains its mixing with the other contribution. So SN contains SN-S, the NI term contains the NI-S contribution and the NI-SN term and I contains I-S, I-SN and I-NI contributions. These are plotted over the same $k$-bins we use in our forecasts for this single redshift bin.}
\label{fig:cov_components}
\end{figure}

As the integrated contributions to the power spectrum have a LOS dependence we need to include them in the full wide-separation framework (\cref{ap:integrated_covariance}), however we examine their impact on the sample variance with respect to the local relativistic contributions in this section. The integrated relativistic contributions to the monopole covariance are greater than the local relativistic contributions (see \cref{fig:cov_components}), and become relevant on large scales, just like the integrated contribution to the power spectrum itself. At higher redshifts the integrated effects are comparatively larger and indeed for MegaMapper the integrated contribution is greater than the Kaiser contribution for the largest scale $k$-bins where $k<0.01 \, [\aMpch]$.

\subsubsection{Wide-Separation Corrections to the Covariance}\label{sec:WS_cov}

Here, we calculate the wide-separations corrections to the covariance of the Yamamoto estimator. Following on from \cref{eq:full_cov}, the full multi-tracer Gaussian multipole covariance can then be written, using \cref{eq:F_est}, as
\begin{equation}
\begin{aligned}
\textsf{C}[P^{ab}_{\ell_i},P^{cd}_{\ell_j}](k)=(2 \ell_i +1)(2\,\ell_j&+1) \int \frac{\dd \Omega_k}{4\,\pi} \frac{\dd \Omega_q}{4\,\pi}  \int_{\bm{x}_1,\bm{x}_2,\bm{x}_3,\bm{x}_4}  {\rm e}^{-i\,\bm{k} \cdot \bm{x}_1}{\rm e}^{i\,\bm{k} \cdot \bm{x}_3} \mathcal{L}_i(\hat{\bm{k}} \cdot \hat{\bm{x}}_1)\\ & \times \bigg[{\rm e}^{i\,\bm{q} \cdot \bm{x}_2} {\rm e}^{-i\,\bm{q} \cdot \bm{x}_4}\langle \Delta^a_g(\bm{x}_1) \Delta^c_g(\bm{x}_2) \rangle \langle \Delta^b_g(\bm{x}_3)\,\Delta^d_g(\bm{x}_4) \rangle\mathcal{L}_j(\hat{\bm{q}} \cdot \hat{\bm{x}}_2)\\  & \quad+{\rm e}^{-i\,\bm{q} \cdot \bm{x}_2} {\rm e}^{i\,\bm{q} \cdot \bm{x}_4}\langle \Delta^a_g(\bm{x}_1) \Delta^d_g(\bm{x}_2) \rangle \langle \Delta^b_g(\bm{x}_3)\,\Delta^c_g(\bm{x}_4) \rangle\mathcal{L}_j(\hat{\bm{q}} \cdot \hat{\bm{x}}_4)\bigg].
\end{aligned}
\end{equation}
for the $t=0$ Yamamoto estimator. If alternatively, we defined the covariance for the $t=1$ estimator, then the $\bm{x}_i$ dependence in the legendre polynomials change: $\mathcal{L}_i(\hat{\bm{k}} \cdot \hat{\bm{x}}_1)\rightarrow \mathcal{L}_i(\hat{\bm{k}} \cdot \hat{\bm{x}}_3)$ and $\mathcal{L}_j(\hat{\bm{k}} \cdot \hat{\bm{x}}_2)\leftrightarrow \mathcal{L}_j(\hat{\bm{k}} \cdot \hat{\bm{x}}_4)$. Our covariance is correlating the density field at $4$ different points in the sky and under the assumption local statistical homogeneity, we can model this with one single LOS and as such our covariance is a function of a single $k$-vector. wide-separation corrections, then arise just as in the case of the power spectrum, from the breaking of this local statistical homogeneity, from RSD (wide-angle effects) and evolution of the signal with redshift (radial evolution effects). To calculate wide-separation effects the covariance we adopt the formalism developed in \cite{Noorikuhani:2022bwc,Addis_2025}.

\begin{figure}
\centering
\includegraphics[width=0.6\linewidth]{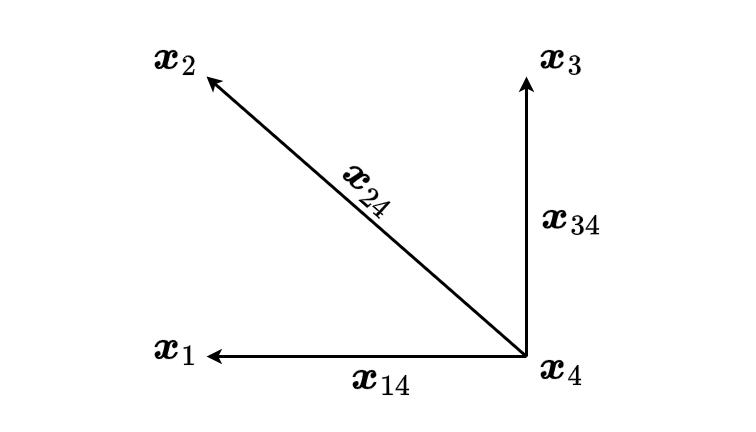}  
\caption{Position space local 4-point function. We define our basis with respect $\bm{x}_4$ which we pick as our LOS to simplify the calculations.} \label{fig:local_tk}
\end{figure}

It is then helpful to define our `local' covariance
\begin{equation}\label{eq:local_cov}
    \textsf{C}[P^{ab}_{\ell_i},P^{cd}_{\ell_j}](k) = \frac{1}{V_s}\int_{\bm{d}} C^{\rm loc}[P^{ab}_{\ell_i},P^{cd}_{\ell_j}](k;d),
\end{equation}
which is defined in a local region in which we perform our wide-separation series expansion in our $4$-point function.

Defining a basis, in our local $4$-point function as shown in \cref{fig:local_tk}, with vectors $\bm{x}_{14},\bm{x}_{24},\bm{x}_{34}$ which simplifies the calculation for $\bm{d}=\bm{x}_4$ which is a natural choice for the LOS. The choice of LOS here is not too important; we do not `choose' a LOS for our covariance like we do in our estimator (either endpoint). The choice of endpoint LOS in our estimator is already encoded in the covariance and as such the LOS choice only denotes the point at which we Taylor expand from to include wide-separation corrections.

Therefore, using $\bm{x}_{12}=\bm{x}_{14}-\bm{x}_{24},\bm{x}_{23}=\bm{x}_{24}-\bm{x}_{34}$ and fixing $\bm{d}=\bm{x}_4$
\begin{equation}\label{eq:local_trispectrum}
\begin{aligned}
    C^{\rm loc}[P^{ab}_{\ell_i},P^{cd}_{\ell_j}](k;d)  = (2 \ell_i +1)(2\,\ell_j+1)&\int \frac{\dd \Omega_k}{4\,\pi} \frac{\dd \Omega_q}{4\,\pi} \int_{\bm{x}_{14},\bm{x}_{24},\bm{x}_{34}}  {\rm e}^{-i\,\bm{k} \cdot \bm{x}_{14}}   {\rm e}^{i\,\bm{k} \cdot \bm{x}_{34}} \mathcal{L}_i(\hat{\bm{k}} \cdot \hat{\bm{x}}_1)
    \\ & \times \bigg[{\rm e}^{i\,\bm{q} \cdot \bm{x}_{24}} \hat{\xi}^{ac}_{\rm loc}(\bm{x}_1, \bm{x}_{2}) \hat{\xi}^{bd}_{\rm loc}(\bm{x}_3, \bm{x}_{4})\,\mathcal{L}_j(\hat{\bm{q}} \cdot \hat{\bm{x}}_2)  \\ & \quad + \,{\rm e}^{-i\,\bm{q} \cdot \bm{x}_{24}}\hat{\xi}^{ad}_{\rm loc}(\bm{x}_1, \bm{x}_{2}) \hat{\xi}^{bc}_{\rm loc}(\bm{x}_3, \bm{x}_{4})\,\mathcal{L}_j(\hat{\bm{q}} \cdot \hat{\bm{x}}_4)\bigg]
\end{aligned}
\end{equation}
where our local $2$-point functions are defined, including shot-noise contributions (for simplicity and visualization we only compute wide-separation corrections on the non-integrated terms, though we could include the full local $2$-point function as in \cref{eq:local_2point}), by:
\begin{equation}\label{eq:local_2point_cov}
    \hat{\xi}^{ab}_{\rm loc}(\bm{x}_i, \bm{x}_{j}) = \int_{\bm{q}'} {\rm e}^{i\bm{q}' \cdot \bm{x}_{ij}} \, \left[\mathcal{K}^a(\bm{q}',\bm{x}_i)\mathcal{K}^b(-\bm{q}',\bm{x}_j) P(\bm{q}') + \frac{\delta^K_{a,b}}{n_a} \right].
\end{equation}
Now our next step is to reparameterise our kernels in terms of $\bm{d},\bm{x}_{14},\bm{x}_{24},\bm{x}_{34}$ such that our endpoint vectors become
\begin{equation}
    \bm{x}_i =  \bm{d} + \bm{x}_{i4},
\end{equation}
and the scalar magnitudes can be rewritten as
\begin{equation} 
    x_i = d \sqrt{1 + 2 \, \mu_{i4} \, \epsilon_{i4} + \epsilon_{i4}^2}.
\end{equation}
Where $\mu_{i4}=\hat{\bm{x}}_{i4}\cdot \hat{\bm{d}}$ and we have defined our three expansion parameters, $\epsilon_i = x_{i4}/d$, for $i=1,2,3$.

The approach is then to Taylor expand any $\bm{x}_1,\bm{x}_2,\bm{x}_3$ dependence in terms of $\epsilon_{14},\epsilon_{24},\epsilon_{34}$ respectively, about the point $\epsilon_{14}=\epsilon_{24}=\epsilon_{34}=0$, to incorporate these corrections from the breaking of statistical homogeneity in our local $4$-point functions, \cref{eq:local_trispectrum}. The `wide-angle' part comes as we Taylor expand the unit vectors $\hat{\bm{x}}_i = \bm{d}+\bm{x}_{i4}/\bm{x}_i$ while the `radial evolution' part arises as we Taylor expand for any redshift dependent function (and therefore comoving distance dependent), $f(x_i)$.

Taylor expanding and truncating at some order, and expressing $\bm{x}_{i4}$ in terms of their Cartesian vector components, we can then express our local $4$-point as a sum over powers of our expansion parameters, $\epsilon_{14},\epsilon_{24},\epsilon_{34}$

\begin{equation}\label{eq:bk_int_expansion}
    \begin{split}
         C^{\rm loc}[P^{ab}_{\ell_i},P^{cd}_{\ell_j}](k;d) = & \frac{(2 \ell_i +1)(2\,\ell_j+1)}{N_k} \int \frac{\dd \Omega_{k}}{4\,\pi} \int_{\bm{q},\bm{x}_{14},\bm{x}_{24},\bm{x}_{34},\bm{q}',\bm{k}'} {\rm e}^{-i (\bm{k}-\bm{k}') \cdot \bm{x}_{14}} {\rm e}^{i\,(\bm{k}+\bm{q}') \cdot \bm{x}_{34}} \\ &\times\sum_{i_x,i_y,i_z}\epsilon_{14}^{(i_x+i_y+i_z)}\left[\left(\frac{x_{14,x}}{x_{14}}\right)^{i_x}\left(\frac{x_{14,y}}{x_{14}}\right)^{i_y}\left(\frac{x_{14,z}}{x_{14}}\right)^{i_z} \right]\\ &\times \sum_{j_x,j_y,j_z}\epsilon_{24}^{(j_x+j_y+j_z)}\left[\left(\frac{x_{24,x}}{x_{24}}\right)^{j_x}\left(\frac{x_{24,y}}{x_{24}}\right)^{j_y}\left(\frac{x_{24,z}}{x_{24}}\right)^{j_z} \right]\\ &\times \sum_{m_x,m_y,m_z}\epsilon_{34}^{(m_x+m_y+m_z)}\left[\left(\frac{x_{34,x}}{x_{34}}\right)^{m_x}\left(\frac{x_{34,y}}{x_{34}}\right)^{m_y}\left(\frac{x_{34,z}}{x_{34}}\right)^{m_z} \right]\\&\times \mathcal{C}^{a,b,c,d}_{i_x,i_y,i_z,j_x,j_y,j_z,m_x,m_y,m_z}(\bm{k},\bm{q},\bm{k}',\bm{q}',\bm{d}),
    \end{split}
\end{equation}
where $\mathcal{C}^{a,b,c,d}_{i_x,i_y,i_z,j_x,j_y,j_z,m_x,m_y,m_z}$ is a coefficient that contains the $\bm{k},\bm{d}$ dependence of the covariance and reduces to
\begin{equation}
\begin{aligned}
    \mathcal{C}^{a,b,c,d}_{0,...,0}&(\bm{k},\bm{q},\bm{k}',\bm{q}',\bm{d}) = \mathcal{K}^{a}(\bm{k}',\bm{d})\mathcal{K}^{b}(\bm{q}',\bm{d})P(\bm{k}')P(\bm{q}')\,\mathcal{L}_i(\hat{\bm{k}} \cdot \hat{\bm{d}})\,\mathcal{L}_j(\hat{\bm{q}} \cdot \hat{\bm{d}}) \\ & \times
    \bigg[\mathcal{K}^{c}(-\bm{k}',\bm{d})\mathcal{K}^{d}(-\bm{q}',\bm{d}){\rm e}^{-i (\bm{k}'-\bm{q}) \cdot \bm{x}_{24}} + \mathcal{K}^{d}(-\bm{k}',\bm{d})\mathcal{K}^{c}(-\bm{q}',\bm{d}) {\rm e}^{-i (\bm{k}'+\bm{q}) \cdot \bm{x}_{24}}\bigg]
    \\ & + \,\mathcal{L}_i(\hat{\bm{k}} \cdot \hat{\bm{d}})\,\mathcal{L}_j(\hat{\bm{q}} \cdot \hat{\bm{d}}) \Bigg[
    \\ & \qquad {\rm e}^{-i (\bm{k}'-\bm{q}) \cdot \bm{x}_{24}} \left( \frac{\delta^K_{a,c}}{n_a}\, \mathcal{K}^{b}(\bm{q}',\bm{d})\mathcal{K}^{d}(-\bm{q}',\bm{d}) P(\bm{q}') + \frac{\delta^K_{b,d}}{n_b}\, \mathcal{K}^{a}(\bm{k}',\bm{d})\mathcal{K}^{c}(-\bm{k}',\bm{d}) P(\bm{k}')\right)
    \\ & \qquad + {\rm e}^{-i (\bm{k}'+\bm{q}) \cdot \bm{x}_{24}} \left( \frac{\delta^K_{a,d}}{n_a}\, \mathcal{K}^{b}(\bm{q}',\bm{d})\mathcal{K}^{c}(-\bm{q}',\bm{d}) P(\bm{q}') + \frac{\delta^K_{b,c}}{n_b}\, \mathcal{K}^{a}(\bm{k}',\bm{d})\mathcal{K}^{d}(-\bm{k}',\bm{d}) P(\bm{k}')\right)
    \\ & \qquad + \left({\rm e}^{-i (\bm{k}'-\bm{q}) \cdot \bm{x}_{24}}\frac{\delta^K_{a,c}}{n_a} \frac{\delta^K_{b,d}}{n_b}+{\rm e}^{-i (\bm{k}'+\bm{q}) \cdot \bm{x}_{24}}\frac{\delta^K_{a,d}}{n_a} \frac{\delta^K_{b,c}}{n_b}\right)\Bigg],
\end{aligned}
\end{equation}
at zeroth-order (in the plane-parallel constant redshift limit). The second square bracket here includes the shot-noise contribution to the covariance. In this case, we get back \cref{eq:cov_component}.

However, to compute the full covariance we then using the Fourier relation,
\begin{equation}
    \frac{\partial}{\partial k_j}F(\bm{k}) = -i \int \dd^3 \bm{x}\,\, {\rm e}^{-i\,\bm{k}\cdot\bm{x}} x_j f(\bm{x}),
\end{equation}
such that we can remove the $x_{i4,j}$ dependence from inside the integral as $x_{14,j} \rightarrow i\partial_{(k-k')_{j}}$, $x_{34,j} \rightarrow -i\partial_{(k+q')_{j}}$ and then $x_{24,j} \rightarrow i\partial_{(k\pm q)_{j}}$, where the $\pm$ depends on which $\bm{x}_{24}$ exponential is being considered. Once there is no $\bm{x}_{i4}$ dependence left in the real space integrals, we form Dirac deltas ($\delta_D(\bm{k}-\bm{k}'), \, \delta_D(\bm{k}+\bm{q}'), \, \delta_D(\bm{k}\pm\bm{q})$) from the $\bm{x}_{i4}$ integrals which close the $\bm{k}',\bm{k}',\bm{q}$ integrals, such that the covariance is dependent on a single $k$-magnitude. Therefore the local covariance can be expressed such that there are $n$ partial derivatives acting on each coefficient for each term in the series
\begin{equation}\label{eq:WSfullexpansion}
    \begin{split}
        C^{\rm loc}[P^{ab}_{\ell_i},P^{cd}_{\ell_j}](k;d) &= \sum_n \left(\frac{i}{d}\right)^{n} \\
        &\quad \times \sum^{i_x+i_y+i_z+j_x+j_y+j_z+m_x+m_y+m_z=n}_{i_x,i_y,i_z,j_x,j_y,j_z,m_x,m_y,m_z} \left(\partial^{i_x+j_x+m_x}_{k_{x}}\partial^{i_y+j_y+m_y}_{k_{y}}\partial^{i_z+j_z+m_z}_{k_{z}} \right)\\
        &\quad \times (-1)^{j_x+j_y+j_z} \mathcal{C}_{i_x,i_y,i_z,j_x,j_y,j_z,m_x,m_y,m_z}\left(\bm{k},\pm \bm{k},\bm{k},\bm{d}\right).
    \end{split}
\end{equation}
The full covariance with wide-separation corrections is then given by averaging over the survey as in \cref{eq:local_cov}. 

\begin{figure}[ht]
\centering
\includegraphics[width=0.49\linewidth]{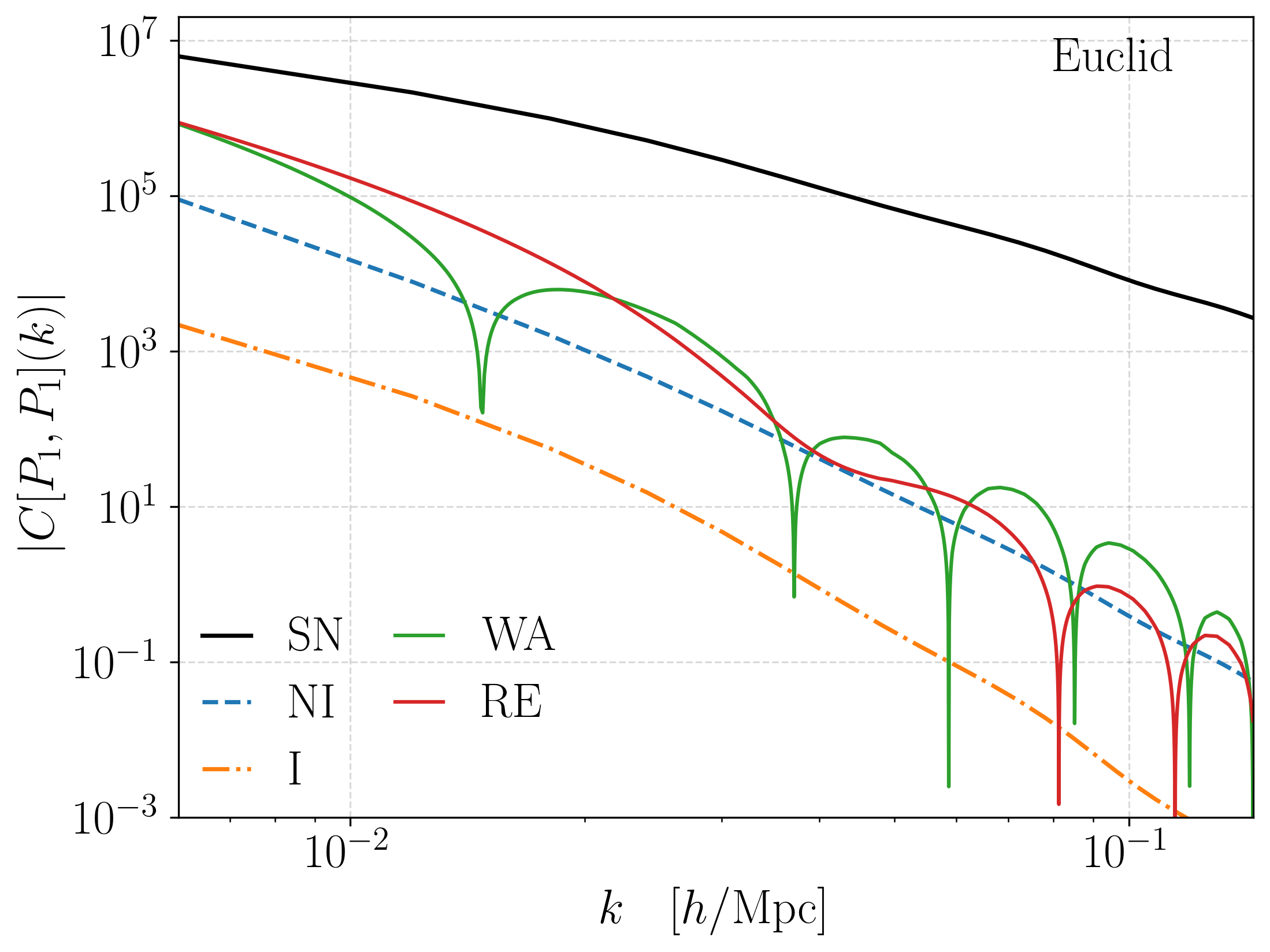}
\includegraphics[width=0.48\linewidth]{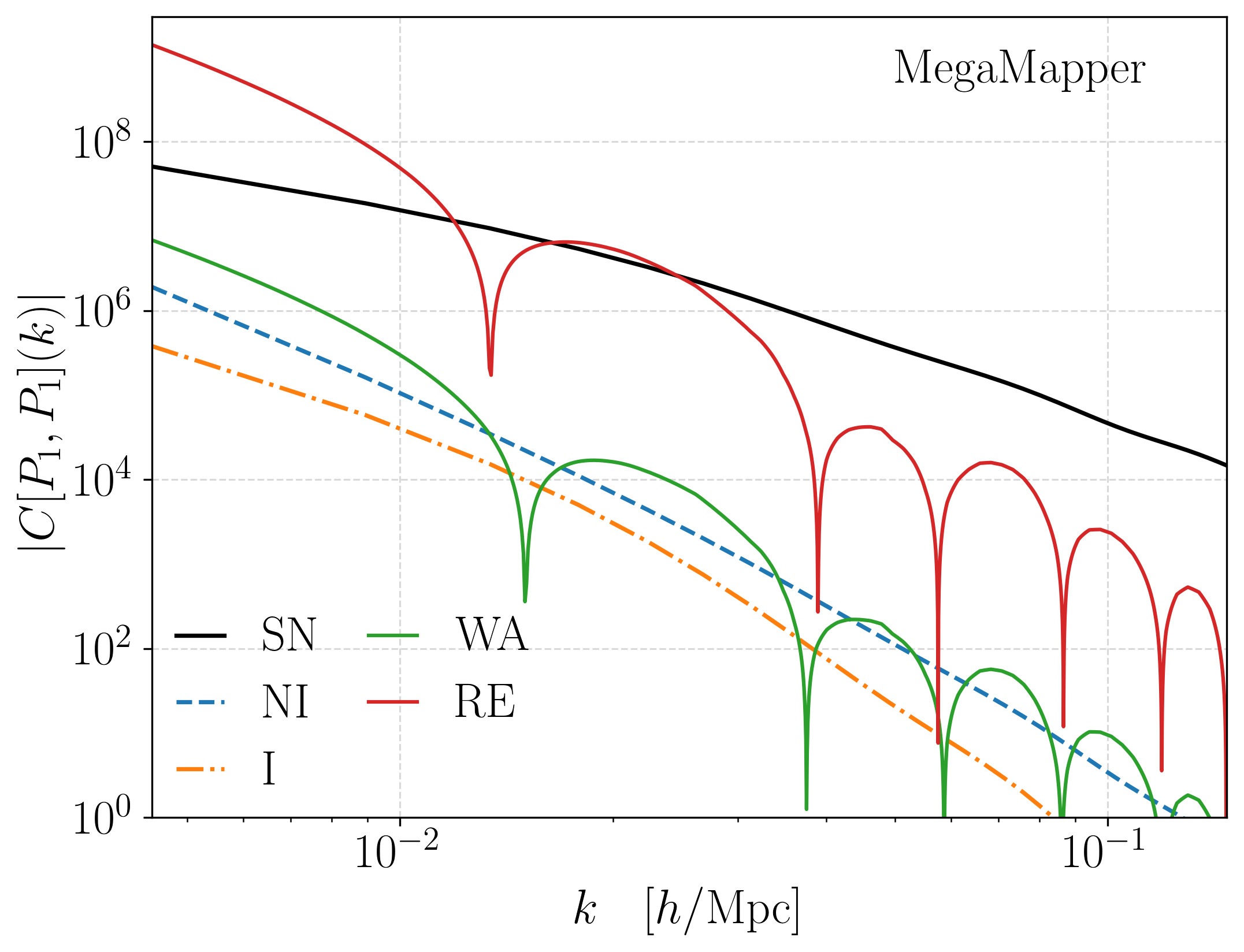}
\caption{Contributions to the bright-faint dipole covariance for a \textit{Euclid}-like redshift bin $(1.26<z<1.44)$ left and a MegaMapper-like redshift bin $(3.5<z<3.8)$, where I represents all integrated contributions and SN is the shot-noise contribution. The wide-angle term (WA) is represented in green and the radial evolution term (RE) is represented in red. These are plotted over the same $k$-bins we use in our forecasts for this single redshift bin.}
\label{fig:cov_components_WS}
\end{figure}

We truncate this expansion at $n=2$ for simplicity in our analysis. While these corrections are unimportant to the even multipoles, they do make a significant impact to the odd multipole covariance, \cref{fig:cov_components_WS}. For the odd multipoles there is no pure Kaiser contribution, instead the leading contribution for the \textit{Euclid} and MegaMapper-like samples is shot-noise. On large scales, the wide-separation, particularly the radial evolution, corrections to the dipole covariance become non-negligible. Notably, for the surveys considered here, these are larger than the contribution from the imaginary part of the power spectrum (\cref{ap:odd_multipole_cov}). Therefore, wide-separation corrections to the odd multipole covariance for a multi-tracer analysis should be considered in future analyses, especially for samples with high number densities. Though generally, the wide-angle corrections will naturally be included if one computes the covariance from a suite of simulations, the radial evolution corrections requires simulations that have the correct non-discrete redshift evolution.

\subsection{Integrated Contributions to the Covariance}\label{ap:integrated_covariance}

Due to their LOS dependence in the local power spectrum, we need to define the integrated contributions with respect to our LOS choice in our covariance, $\bm{d}=\bm{x}_4$. In this case, this is equivalent to computing integrated power spectrum with $t=1$. We outline this for the I\texttimes I contribution below.

The integrated contribution to the local 2-point function (this is an additional term to \cref{eq:local_2point_cov}) can be written as,
\begin{equation}
    \hat{\xi}^{ab, \text{I}\times \text{I}}_{\rm loc}(\bm{x}_i, \bm{x}_{j}) = \int_0^1\dd y_i \int_0^1\dd y_j\int_{\bm{q}'} {\rm e}^{i\bm{q}' \cdot \bm{r}_{ij}} \, \left[ x_i \, x_j\, \mathcal{K}^a(\bm{q}',\bm{x}_i,r)\mathcal{K}^b(-\bm{q}',\bm{x}_j,r) P(\bm{q}') \right],
\end{equation}
and then re-expressing the exponential using $\bm{x}_i=\bm{d}+\bm{x}_{i4}$
\begin{equation}
\begin{aligned}
    \hat{\xi}^{ab, \text{I}\times \text{I}}_{\rm loc}(\bm{x}_i, \bm{x}_{j}) = \int_0^1\dd y_i \int_0^1\dd y_j\int_{\bm{q}'} {\rm e}^{i\, y_i \,\bm{q}' \cdot \bm{x}_{i4}} & {\rm e}^{-i\, y_j\,\bm{q}' \cdot \bm{x}_{j4}} \, {\rm e}^{i\, (y_i-y_j)\,\bm{q}' \cdot \bm{d}}\\ &\times x_i \, x_j\,\mathcal{K}^a(\bm{q}',\bm{x}_i,r)\mathcal{K}^b(-\bm{q}',\bm{x}_j,r) P(\bm{q}').
\end{aligned}
\end{equation}
Therefore, substituting these expressions into \cref{eq:local_trispectrum} we can see we will end up with the Dirac-deltas ($\delta_D(\bm{k}-y_1\bm{k}'), \, \delta_D(\bm{k}+y_3\bm{q}'), \, \delta_D(y_2/y_1 \,\bm{k}\pm \bm{q})$). As the only $\bm{q}$ dependence is contained in a unit vector in the legendre polynomial, this last $y_2/y_1$ factor in the Dirac-delta is inconsequential and we can compute the integrated contributions in the manner described in the previous sections. Conveniently, this then corresponds to $t=1$ case, such that we can write an example integrated contribution to the covariance as
\begin{equation}
\begin{aligned}
\textsf{C}[P^{ab, \text{I}\times \text{I}}_{\ell_i},P^{cd}_{\ell_j}](k) = \frac{(2 \ell_i +1)(2\,\ell_j+1)}{N_k} \int \frac{\dd \Omega_{k}}{4\,\pi} \, \mathcal{L}_i (\mu) \big[ & \mathcal{L}_j (\mu) \hat{P}_{t=1}^{ac, \text{I}\times \text{S}}(k,\mu)\hat{P}_{t=1}^{bd, \text{S}\times \text{I}}(k,\mu)^* \\ &+ \mathcal{L}_j(-\mu) \hat{P}_{t=1}^{ad, \text{I}\times \text{S}}(k,\mu)\hat{P}_{t=1}^{bc, \text{S}\times \text{I}}(k,\mu)^* \big],
\end{aligned}
\end{equation}
for the zeroth order term of the wide-separation expansion.

\section{Integrated \texttimes\ Local Contribution to the Local Power Spectrum}\label{ap:local_IS}

Here we give a brief summary of the of mixing contribution between the integrated and local terms in power spectrum, which has also been calculated in \cite{Noorikuhani:2022bwc,Guedezounme:2024pbj}. The I\texttimes S contribution to the local $2$-point correlation, from the expansion in \cref{eq:local_2point}, is composed of $2$ asymmetric cross-terms

\begin{equation}
\begin{split}
\xi_{\rm loc}(\bm{x}_1,\bm{x}_2) = \int_{\bm{q}} \bigg[\int_0^{x_1} \dd r_1 \, &{\rm e}^{ i\,\bm{q} \cdot (\bm{r}_1-\bm{x}_2)} \mathcal{K}^{\rm I}(\bm{q},\bm{x}_1, r_1) \mathcal{K}(-\bm{q},\bm{x}_2) \\& +\int_0^{x_2} \dd r_2 \, {\rm e}^{ i\,\bm{q} \cdot (\bm{x}_1-\bm{r}_2)} \mathcal{K}(\bm{q},\bm{x}_1) \mathcal{K}^{\rm I}(-\bm{q},\bm{x}_2,r_2) \bigg]P (q)
\end{split}
\end{equation}
with two integrals separate 1D integrals, where $ \mathcal{K}(\bm{q},\bm{x})$ represents any combination of the local kernels -- though we still refer to this as the I\texttimes S term.

Then as with the I\texttimes I case, we can rearrange the exponentials, $\bm{r}_1=\frac{r_1}{x_1}\bm{x}_1, \, \bm{r}_2=\frac{r_2}{x_2}\bm{x}_2$ and $\bm{x}_1 = \bm{d} + t\bm{x}_{12}$, $\bm{x}_2 = \bm{d} - (1-t) \bm{x}_{12}$, such that
\begin{subequations}
\begin{align}
    {\rm e}^{i\,\bm{q} \cdot (\bm{r}_{1}-\bm{x}_{2})} = &\,{\rm e}^{i\,(\frac{r_1}{x_1}-1) (\bm{q} \cdot \bm{d})}{\rm e}^{i\,(\frac{r_1}{x_1}t+(1-t)) (\bm{q} \cdot \bm{x}_{12})}\\ {\rm e}^{i\,\bm{q} \cdot (\bm{x}_{1}-\bm{r}_{2})} = &\,
    {\rm e}^{i\,(1-\frac{r_2}{x_2}) (\bm{q} \cdot \bm{d})}{\rm e}^{i\,(t+(1-t)\frac{r_2}{x_2}) (\bm{q} \cdot \bm{x}_{12})}.
\end{align}
\end{subequations}
Taking the zeroth order term in our perturbative wide-separation corrections, and again redefining the integration variables $y_1=r_1/x_1, \, y_2 = r_2/x_2$, the I\texttimes S contribution to the local power spectrum can be written as
\begin{equation}
\begin{split}
P^{\rm I\times S}_{\text{loc}}(\bm{k};\bm{d})& = \int_{\bm{q},\bm{x}_{12}} \bigg[\int_0^{1} \dd y_1 \, {\rm e}^{ i\,\left(y_1-1 \right) (\bm{q} \cdot \bm{d})} \, {\rm e}^{ -i\,\left(\bm{k}-(1+t(y_1-1)) \bm{q} \right)\cdot \bm{x}_{12}}\, d \,\mathcal{K}^{\rm I}(\bm{q}, \bm{d}, d \, y_1) \mathcal{K}(-\bm{q}, \bm{d})\\
&+ \int_0^{1} \dd y_2 \, {\rm e}^{ i\,\left(1-y_2\right) (\bm{q} \cdot \bm{d})} \, {\rm e}^{ -i\,\left(\bm{k}-(y_2 + t(1-y_2)) \bm{q} \right) \cdot \bm{x}_{12}} d \,\mathcal{K}(\bm{q}, \bm{d}) \mathcal{K}^{\rm I}(-\bm{q}, \bm{d},d \, y_2)\bigg]P (q),
\end{split}
\end{equation}
Then, as there is now no $\bm{x}_{12}$ dependence left in the kernels, the $\bm{x}_{12}$ integral therefore becomes a Dirac-delta. Using the shifting property of the Dirac-delta and defining two functions $G_1(y_1) = (1 + t(y_1-1))$ and $G_2(y_2) = (y_2 + t(1-y_2))$, we can then simplify such that
\begin{equation}
\begin{split}
P^{\rm I\times S}_{\text{loc}}(k,\mu;d)& =  \int_0^{1} \dd y_1 \, {\rm e}^{ i\,\left(y_1-1 \right) k\, \mu\, d/G_1(y_1)}\, d \,G_1(y_1)^{-3} \mathcal{K}^{\rm I}(G_1(y_1)^{-1} k, \mu, d, y_1) \\ & \hspace{12em}\times\mathcal{K}(G_1(y_1)^{-1} k, -\mu) \, P (G_1(y_1)^{-1} k)\\
 &\qquad+ \int_0^{1} \dd y_2 \, {\rm e}^{ i\,\left(1-y_2 \right) k\, \mu\, d/G_2(y_2)} \, d \, G_2(y_2)^{-3} \mathcal{K}^{\rm I}(G_2(y_2)^{-1} k, -\mu, d, y_2)\\ & \hspace{12em}\times \mathcal{K}(G_2(y_2)^{-1} k,\mu) \, P (G_2(y_2)^{-1} k).
\end{split}\label{eq:IS_final}
\end{equation}

The numerical computation of these integrals is discussed in \cref{ap:numerics}.

\section{Local Relativistic Kernels}\label{ap:more_kernels}
The linear standard `Newtonian' Kaiser kernel is given by
\begin{equation}
\mathcal{K}_{\rm N}(\bm{q},\bm{x}) = D \left[b_1 + f (\hat{\bm{q}} \cdot \hat{\bm{x}})^2\right],
\end{equation}
and the linear local relativistic correction is given by 
\begin{equation}
\mathcal{K}_{\rm NI}(\bm{q},\bm{x})=D\left[i \, (\bm{q} \cdot \hat{\bm{x}}) \, \frac{\gamma_1}{q_1^2}+\frac{\gamma_2}{q_1^2}\right]
\end{equation}
where $\gamma_1$ and $\gamma_2$ contain the redshift dependence of the projection effects such that
\begin{subequations}\label{eq:gamma1gamma2}
\begin{align}
    \frac{\gamma_1}{\mathcal{H}} &= D\,f\left[b_e - 2\mathcal{Q}-\frac{2(1-\mathcal{Q})}{x_1 \, \mathcal{H}}-\frac{-(1+z)\mathcal{H}_z}{\mathcal{H}^2}\right],\\
    \begin{split}
        \frac{\gamma_2}{\mathcal{H}^2} &= D\,f(3-b_e) + \frac{3}{2}D\,\Omega_m \bigg[2 + b_e - f - 4 \mathcal{Q} -\frac{2(1-\mathcal{Q})}{x_1 \, \mathcal{H}}-\frac{-(1+z)\mathcal{H}_z}{\mathcal{H}^2}\bigg].
    \end{split}
\end{align}
\end{subequations}

\section{Bias from Non-linear Contributions to the Integrated Effects}\label{ap:nonlinear_effects}

\begin{figure}
\centering
\includegraphics[width=0.7\linewidth]{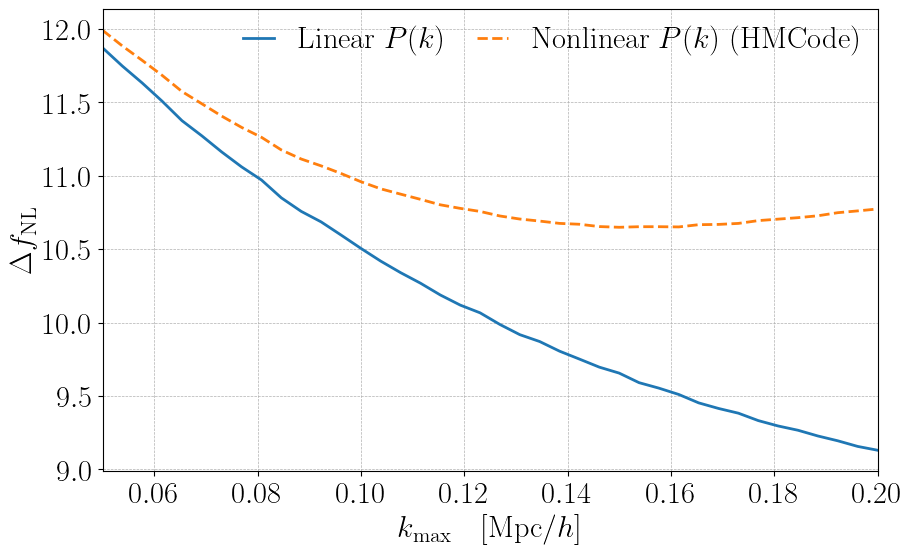} 
\caption{Forecasted bias from neglecting solely integrated (I\texttimes S and I\texttimes I) for a \textit{Euclid}-like analysis, for varying $k_{\rm max}$, on the best fit measurement of $\fNL$. For the dashed line we use the non-linear power spectrum instead of the linear power spectrum in the forecast.}
\centering
\label{fig:nonlin_bias}
\end{figure}

To quantify the impact of non-linear effects, \cref{sec:nonlinear_effects}, in our typical $\fNL$ analysis, \cref{fig:nonlin_bias} shows effect of this simple non-linear modelling on the bias on local type $\fNL$ from neglecting purely integrated effects (here we also use the non-linear power spectrum from the covariances). We can see compared to the linear case, there is an increase to the bias that increases with $k_{\rm max}$. Even though there is relatively modest constraining power on $\fNL$ at these small scales, we can see that the large increase in the amplitude of the integrated effects at these scales causes the bias on the measurement of $\fNL$ to increase.

In our forecasts, we also neglect the impact of fingers of god damping, however we note this can be trivially added to our pipeline; though it would have relatively small impact on the forecasts considered here due to our focus on large scale effects. 

\section{LOS Dependence of the Local Power Spectrum}\label{ap:los_depend}

\begin{figure}
\centering
\includegraphics[width=0.7\linewidth]{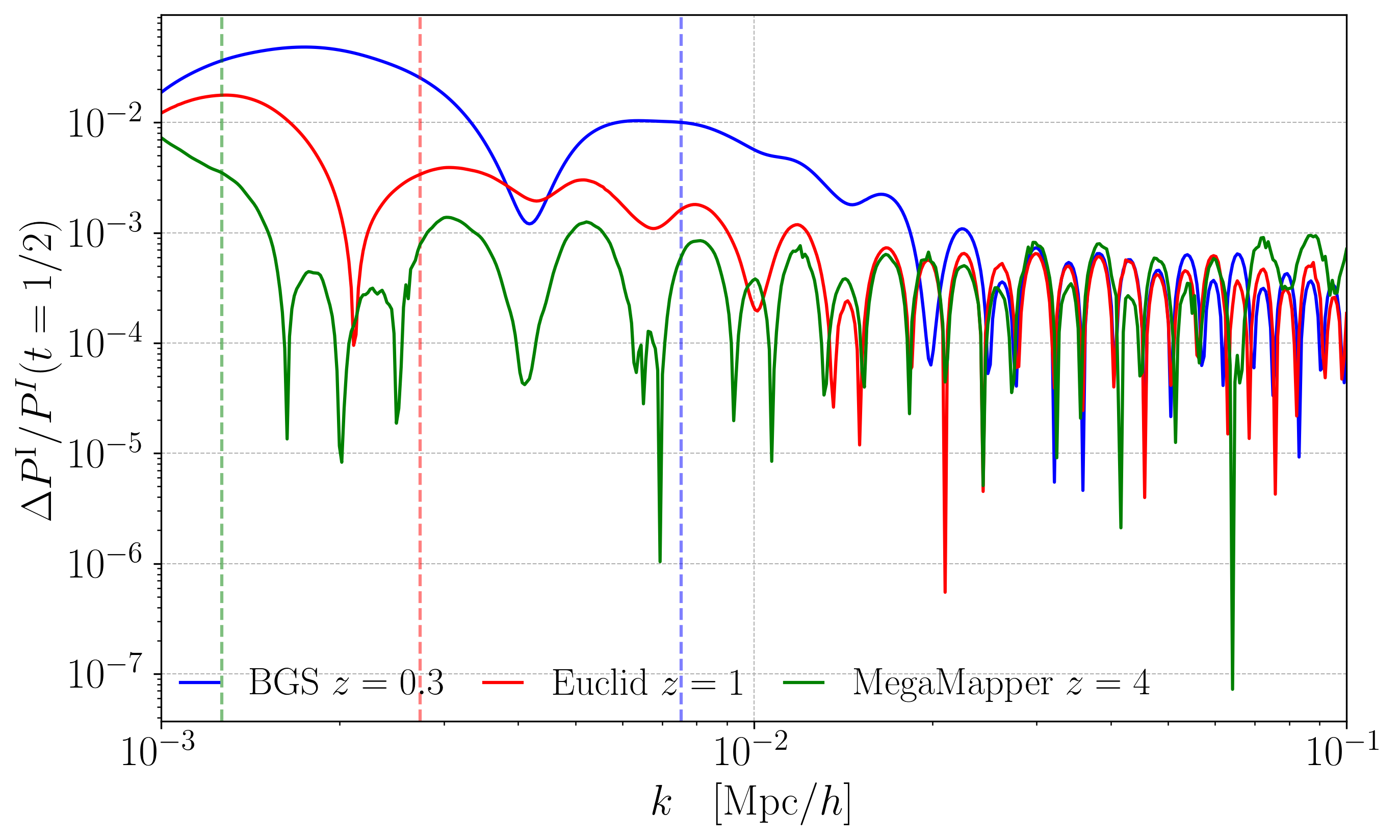} 
\caption{Fractional difference of the I\texttimes I contribution to the power spectrum monopole between $t=1/2$ and $t=0$, for three different surveys specifications. The dashed vertical lines represent the scale which corresponds to a separation such that $x_{12}/d=1$, for each survey. Thus, as we approach these scales, we expect the wide-separation expansion to break down.}
\centering
\label{fig:t_depend}
\end{figure}

Here, we reiterate the key points on the properties of the local power spectrum, and in particular the integrated contributions to it, on large scales:

\begin{enumerate}
\item Contributions that break translation invariance in our local power spectrum are $t$-dependent; these are both consequences of $\bm{d}$-dependence in the local power spectrum.
\item At zeroth order in our wide-separation $x_{12}/d$ expansion, local contributions to the local power spectrum are translation invariant; however, outside the Limber approximation, the integrated effects break translation invariance. Consequently, the non-diagonal parts (which are neglected in a Limber-type approximation) of the I\texttimes I term (and the non source-source correlation for the I\texttimes S term) can be considered a wide-separation type effect—analogous to the non-zeroth order terms in the $x_{12}/d$ expansion.
\item These wide-separation type corrections to the local power spectrum depend on the choice of LOS (they are $t$-dependent), and one can therefore observe their relevance by examining the $t$-dependence in the monopole of the local power spectrum (\cref{fig:t_depend}). In general, they scale as $(1/k \,d)^n$, behaving similarly to the local wide-separation terms.
\item Thus, outside both the Limber approximation and the plane-parallel, constant redshift limit, we have a $t$-dependent integrated local power spectrum. However, when one averages over the survey volume, the power spectrum monopole becomes independent of the LOS choice\footnote{While this may initially appear unconvincing, if one considers a correlation function that is simply a polynomial in $\bm{x}_1$ and $\bm{x}_2$ (e.g., $\xi(x_1,x_2) = x_1^2 \, x_2^2$, which has a finite expansion in $x_{12}/d$), one can see that the survey-averaged monopole is independent of $t$. This can be computed via pair counting: for each pair, the comoving distance $d$ at which we evaluate the pair is affected by our LOS choice. For an endpoint LOS, one samples slightly more extremal values of $d$ than in the midpoint case.} in the local 2-point function, as expected when considering the Yamamoto estimator. For higher multipoles, this is no longer the case due to the additional LOS dependence introduced when decomposing the multipoles.
\end{enumerate}

\subsection{Mixing in the Integrated terms beyond the Limber approximation}

As the off-diagonal parts of the integrand correlate two different redshifts, this breaks the $y_1 \leftrightarrow y_2$ symmetry and thus the integrated contribution is dependent on the LOS choice even in the plane-parallel constant redshift limit. Therefore, these terms are dependent on the LOS choice in our estimator. We can view this effect as mixing the multipoles for different choices of t.

If we return to \cref{eq:II_final}, we can express the I\texttimes I contribution as
\begin{equation}
P^{\rm I\times I}_{\rm loc}(k,\mu,d;t)=
\int_0^1\!\dd y_1\!\int_0^1\!\dd y_2\;
{\rm e}^{i\,\mu\,A_t(y_1,y_2)}
\,\mathcal{F}\!\left(\frac{k}{G_t},\mu;d,y_1,y_2\right),
\label{eq:Pint_generic}
\end{equation}
where
\begin{equation}
G_t(y_1,y_2)\equiv y_2+t(y_1-y_2),\qquad
A_t(y_1,y_2)\equiv (y_1-y_2)\,\frac{k\,d}{G_t(y_1,y_2)},
\end{equation}
and $\mathcal{F}$ denotes the remaining (non-exponential) factor
\[
\mathcal{F}\!\left(\frac{k}{G_t},\mu;d,y_1,y_2\right)
\equiv d^2\,G_t^{-3}\,
\mathcal{K}^{\rm I}\!\left(\frac{k}{G_t},\mu,d,y_1\right)
\mathcal{K}^{\rm I}\!\left(\frac{k}{G_t},-\mu,d,y_2\right)
P\!\left(\frac{k}{G_t}\right).
\]
The key point is that the $t$-dependence appears in \emph{both} the oscillatory factor ${\rm e}^{i\mu A_t}$ and in the rescaling $k\to k/G_t$ inside $\mathcal{F}$. This changes the full $\mu$-dependence of the integrand, so projecting onto Legendre multipoles necessarily mixes different $\ell$.

A simple way to see the mixing is to expand the oscillatory factor in Legendre polynomials,
\begin{equation}
{\rm e}^{i\mu A}
=\sum_{n=0}^\infty (2n+1)\,i^n\,j_n(A)\,\mathcal{L}_n(\mu),
\label{eq:plane_wave_legendre}
\end{equation}
and (for fixed $y_1,y_2$) expand the remaining $\mu$-dependence as
\begin{equation}
\mathcal{F}\!\left(\frac{k}{G_t},\mu;d,y_1,y_2\right)
=\sum_{\ell'} \mathcal{F}_{\ell'}\!\left(\frac{k}{G_t};d,y_1,y_2\right)\mathcal{L}_{\ell'}(\mu).
\end{equation}
Then the local multipoles are
\begin{equation}
P^{\rm I\times I}_{\ell,{\rm loc}}(k,d;t)
=\frac{2\ell+1}{2}\int_{-1}^{1}\dd\mu\;
P^{\rm I\times I}_{\rm loc}(k,\mu,d;t)\,\mathcal{L}_\ell(\mu),
\end{equation}
and inserting \cref{eq:Pint_generic}--\cref{eq:plane_wave_legendre} gives the schematic coupling
\begin{equation}
P^{\rm I\times I}_{\ell,{\rm loc}}(k,d;t)
=\int_0^1\!\dd y_1\!\int_0^1\!\dd y_2\;
\sum_{\ell',n}\;
\mathcal{M}_{\ell\ell'n}\!\left[A_t(y_1,y_2)\right]\;
\mathcal{F}_{\ell'}\!\left(\frac{k}{G_t};d,y_1,y_2\right),
\label{eq:mixing_generic}
\end{equation}
with mixing coefficients
\begin{equation}
\mathcal{M}_{\ell\ell'n}(A)
\equiv (2n+1)\,i^n\,j_n(A)\,
\frac{2\ell+1}{2}\int_{-1}^{1}\dd\mu\;
\mathcal{L}_\ell(\mu)\mathcal{L}_{\ell'}(\mu)\mathcal{L}_n(\mu).
\label{eq:Gaunt}
\end{equation}
The triple-Legendre integral in \cref{eq:Gaunt} enforces the usual selection rules (triangle conditions and $\ell+\ell'+n$ even), so the monopole at a given $t$ generically receives contributions from multiple even $\ell'$ once $n\ge 2$ is present. Since $A_t$ (and $k/G_t$) depends on $t$, the weights multiplying each $\ell'$ differ between $t=0$ and $t=1/2$.

This implies a non-trivial linear relation of the form
\begin{equation}
P^{\rm I\times I}_{\ell,{\rm loc}}(k,d;t=0)
=\sum_{\ell'} {T}_{\ell\ell'}(k,d)\;
P^{\rm I\times I}_{\ell',{\rm loc}}(k,d;t=\tfrac12),
\label{eq:mixing_matrix_statement}
\end{equation}
where ${T}_{\ell\ell'}$ is induced by \cref{eq:mixing_generic}--\cref{eq:Gaunt} and the $t$-dependence of $G_t$ and $A_t$, and is given by
\begin{align}
T_{\ell\ell'}(A_0,A_{1/2})
&= \frac{2\ell+1}{2}\int_{-1}^{1}\dd\mu\;
\mathcal{L}_{\ell}(\mu)\,\mathcal{L}_{\ell'}(\mu)\,
\exp\!\Big[i\mu\big(A_0-A_{1/2}\big)\Big]\nonumber\\ &
=(2\ell+1)\sum_{n}(2n+1)i^{\,n}j_n(A_0-A_{1/2})
\begin{pmatrix}\ell&\ell'&n\\0&0&0\end{pmatrix}^{\!2}\,.
\end{align}

In particular the monopole at $t=0$ contains (in general) pieces that would project onto $\ell'=2,4,\ldots$ at $t=1/2$, and vice versa. Note that the non-integrated limit is equivalent to A=0 which implies $n=0$ and $\ell'=\ell$.\footnote{Starting from \cref{eq:II_final}, compare the endpoint and midpoint choices by factorising the phase in the
integrand,
\begin{equation}
{\rm e}^{i\mu A_{0}(y_1,y_2)}={\rm e}^{i\mu A_{1/2}(y_1,y_2)}\,{\rm e}^{i\mu\Delta A(y_1,y_2)},
\qquad
\Delta A\equiv A_{0}-A_{1/2},
\label{eq:phase_factorisation}
\end{equation}

where \(A_t\) is defined below \cref{eq:II_final}. Inserting \cref{eq:phase_factorisation} into \cref{eq:II_final}, and
using the plane-wave/Legendre expansion \cref{eq:plane_wave_legendre} for the midpoint factor,
\begin{equation}
{\rm e}^{i\mu A_{1/2}}=\sum_{n=0}^{\infty}(2n+1)\,i^{\,n}\,j_n(A_{1/2})\,\mathcal{L}_n(\mu),
\end{equation}
together with a Legendre expansion of the remaining bracket
\begin{equation}
{\rm e}^{i\mu\Delta A}\,\mathcal{F}_0(\mu)
=\sum_{\ell'}\Big[{\rm e}^{i\mu\Delta A}\mathcal{F}_0\Big]_{\ell'}\,\mathcal{L}_{\ell'}(\mu),
\label{eq:bracket_legendre_F7}
\end{equation}
(where \(\mathcal{F}_0\) denotes the non-exponential part of the $t=0$ integrand in \cref{eq:II_final}), the multipole
projection gives a coupled sum over \(\ell'\) and \(n\) involving the Gaunt integral
\cref{eq:Gaunt}. This yields the mixing kernel
\begin{equation}
P^{\rm I\times I}_{\ell,{\rm loc}}(k,d;0)
=\int_0^1\!\dd y_1\!\int_0^1\!\dd y_2\;
\sum_{\ell'} T_{\ell\ell'}\!\big(A_{1/2}(y_1,y_2)\big)\,
\Big[{\rm e}^{i\mu\Delta A}\mathcal{F}_0\Big]_{\ell'}(k,d;y_1,y_2),
\label{eq:mixing_from_F7}
\end{equation}
with
\begin{equation}
T_{\ell\ell'}(A)\equiv(2\ell+1)\sum_{n=0}^{\infty}(2n+1)\,i^{\,n}\,j_n(A)\,
\begin{pmatrix}\ell&\ell'&n\\0&0&0\end{pmatrix}^{\!2}.
\label{eq:Tll_def}
\end{equation}
For non-integrated local terms there is no $t$-dependent phase analogous to \(A_t\) in \cref{eq:II_final} (so
effectively \(\Delta A=0\)); since \(j_n(0)=\delta_{n0}\), \cref{eq:Tll_def} reduces to \(T_{\ell\ell'}=\delta_{\ell\ell'}\),
i.e.\ no multipole mixing.}

Finally, the special role of $t=\tfrac12$ is that for single-tracer auto-correlations the I$\times$I integrand has a $y_1\leftrightarrow y_2$ symmetry which makes the local power spectrum even in $\mu$ at leading order (so odd multipoles vanish), whereas this symmetry is broken for endpoint choices (e.g.\ $t=0$ where $G_t=y_2$), allowing an odd/imaginary part to appear. This is the same origin as the $t$-dependent odd-parity signal discussed in \cref{sec:los_depend}: it is a wide-separation type effect caused by the breaking of local translation invariance by the integrated contributions, rather than by the usual polynomial $\mu$-structure of the local terms.

\section{Properties of the Integrands}\label{ap:numerics}

\begin{figure}
\centering
\includegraphics[width=0.6\linewidth]{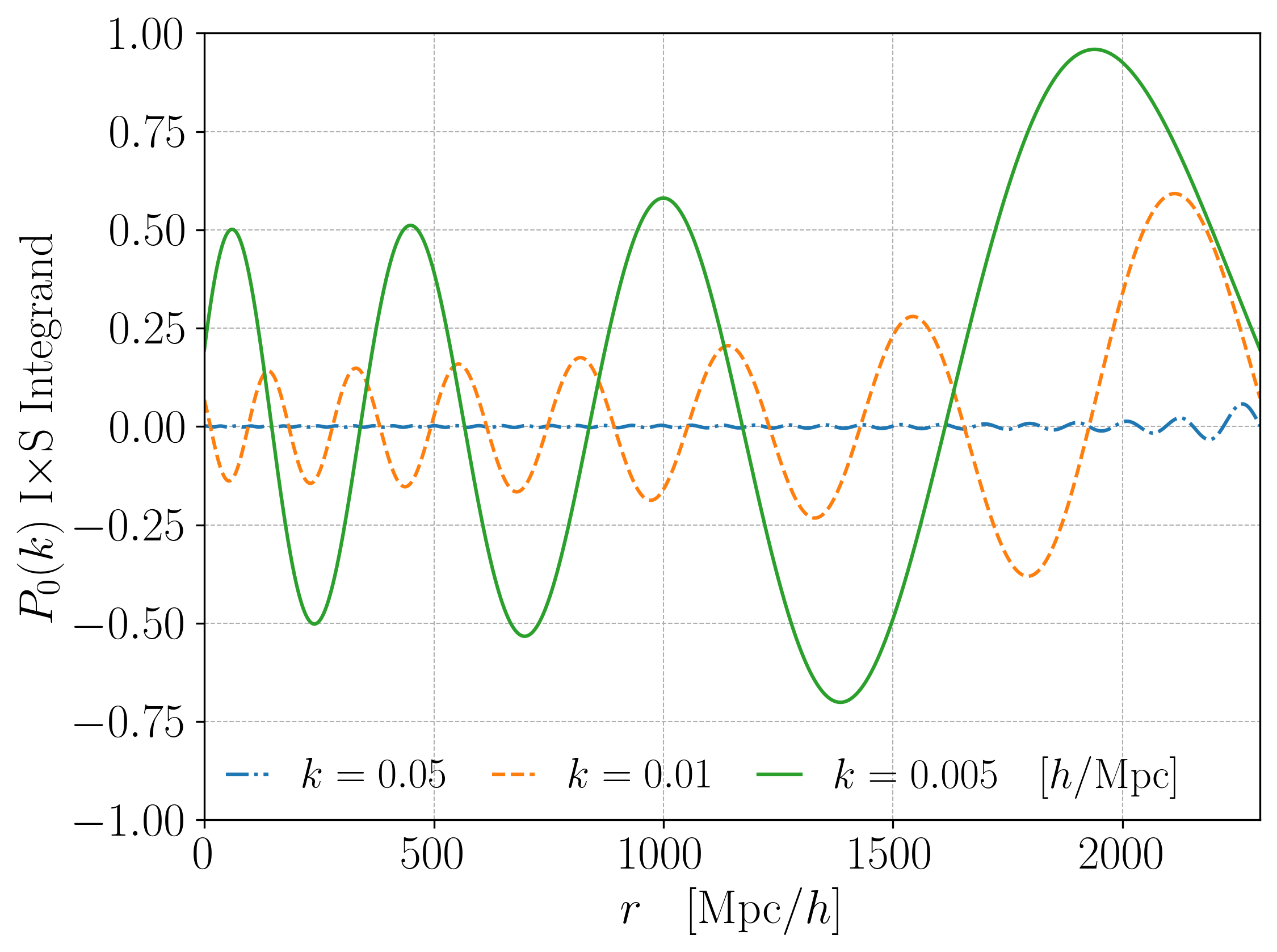} 
\caption{Integrand of the I\texttimes S term from observer to source at $z=1$ for a \textit{Euclid}-like H$\alpha$ survey for a few different $k$-values, for $t=1/2$.}
\centering
\label{fig:IS_integrand}
\end{figure}

The I\texttimes S term, \cref{ap:local_IS}, is computed by a oscillatory 1D integrand, \cref{fig:IS_integrand}, and, ignoring the $\mu$ dependence in the kernels, has the general form of $\sinc (k \frac{r-d}{d+r})$, weighted by the $r$ dependence in the integrated kernels. Therefore, as this behaves like a sine integral, the I \texttimes S term is largely determined by its behaviour close to the source as the correlation of the source density fields with the density fields along the path largely cancel. Therefore, the value of the integral, for a fixed comoving distance $d$, is dependent of the frequency of the sine wave, set by the $k$ scale; thus one can see that this produces the characteristic wiggles in the I\texttimes S contribution when plotted as a function of $k$, \cref{fig:mono_quad}. 

Alternatively, one can consider the I\texttimes S, \cref{eq:IS_final} (or I\texttimes I - \cref{eq:II_final}) contribution as a Fourier transform over $y$ ($y_1$ and $y_2$) with a sharp window function at the source and observer. Therefore, the wiggles are a Gibbs phenomenon due to sharp cut-off in the integrand.

These oscillations in $k$ are damped in the lensing contribution, as the real part of the lensing kernel peaks halfway between the observer and source, due to the $(d-r)(d)$ dependence, and therefore the sharp cut-offs are partially smoothed. The imaginary part however just has a $(d-r)$ dependence and therefore has a sharp cut-off at the observer. Furthermore, as the lensing kernel is comparatively small close to the source, this therefore reduces the relative size of the lensing contribution to the I\texttimes S term. 

\begin{figure}[h]
\centering
\begin{subfigure}{\linewidth}
    \centering \includegraphics[width=0.49\linewidth]{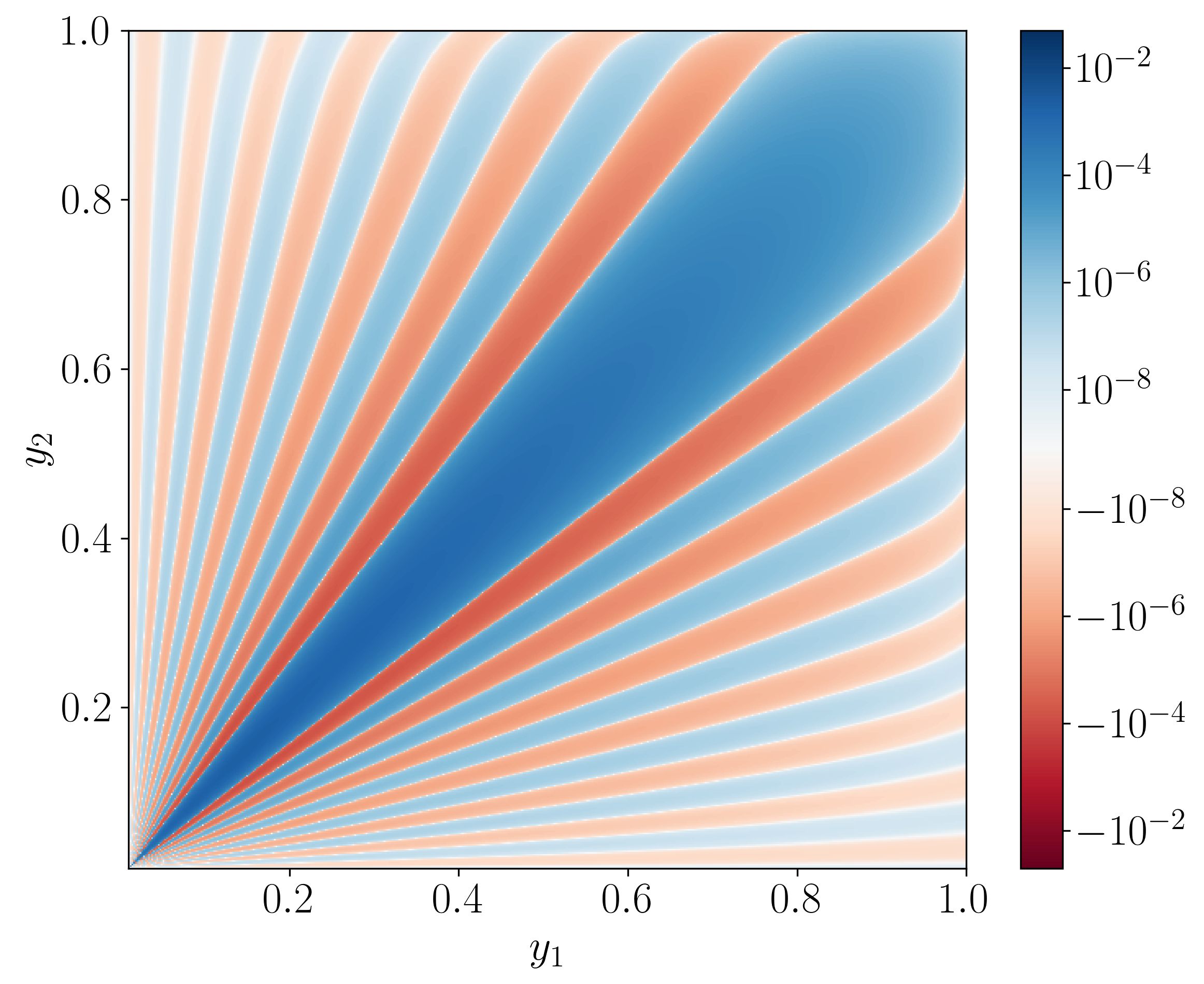}
    \includegraphics[width=0.49\linewidth]{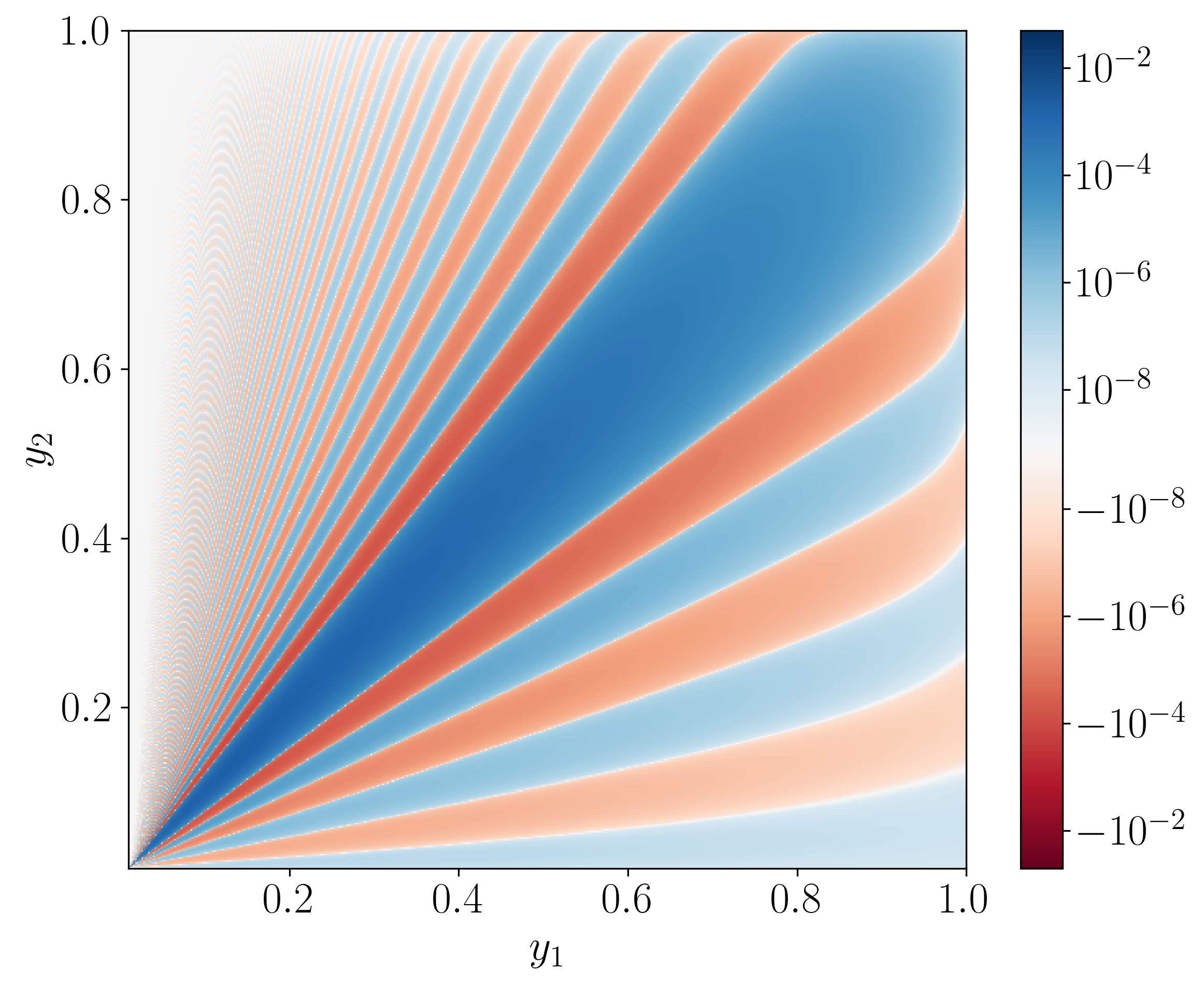}
\end{subfigure}
\caption{Real single tracer integrand of the monopole for the I\texttimes I contribution with $t=1/2$ (left) and $t=0$ (right) for a \textit{Euclid}-like H$\alpha$ survey for $k=0.01$ at $z=1.5$. Bottom left corner, $y_1=y_2=0$, represents the observer and top right, $y_1=y_2=1$, represents the source.}
\label{fig:II_integrand_ream}
\end{figure}

\begin{figure}[h]
\centering
\begin{subfigure}{\linewidth}
    \centering \includegraphics[width=0.49\linewidth]{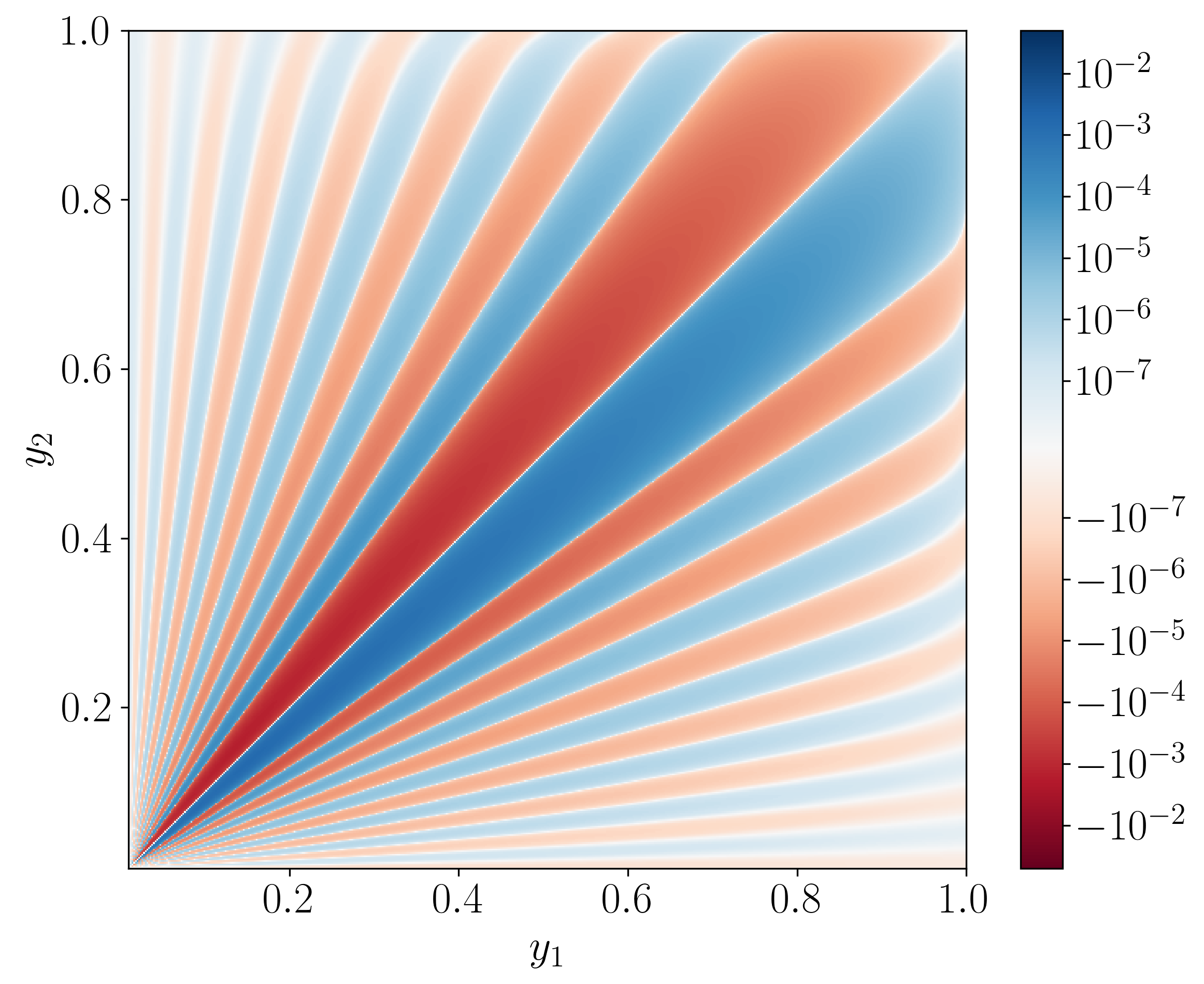}
    \includegraphics[width=0.49\linewidth]{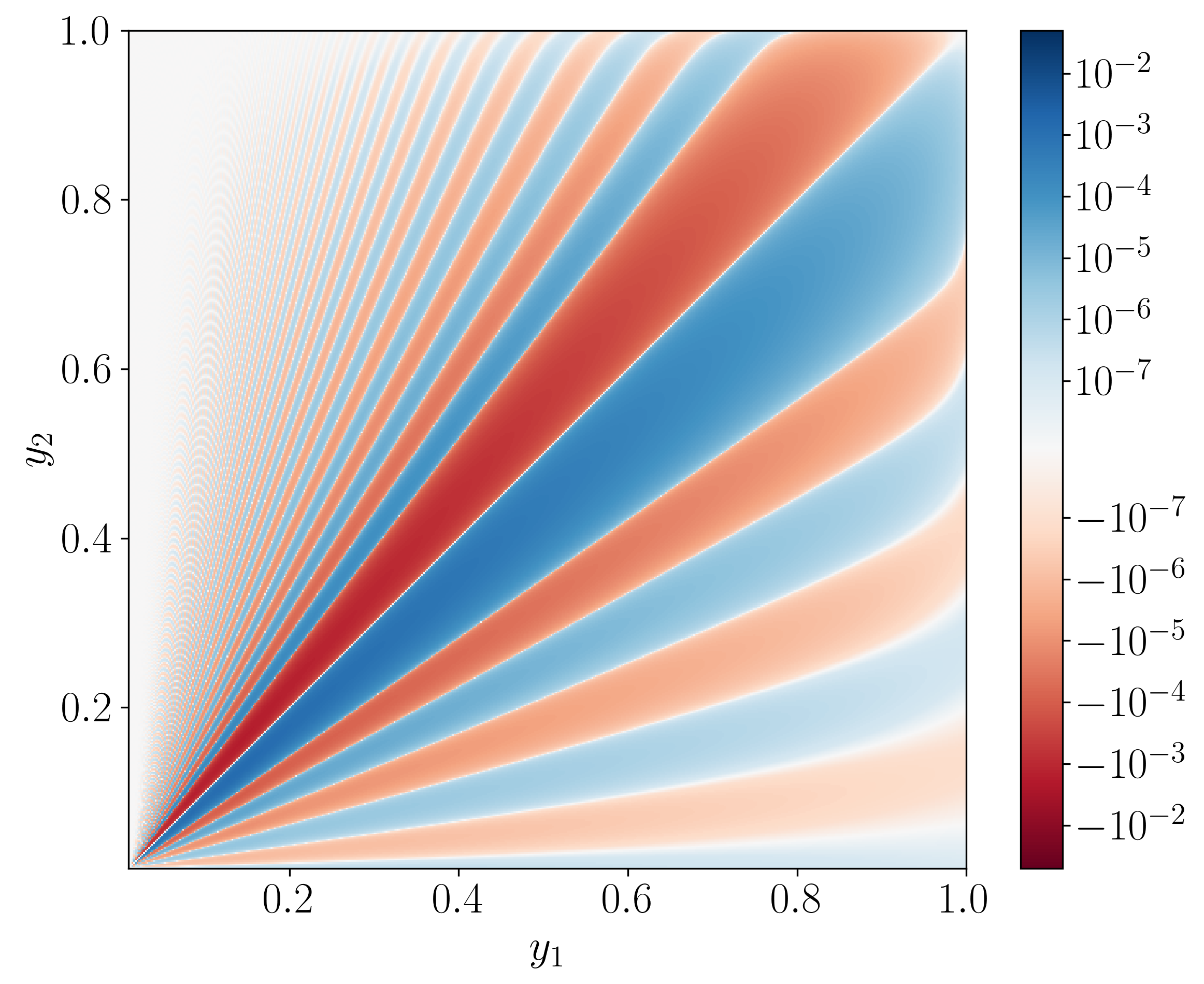}
\end{subfigure}
\caption{Imaginary single tracer integrand of the dipole for the I\texttimes I contribution with $t=1/2$ (left) and $t=0$ (right) for a \textit{Euclid}-like H$\alpha$ survey for $k=0.01$ at $z=1.5$. Bottom left corner represents the observer and top right represents the source.}
\label{fig:II_integrand_imag}
\end{figure}

The single tracer 2D I\texttimes I integrand, \cref{fig:II_integrand_ream}, is symmetric (anti-symmetric) for the even (odd) multipoles about $y_1=y_2$ for $t=1/2$ but this symmetry is broken for other LOS choices. This, therefore generates the imaginary dipole signal for an endpoint LOS as in the Yamamoto estimator, \cref{eq:yamamoto}. Similarly, for the I\texttimes S term, the two integrals (IS and SI) only cancel in the single tracer odd parity multipoles for $t=1/2$. However, when we consider the cross-power spectrum of two different tracers, these symmetries are broken just like the local relativistic contributions case.

\subsection{Convergence and Divergence and Computation Time}

\begin{figure}
\centering
\includegraphics[width=0.7\textwidth]{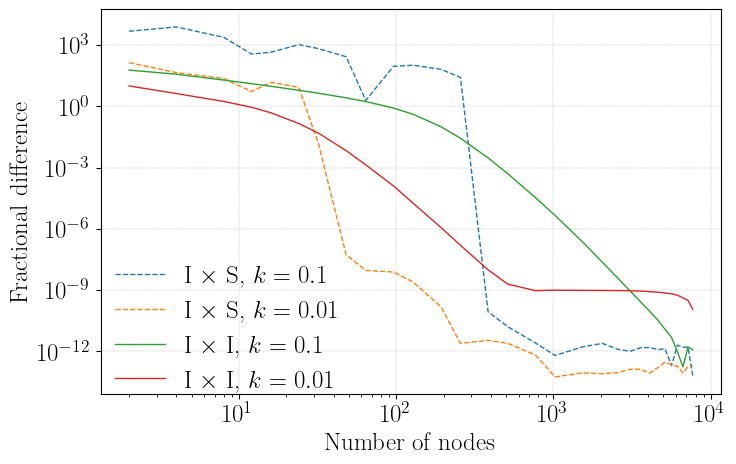}
\caption{Convergence of the power spectrum monopole ($t=1/2$) for different numbers of Legendre-Gauss nodes compared to a reference value of $n=8192$ for two different $k$-values, plotted for a \textit{Euclid}-like H$\alpha$ survey at $z=1.5$.}\label{fig:convergence}
\end{figure}

For the I\texttimes S (\cref{eq:IS_final}) and I\texttimes I (\cref{eq:II_final}) contributions, we can do the $\mu$ integration analytically, which leaves a single integral over $r$ and a double integral over $r_1, \, r_2$ to compute, respectively. To implement this in \textsc{CosmoWAP} \href{https://github.com/craddis1/CosmoWAP}{\faGithub}, we use Gauss-Legendre quadrature. The convergence of these integrals for different number of Gauss-Legendre nodes, $n$, is shown in \cref{fig:convergence}. These integrals converge slower for larger $k$ values, as the frequency of the oscillations in $r_1, \, r_2$ increase with $k$, and therefore, one needs more samples to accurately resolve these oscillations. For our forecasts, we use $n=256$ as, at that number of nodes, the I\texttimes I term has converged to percentage level at $k=0.1 \, \aMpch$. Beyond these scales, while the contribution is not as well converged, there is limited impact on our forecasts, and indeed, we have verified that for our forecasts on $\fNL$, the change in constraints and bias from $n=256$ to $n=512$ is sub-percentage level.

Here, therefore the computation time for the I\texttimes I term scales as $n^2 \, n_k$ where $n_k$ is the number of $k$ modes. In the single tracer, for $t=1/2$, this is halved due to the aforementioned symmetry in the integrand. For the integrated contribution to the covariance, it is however simpler to work with the $\mu$--dependent expressions such that this scales as $n^2 \, n_k \, n_{\mu}$. 

\subsubsection{Speeding up the Integral Computation}

We can however compute these integrals more efficiently if we more closely consider the $y_1, \, y_2, \, k \text{ and } \mu$ dependences. In particular, in the case where we consider an endpoint LOS we can reduce the amount of required computation for a given analysis. For example, if we consider the I\texttimes I contribution \cref{eq:II_final} for $t=0$ (thus $G(y_1,y_2)=y_2$) then it is useful to compute the following integral
\begin{equation}
\begin{split}\label{eq:I_integral}
I_{mn}(p) = 
\int_0^{1} \dd y_1 \, {\rm e}^{i\,\left(y_1 \right)\,p}\, d \,\mathcal{Z}^{\rm I}_{mn}(d,y_1),
\end{split}
\end{equation}
where $\mathcal{Z}^{\rm I}_{mn}(d,y_1)$ are the coefficients for the given $\mu$ and $k$ dependences in an integrated kernel $\mathcal{K}^{\text{I}}(k,\mu,d,r)$
\begin{equation}
    \mathcal{K}(k,\mu,d,r) = \sum_n \sum_m k^n \mu^m \mathcal{Z}^{\rm I}_{mn}(\mu,d,y_1).
\end{equation}
The $y_1$ integral can be written as 
\begin{equation}
    \int_0^{1} \dd y_1 \, {\rm e}^{ i \, y_1\,k\,\mu\, d/y_2 } \, d\, \mathcal{K}(k/y_2,\mu,d,r) =  \sum_n \sum_m (k/y_2)^n \mu^m I_{m n}(p)
\end{equation}
where we define $p = k\,\mu\, d/y_2$. Interpolating this function there allows us to reduce the dimensions of the integral we need to compute. So therefore the I\texttimes I integral involves computing
\begin{equation}
\begin{split}
P^{\rm I \times I}_{\text{loc}}(k,\mu, d) = {\rm e}^{ -i k \mu d }  \sum_n \sum_m \mu^m\int_0^{1} \dd y_2 \, (k/y_2)^n &\, d\, \, y_2^{-3}\mathcal{K}^{\rm I}(k/y_2,-\mu,d,r_2)\\ & \quad \, \times P (k/y_2) I_{mn}(k \mu /y_2).
\end{split}
\end{equation}
This $r_2$ integral is quite oscillatory close to the source and therefore we find Filon quadrature to be useful. Note that for \cref{eq:I_integral}, we compute this oscillatory integral over a large domain of $p$ and therefore we found fitting the simple $y_1$ dependence in the kernel coefficients, $\mathcal{Z}^{\rm I}_{mn}(\mu,d,y_1)$, with a polynomial and then using integration by parts to compute the integral analytically, achieves good results. Similar principles can be implemented for the I\texttimes S method. We keep things more flexible by doing the $\mu$ integral numerically which allows us to compute the full $\mu$ dependent power spectrum once for each redshift bin from which we can trivially obtain the multipoles. This gives $\mathcal{O}(10)$ speed up in computation time compared to the initial analytical $\mu$ integration and Gauss-Legendre quadrature, however there is far greater speed up in the computation of the covariances, with a speed up of $\mathcal{O}(1000)$ or greater for an analysis similar to the ones considered here.

\section{ISW, Time Delay and Lensing Contributions}\label{ap:components}

\begin{figure}
\centering
\includegraphics[width=1\textwidth]{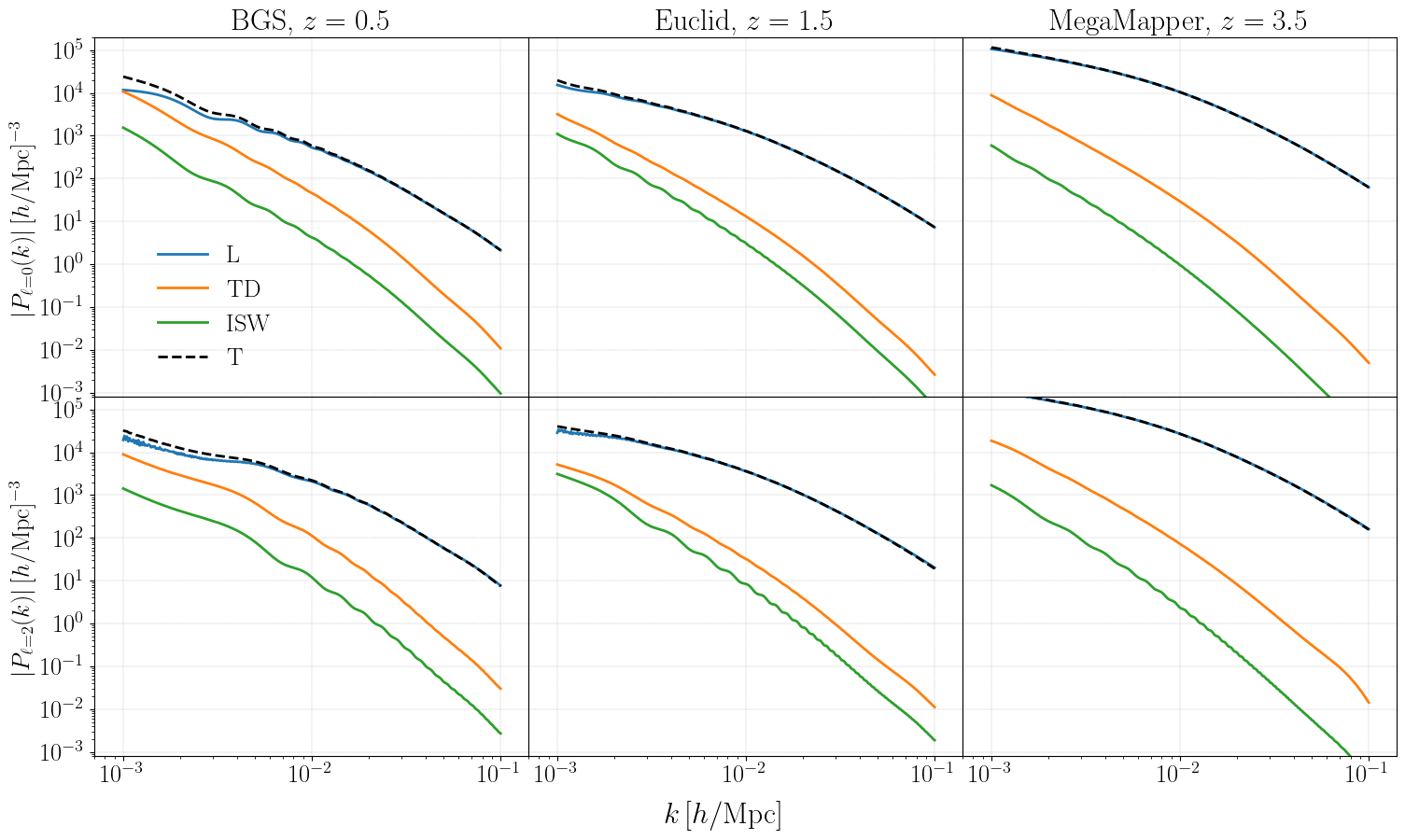} 
\caption{Integrated component contributions of the lensing (L), time delay (TD) and ISW effects to the monopole and quadrupole of the power spectrum. Each contribution contains all the mixing terms with the previous term -- e.g. TD contains the mixing terms TD\texttimes S and TD\texttimes L.}
\label{fig:even_components}
\end{figure}

\begin{figure}
\centering
\includegraphics[width=0.8\textwidth]{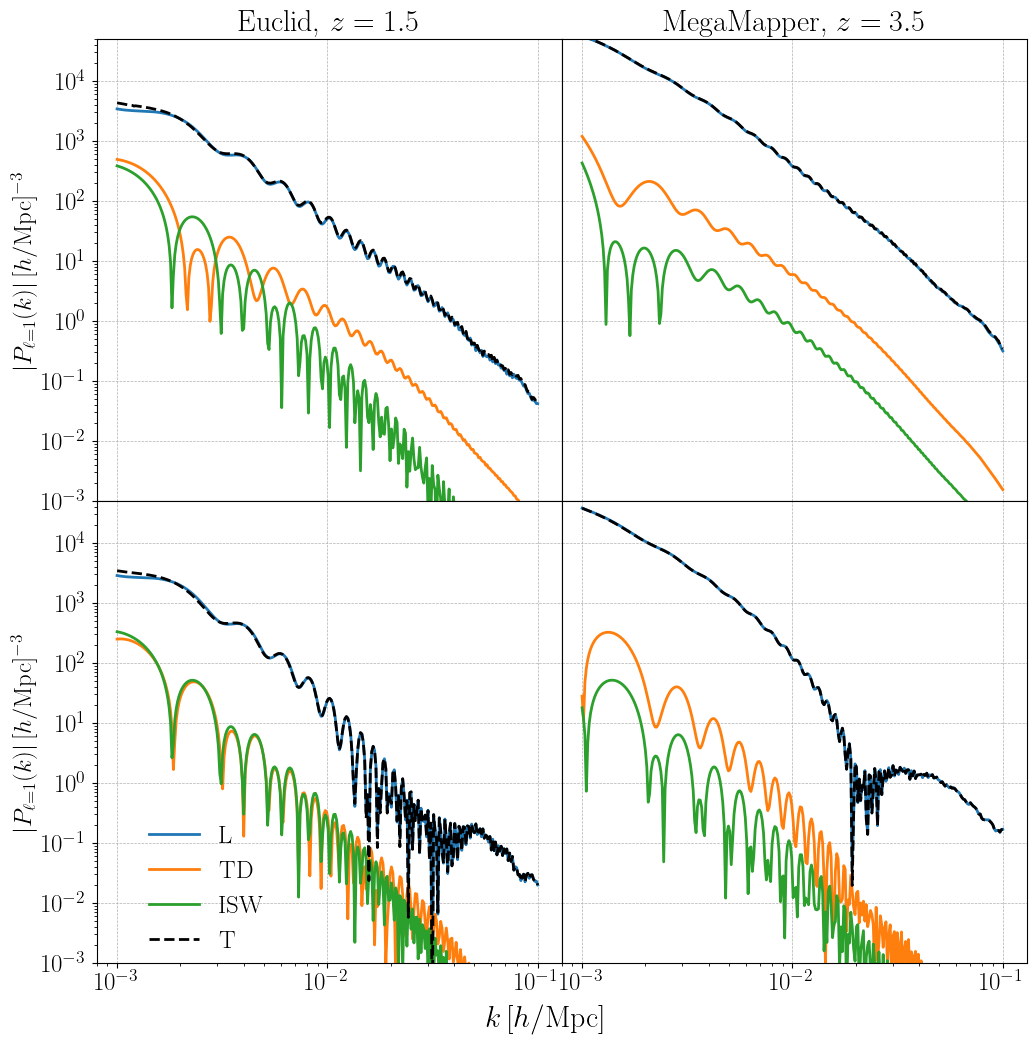}
\caption{Imaginary integrated component contributions to the dipole of the power spectrum for lensing (L), time delay (TD) and ISW effects for \textit{Euclid} and MegaMapper-like surveys, in the single tracer case (bottom panel) and for bright-faint split cross-power spectra (top panel). Each contribution contains all the mixing terms with the previous term -- e.g. TD contains the mixing terms TD\texttimes S and TD\texttimes L.}\label{fig:dipole_components}
\end{figure}

\cref{fig:even_components} and \cref{fig:dipole_components} show the lensing, time delay and ISW contributions separately, for the even and odd contributions to the power spectrum, respectively. Lensing is the dominant integrated effect, particularly at higher redshifts. At smaller redshifts (and particularly for the quadrupole), the imaginary part of the lensing kernel, which has a $(d-r)$ dependence, is more important, and this generates the oscillations in the contribution from the sharp cut-off in the Fourier transform at the observer. The time delay and ISW are subdominant to the lensing contribution but are comparatively larger at smaller redshifts and for smaller $k$ -- due to their additional $1/q^2$ dependence. The ISW contribution in particular peaks at very low redshifts as it sensitive to the late-time acceleration in expansion. 

\begin{figure}[ht]
\centering
\includegraphics[width=0.49\linewidth]{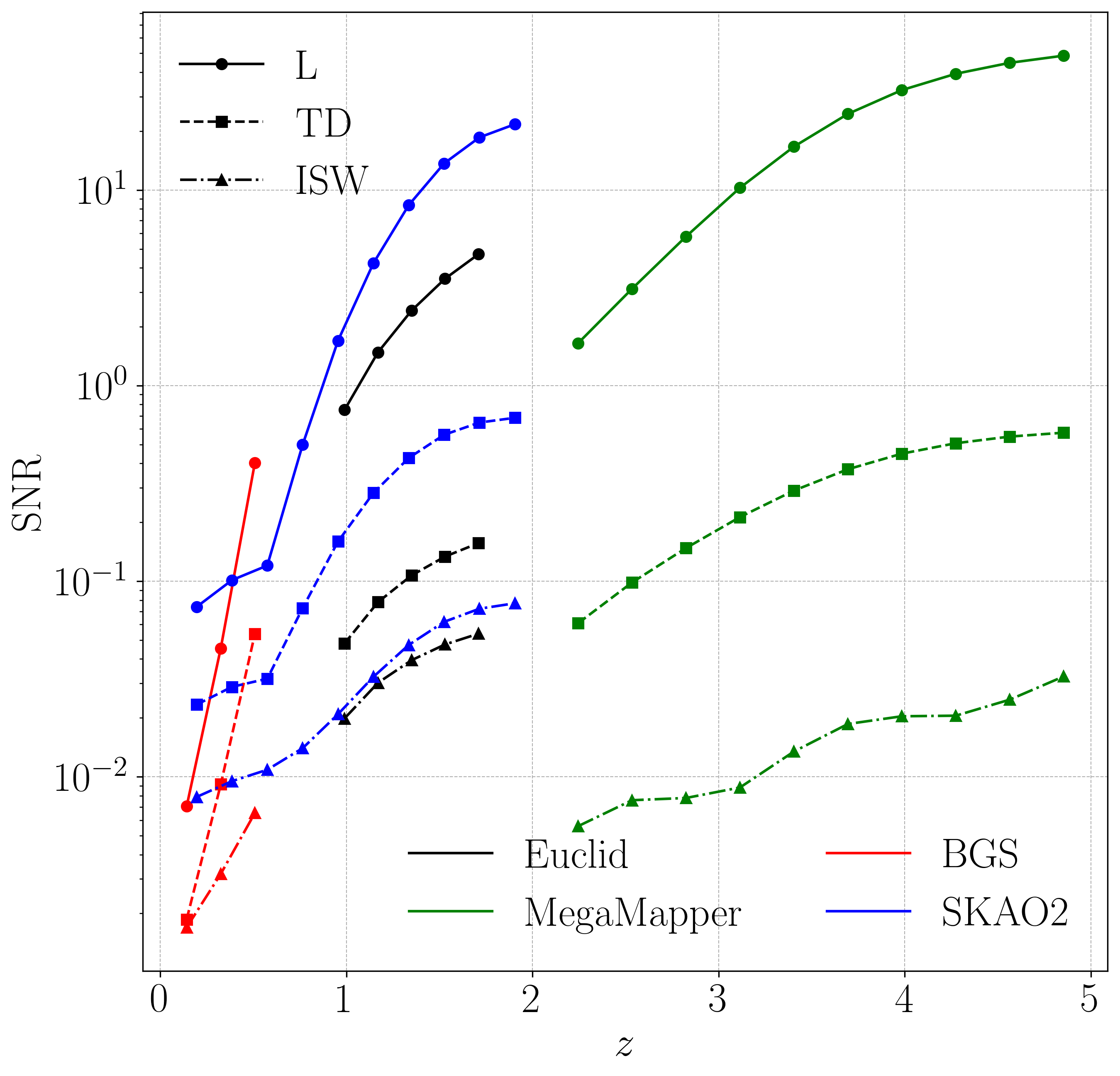}
\includegraphics[width=0.49\linewidth]{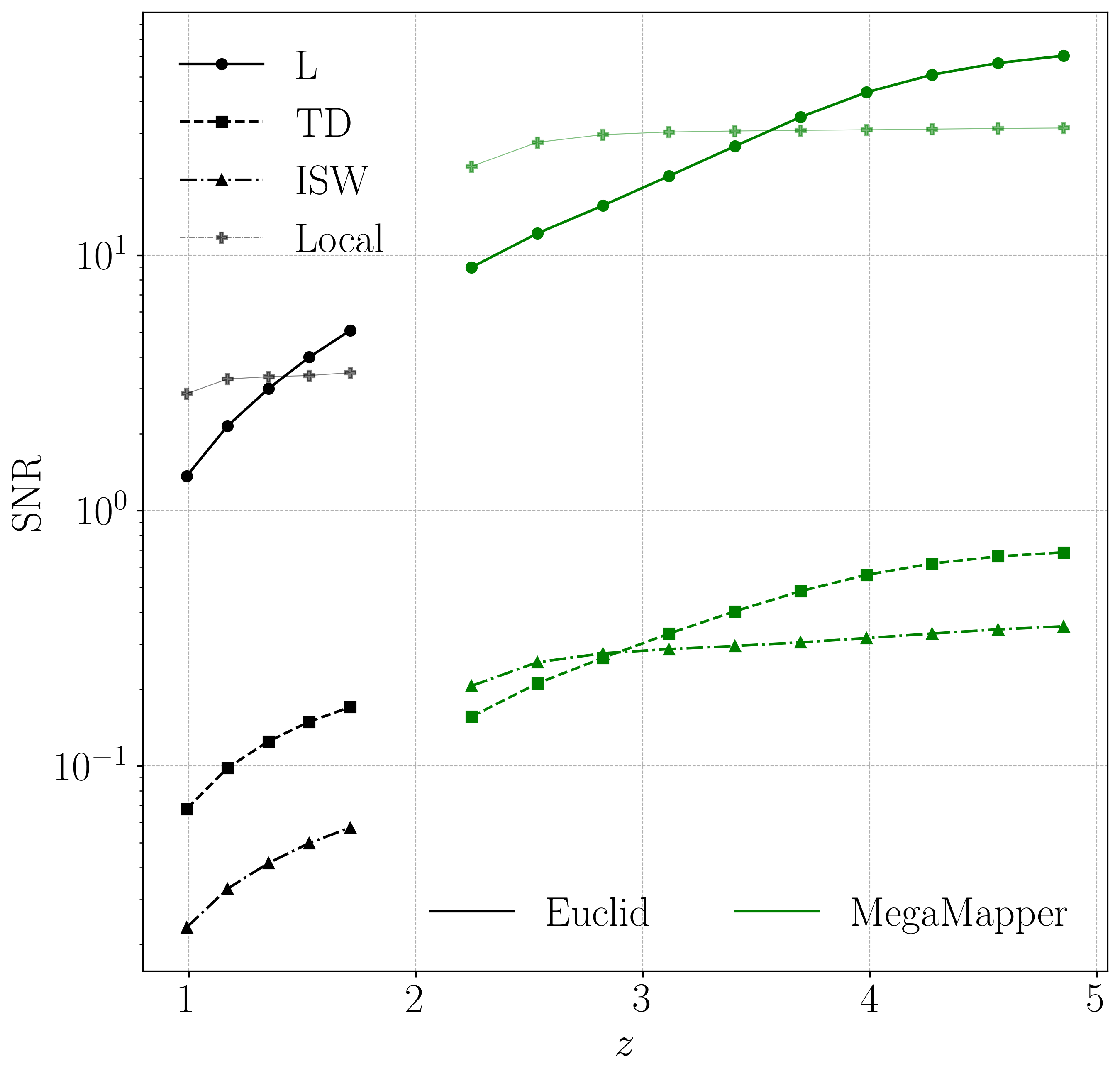}
\caption{Cumulative SNRs for a single-tracer analysis (left) and a bright-faint split approach (right), for the different integrated terms, plotted over redshift bins. The SNR of the local relativistic term is included for reference in the multi-tracer case. Note that we do not marginalise over any parameters for these SNRs.}
\label{fig:SNRs}
\end{figure}

The cumulative signal-to-noise ratio (SNR) is plotted for each integrated component term for a single-tracer (left) and multi-tracer (right) analysis in \cref{fig:SNRs}. We forecast the lensing term should be detectable for \textit{Euclid}, SKAO2 and MegaMapper in the single-tracer case and indeed, there is limited improvement in the multi-tracer case. However, the effects from time delay and ISW are unlikely to be detectable in the 3D matter power spectrum. Some constraining power is lost on large scales and at low redshifts due to the $k < 2 \pi/x(z_{\rm min})$ cut we impose due to the perturbative wide-separation expansion breaking down.

\section{Additional Plots}\label{ap:more_plots}

\begin{figure}[ht]
\centering
\includegraphics[width=1\textwidth]{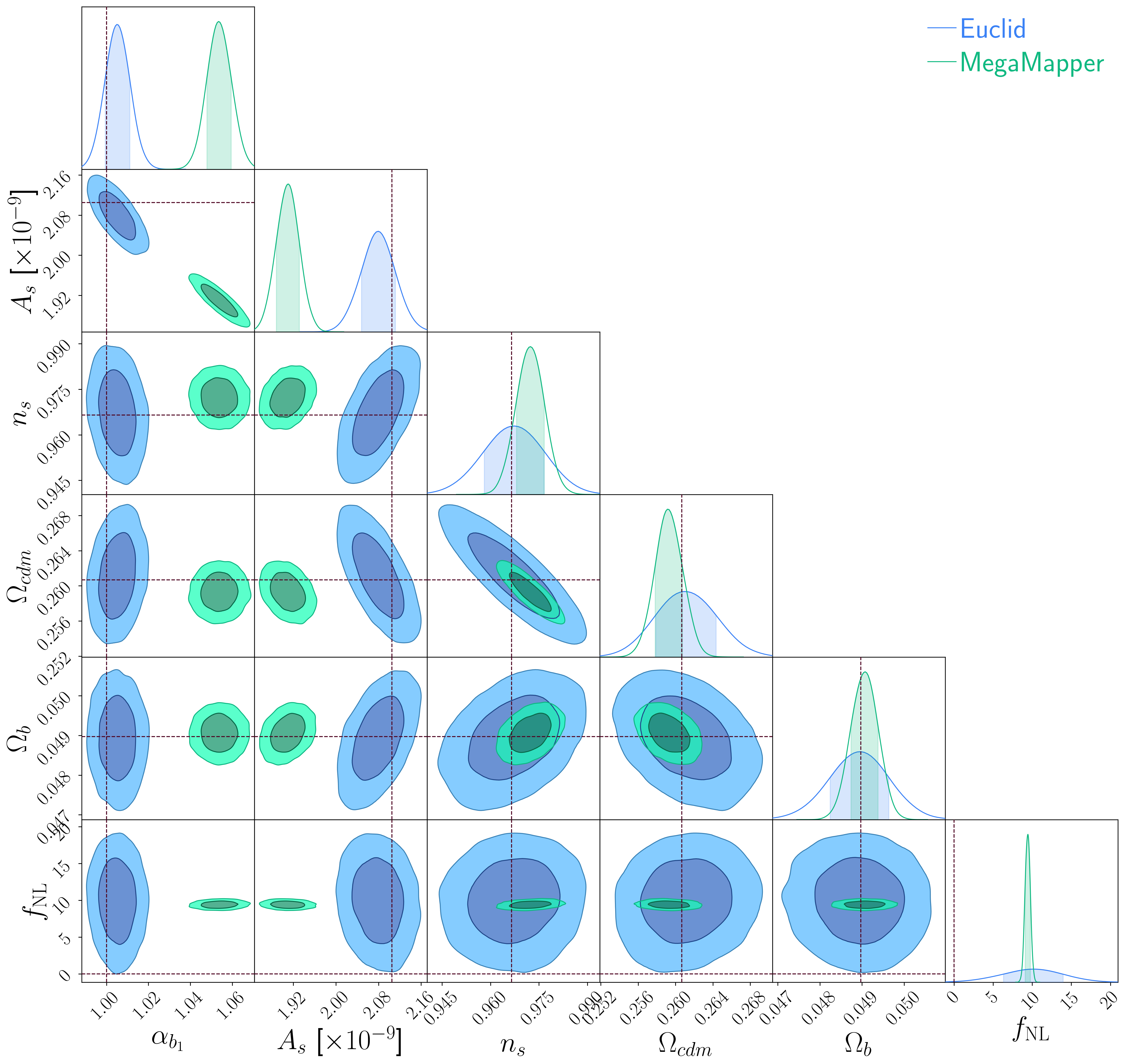} 
\caption{Forecasted marginalised and joint parameter constraints for a \textit{Euclid}-like H$\alpha$ survey (blue) and a MegaMapper-like LBG survey (green), if we neglect integrated, local relativistic and wide-separation effects in our analysis. This is equivalent to \cref{fig:full_MCMC} but without the assumption of cosmological priors from \textit{Planck}. Solid colour representing the $1\sigma$ constraints and the lighter region denoting the $2\sigma$ constraints.}
\label{fig:full_MCMC_noprior}
\end{figure}



\clearpage

\bibliographystyle{jhep}
\bibliography{main}

\end{document}